\documentclass[twocolumn]{aastex63}

\usepackage{amsmath}
\usepackage{amssymb}
\usepackage{hyperref}
\usepackage{CJK}
\usepackage{soul}
\newcommand\sersic{S\'ersic}
\newcommand\gamornet{G\textsc{a}M\textsc{or}N\textsc{et}}

\shorttitle{Machine Learning Framework to Determine AGN Host Galaxy Morphologies}
\shortauthors{Tian et al.}

\let\OLDthebibliography\thebibliography
\renewcommand\thebibliography[1]{
  \OLDthebibliography{#1}
  \setlength{\parskip}{0pt}
  \setlength{\itemsep}{0pt plus 0.3ex}
}

\begin{document}
\begin{CJK*}{UTF8}{gbsn}

\title{Using Machine Learning to Determine Morphologies of $z<1$ AGN Host Galaxies in the Hyper Suprime-Cam Wide Survey}

\author[0000-0003-4056-7071]{Chuan Tian (田川)}
\affiliation{Department of Physics, Yale University, New Haven, CT, USA}
\email{chuan.tian@yale.edu; ufsccosmos@gmail.com}

\author[0000-0002-0745-9792]{C. Megan Urry}
\affiliation{Department of Physics, Yale University, New Haven, CT, USA}
\affiliation{Department of Astronomy, Yale University, New Haven, CT, USA}
\affiliation{Yale Center for Astronomy and Astrophysics, Yale University, New Haven, CT, USA}

\author[0000-0002-2525-9647]{Aritra Ghosh}
\affiliation{Department of Astronomy, Yale University, New Haven, CT, USA}
\affiliation{Yale Center for Astronomy and Astrophysics, Yale University, New Haven, CT, USA}

\author{Ryan Ofman}
\affiliation{Department of Astronomy, Yale University, New Haven, CT, USA}

\author[0000-0001-8211-3807]{Tonima Tasnim Ananna}
\affiliation{Department of Physics and Astronomy, Dartmouth College, 6127 Wilder Laboratory, Hanover, NH, USA}

\author[0000-0002-5504-8752]{Connor Auge}
\affiliation{
Institute for Astronomy, University of Hawaii, Honolulu, HI, USA}

\author[0000-0002-1697-186X]{Nico Cappelluti}
\affiliation{Department of Physics, University of Miami, Coral Gables, FL, USA}
\affiliation{INAF - Osservatorio di Astrofisica e Scienza dello Spazio di Bologna, Bologna, Italy}

\author[0000-0003-2284-8603]{Meredith C. Powell}
\affil{Kavli Institute for Particle Astrophysics and Cosmology, Stanford University, Stanford, CA, USA}

\author[0000-0002-1233-9998]{David B. Sanders}
\affiliation{
Institute for Astronomy, University of Hawaii, Honolulu, HI, USA}

\author[0000-0001-5464-0888]{Kevin Schawinski}
\affiliation{Modulos AG, Technoparkstr. 1, CH-8005, Zurich, Switzerland}

\author{Dominic Stark}
\affiliation{Modulos AG, Technoparkstr. 1, CH-8005, Zurich, Switzerland}

\author[0000-0002-5445-5401]{Grant R. Tremblay}
\affiliation{Center for Astrophysics \textbar Harvard \& Smithsonian, 60 Gardent St., Cambridge, MA 02138, USA}

\begin{abstract}
We present a machine-learning framework to accurately characterize morphologies of Active Galactic Nucleus (AGN) host galaxies within $z<1$. 
We first use PSFGAN to decouple host galaxy light from the central point source, then we invoke the Galaxy Morphology Network (\gamornet{}) to estimate whether the host galaxy is disk-dominated, bulge-dominated, or indeterminate. 
Using optical images from five bands of the HSC Wide Survey, we build models independently in three redshift bins: low ($0<z<0.25$), medium ($0.25<z<0.5$), and high ($0.5<z<1.0$).
By first training on a large number of simulated galaxies, then fine-tuning using far fewer classified real galaxies, our framework predicts the actual morphology for $\sim$ 60\% - 70\% host galaxies from test sets, with a classification precision of $\sim$ 80\% - 95\%, depending on redshift bin.
Specifically, our models achieve disk precision of 96\%/82\%/79\% and bulge precision of 90\%/90\%/80\% (for the 3 redshift bins), at thresholds corresponding to indeterminate fractions of 30\%/43\%/42\%.
The classification precision of our models has a noticeable dependency on host galaxy radius and magnitude. 
No strong dependency is observed on contrast ratio.
Comparing classifications of real AGNs, our models agree well with traditional 2D fitting with GALFIT. 
The PSFGAN+GaMorNet framework does not depend on the choice of fitting functions or galaxy-related input parameters, runs orders of magnitude faster than GALFIT, and is easily generalizable via transfer learning, making it an ideal tool for studying AGN host galaxy morphology in forthcoming large imaging survey.
\end{abstract}


\section{Introduction}
\label{section:introduction} 
Two decades ago, it was suggested that AGN feedback could play a pivotal role in galaxy evolution, based on observed correlations between the mass of the central supermassive black hole (SMBH) and the stellar velocity dispersion in the host galaxy, or with the host galaxy mass, luminosity and other properties \citep{2000ApJ...539L...9F, 2000ApJ...539L..13G, 2013ARA&A..51..511K}.
The impact of an AGN on its host galaxy has also been incorporated into galaxy evolution models to account for the inefficiency of star formation in massive galaxies  \citep[e.g.][]{2005Natur.433..604D, 2006MNRAS.365...11C, 10.1111/j.1745-3933.2006.00234.x, 2007ApJ...654..731H, 10.1111/j.1745-3933.2008.00430.x, 2017NatAs...1E.165H}. 
AGN emits radiation across the electromagnetic spectrum as matter accretes onto the SMBH, which can effectively heat or expel the surrounding gas, suppressing star formation in the host galaxy and possibly fueling of the AGN itself \citep{2012ARA&A..50..455F, 2014ARA&A..52..589H, 2017NatAs...1E.165H}.

Because of the potentially large impact on galaxy evolution, it is important to understand how AGNs are triggered. Current theory suggests two triggering modes: (1) a violent merger between two gas-rich disk-dominated galaxies, or (2) a much slower secular process in the host galaxy that is unrelated to mergers \citep[e.g.][]{1988ApJ...325...74S, 2005ApJ...630..705H, 2006ApJS..163....1H, 2014MNRAS.440..889S, 2017MNRAS.468.1273R}.
Since disk-dominated galaxies can often (but not always) transform into bulge-dominated ones in the first scenario \citep[e.g.][]{2008ApJ...683..597S, 2008MNRAS.391.1137L, 2013MNRAS.433.2986W}, morphology can be a good indicator of a galaxy's recent merger history. 
It is unclear from the published literature whether AGNs appear preferentially in recently merged galaxies
\citep[e.g.][]{2011ApJ...741L..11C, 2012MNRAS.425L..61S, 10.1093/mnras/stu173, 2017MNRAS.470..755H, 2019MNRAS.487.2491E, 2019ApJ...882..141M}, although there is evidence that the most luminous AGNs have undergone recent major mergers \citep[e.g.][]{2008ApJ...674...80U, 2012ApJ...758L..39T}. 
In any case, determining the morphologies of AGN host galaxies is very important for understanding AGN triggering.

AGN activity and star formation increase substantially with redshift \citep{1998ApJ...498..106M, 2003ApJ...582..559V, 2009ApJ...690...20S, doi:10.1146/annurev-astro-081811-125615}---it is likely that AGN feedback also evolves strongly---and luminous AGNs are rare enough that they are found preferentially at higher redshifts.
This means we need to determine host galaxy morphologies beyond the local universe.
With space-based Hubble Space Telescope imaging, it is possible to study AGN host galaxies out to redshifts $z\sim 2$ \citep[e.g.][]{2011ApJ...727L..31S, 2012MNRAS.425L..61S, v2012ApJ...744..148K, 2014ApJ...786..104U, 2015ApJ...806..218G, 2017ApJ...835...22P, 2020ApJ...903...85A}. 
For ground-based imaging, the task is more challenging but in the best cases we can reach $z\sim1$, a limit we explore in this paper.

In the past, studying morphologies of AGN host galaxies generally involved fitting two-dimension light profiles with tools such as GIM2D \citep{1998ASPC..145..108S}, GALFIT \citep{2002AJ....124..266P}, and automated versions thereof (e.g., GALAPAGOS; \citealp{2012MNRAS.422..449B}).
While these methods work well, they are difficult to scale up to large data sets. 
Furthermore, AGN host galaxies are complicated by the presence of an often bright central point source. Even if one tries to fit the point source as an additional component, the parameters can be degenerate. 
And clearly, these methods are not ideal when dealing with larger datasets that we face today.

In contrast, machine learning (ML) has proved useful in measuring galaxy morphologies at scale \citep[e.g.][]{2015MNRAS.450.1441D, 2015ApJS..221....8H, 2018MNRAS.475..894T, 2018MNRAS.478.5410D, 2018MNRAS.476.3661D, 2020A&C....3000334B, 2020MNRAS.491.1554W, 2021MNRAS.506.1927V, 2021MNRAS.503.4446C, 2021MNRAS.507.4425C}, which can be done both qualitatively \citep{2020ApJ...895..112G} and quantitatively \citep{2022ApJ...935..138G}. 
Specifically, \gamornet, a Convolutional Neural Network (\citealp{726791, Imagenet}), classifies galaxies as either disk-dominated, bulge-dominated, or indeterminate, with a precision of $99.7\% $($94.8\%$) for disk(bulge)-dominated SDSS \citep{2000AJ....120.1579Y} galaxies and a precision of $91.8\% $($78.6\%$) for disk(bulge)-dominated CANDELS \citep{2011ApJS..197...36K} galaxies  
\citep{2020ApJ...895..112G}.
\gamornet\ can be easily trained first on a large amount of simulated data then on a small amount of real data via transfer learning \citep[e.g.][]{2018MNRAS.479..415A, 2019MNRAS.484...93D, 9709840}.
Here we develop and implement ML techniques to address AGN host galaxies in particular. 
First, to remove the AGN point source we utilize a modified implementation of PSFGAN, a Generative Adversarial Network (GAN;  \citealp{goodfellow2014generative}) designed for that purpose
\citep{2018MNRAS.477.2513S}. 
An advantage over fitting programs is that PSFGAN does not require explicit knowledge of the point spread function (PSF), and the data processing is orders of magnitude faster. Then, we analyze the residual host galaxy light to determine whether the galaxy is disk-dominated or bulge-dominated, using a specially trained version of \gamornet, in order to provide hints of whether or not a galaxy has undergone a recent major merger. 
Although our approach is completely general, we optimize the particular networks for the high-quality imaging data available from the Hyper Suprime-Cam (HSC) Wide survey \citep{2018PASJ...70S...4A, 2019PASJ...71..114A, 2021arXiv210813045A}, which covers about $1400$ deg$^2$ of sky in five broad-band filters ($grizy$) with a depth at about 26th magnitude and a pixel scale of 0.168 arcsec per pixel.
 
This paper is organized as follows.
In \S~\ref{subsection:data used} we describe the HSC data and simulated data used for neural network training, and in \S~\ref{subsection:methods} we outline the process of building a ML model for AGN host galaxy morphology classification. 
The details of training the PSFGAN and \gamornet\ networks, including transfer learning, are described in \S~\ref{section:training}.
Model performance is analyzed in \S~\ref{section:performance}. In \S~\ref{section:summary}, we summarize our results and main conclusions, and discuss additional implementations in the near future. 
We adopt a $\Lambda$CDM Cosmology with $h_0 = 0.7$, $\Omega_m = 0.3$, and $\Omega_{\Lambda} = 0.7$ throughout this paper.

\section{Data and Methods}
\label{section:data and methods} 
\subsection{HSC Data}
\label{subsection:data used} 
We use publicly available imaging data from the seven fields in the Subaru HSC Wide survey, in its five wide field bands: $g$, $r$, $i$, $z$, and $y$, all from HSC Second Public Data Release \citep{2019PASJ...71..114A}. 
We consider only galaxies with their photometric redshift (\citealp{2020arXiv200301511N}) less than one,
the precision for more distant galaxies being unacceptably low, and we divide them 
into three redshift bins: $0<z<0.25$, $0.25<z<0.5$, and $0.5<z<1.0$\footnote{After training models (Section \ref{subsection:tl}) and choosing mapping thresholds (Section \ref{subsection:tradeoff}), we further analyzed how models for different redshift bins compare, and specifically, whether they create discontinuities at redshift bin boundaries. See Appendix \ref{section:appendix:aja_cnty} for details.}.
We use the $g$ band ($400-550$ nm), $r$ band ($550-695$ nm), and $i$ band ($695-845$ nm) for the low, medium and high redshift bins, respectively, to approximate similar rest-frame wavelengths.
We then develop machine learning models to characterize the galaxy morphologies. 
The division in redshift bins allows separate ML models to each focus on smaller ranges of galaxy apparent magnitude and size, minimizing the parameter space they have to optimize over, and improving their overall precision.

\begin{figure}
\figurenum{1a}
\gridline{\fig{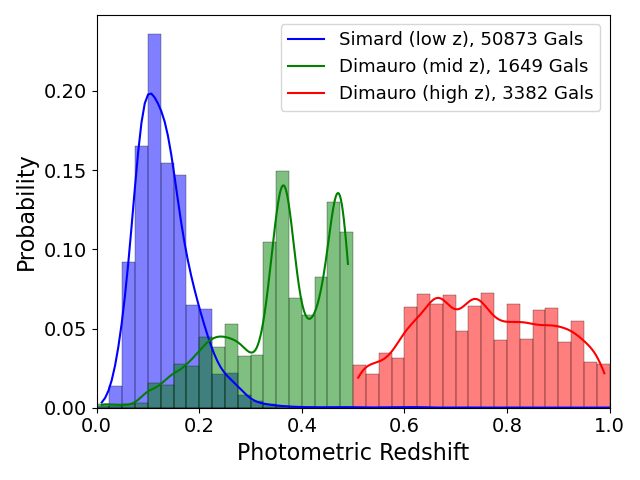}{0.45\textwidth}{}}
\vspace*{-0.6cm}
\caption{Photometric redshift distribution of HSC galaxies used to train and test our ML models: overlapping 50,873 low-redshift SDSS galaxies ({\it blue}, $z<0.25$) analyzed by \citet{2011ApJS..196...11S}, and 1,649 medium-redshift ({\it green}, mostly $0.25 < z < 0.5$) and 3,382 high-redshift ({\it red}, $0.5 < z < 1$) CANDELS galaxies analyzed by \citet{2018MNRAS.478.5410D}. Each histogram is normalized separately to unity. Colored lines show the estimated probability density curves using a Kernel Density Estimator (see \S~\ref{subsection:data used} for details).
\label{fig:training_photoz}}
\end{figure}

\begin{figure}
\figurenum{1b}
\gridline{\fig{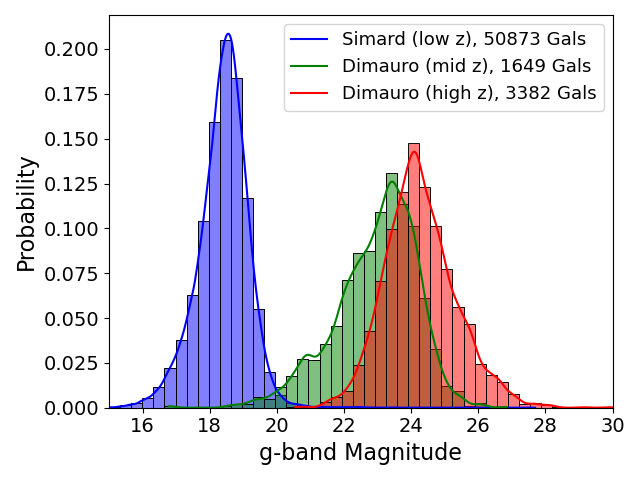}{0.45\textwidth}{}}
\vspace*{-0.6cm}
\caption{Magnitude distribution (g-band) of HSC galaxies used to train and test our ML models (colors as in Fig. \ref{fig:training_photoz}). Each histogram is normalized separately to unity.
\label{fig:training_gmag}}
\end{figure}

\begin{figure}
\figurenum{1c}
\gridline{\fig{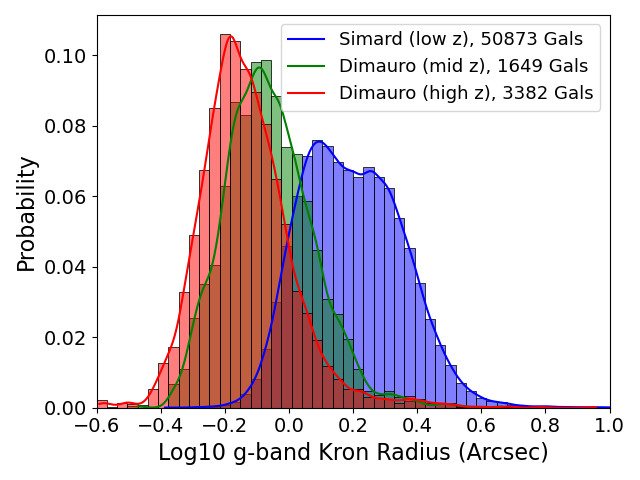}{0.45\textwidth}{}}
\vspace*{-0.6cm}
\caption{Kron radius distribution (g-band) of HSC galaxies used to train and test our ML models (from \href{https://hsc-release.mtk.nao.ac.jp/doc/index.php/tools-2/}{HSC DR2 Public Data Access Page}; colors as in Fig. \ref{fig:training_photoz}). Each histogram is normalized separately to unity.
\label{fig:training_gradius_kron}}
\end{figure}

When training the ML models, we use both simulated galaxies (see \S~\ref{subsection:simgal}) and previously morphologically-classified galaxies.
The number of simulated galaxies can be arbitrarily large, even while the number of previously classified galaxies is generally small.
Concerning the latter, no bulge-disk decomposition has been done specifically for HSC-imaged galaxies. Therefore, we used TOPCAT \citep{2005ASPC..347...29T} to find HSC galaxies that have been classified in other imaging data. 
We first searched for all bona fide galaxies in HSC Wide, resulting in 151,534,486 primary sources. We did this by requiring the source's $\{filter\}\_extendedness\_value$ (from table $wide.forced$), which is based on a magnitude difference between PSF and CModel photometry, to be $1$ (extended source) in all 5 bands.

We then cross-matched these sources to published results from two papers, \citet{2011ApJS..196...11S} and \citet{2018MNRAS.478.5410D}, which we used to train and test our ML models. 
\citet{2011ApJS..196...11S} reported bulge plus disk fits for 1,116,369 galaxies with SDSS DR7 $g$- and $r$-band images. 
Cross-matching the HSC galaxies to their Table\,1, using a positional uncertainty of 0.336\,arcsec (2 HSC pixels), we found 50,873 HSC galaxies, for which we prepared 239-pixel (about 40 arcsec in HSC Wide) square cutouts in all five bands using the \href{https://hsc-release.mtk.nao.ac.jp/das_cutout/pdr2/}{Image Cutout tool} provided by HSC.
For higher redshifts, the Gold catalog of \citet{2018MNRAS.478.5410D} contains single \sersic\ fits to CANDELS images in 4 bands ($F606W, F814W, F125W, F160W$) for 17,327 galaxies with H-band flux $F160W$ brighter than 23\,mag. 
Cross matching these to HSC galaxies (using the same positional uncertainty, $0.336$ arcsec), we found 5,082 sources in the redshift range $0<z<1$, for which we again prepared 239-pixel square cutouts.
The 3,382 galaxies at $0.5<z<1.0$ were used for the high-redshift bin, and the 1,649 galaxies at $0<z<0.5$ (including 291 galaxies at $z<0.25$) were used for the medium-redshift bin.
Figure\,\ref{fig:training_photoz}, Figure\,\ref{fig:training_gmag} and Figure\,\ref{fig:training_gradius_kron} show the redshift, g band magnitude, and g band kron radius distributions of the matched HSC galaxies in each bin using normalized histograms, respectively. We also use the Kernel Density Estimator (KDE)\footnote{KDE assumes our data are samples drawn from an underlying probability distribution, and it tries to estimate this distribution using kernel functions. See \citet{2017arXiv170403924C} for its definition and an overview.} to visualize the data distribution as a continuous variable.

The medium- and high-redshift calibration samples are quite a bit smaller than the low-redshift sample, so we use a standard approach of augmenting the training set.
For the medium-redshift bin, we make 8 copies of each image by a possible rotation of $90$, $180$, or $270$ degrees and a possible reflection with respect to the x-axis. 
For the high-redshift bin, we make 4 copies of each via rotation, omitting the reflection for a comparable number of samples as in the mid-redshift bin. 
The final calibration samples have 13,192 and 13,528
galaxies for the medium- and high-redshift bins, respectively. 
We also prepared 239-pixel square cutouts in the five HSC bands for $\sim 400$ local stars, in order to create multiple PSFs,
for use in creating artificial AGNs (see \S~\ref{section:training}).

\subsection{Overview of PSFGAN and GaMorNet Training }
\label{subsection:methods} 

In this section, we give an overview of how we use PSFGAN and \gamornet\ to build a model that can accurately classify AGN host galaxy morphologies in HSC Wide fields. 
In our initial approach, we tested the practicality of training PSFGAN and \gamornet\ separately---that is, first using PSFGAN for AGN point source removal and then sending PSFGAN-recovered host galaxies to a \gamornet\ trained on ordinary (inactive) galaxies for classification. 
This test resulted in a fairly low precision, which never exceeded $70\%$.
This is because PSFGAN introduces subtle patterns (at very low amplitude) not seen in the original (inactive) galaxies used for training, which therefore confused \gamornet{}. 
We found that the classification precision was much higher when using these networks as a linked pair---that is, after training \gamornet\ on PSFGAN-recovered galaxies rather than inactive galaxies. This section describes the linked training approach. 

Traditionally, there are two main challenges to classifying AGN host galaxies morphologically using ML tools: (1) the impact of the luminous AGN point source and (2) insufficient training data with classified morphological labels. 
We show explicitly that these obstacles can be overcome using PSFGAN and \gamornet{} because those results agree well with results from legacy 2D fitting methods.

\begin{figure*}
\figurenum{2}
\epsscale{1.15}
\plotone{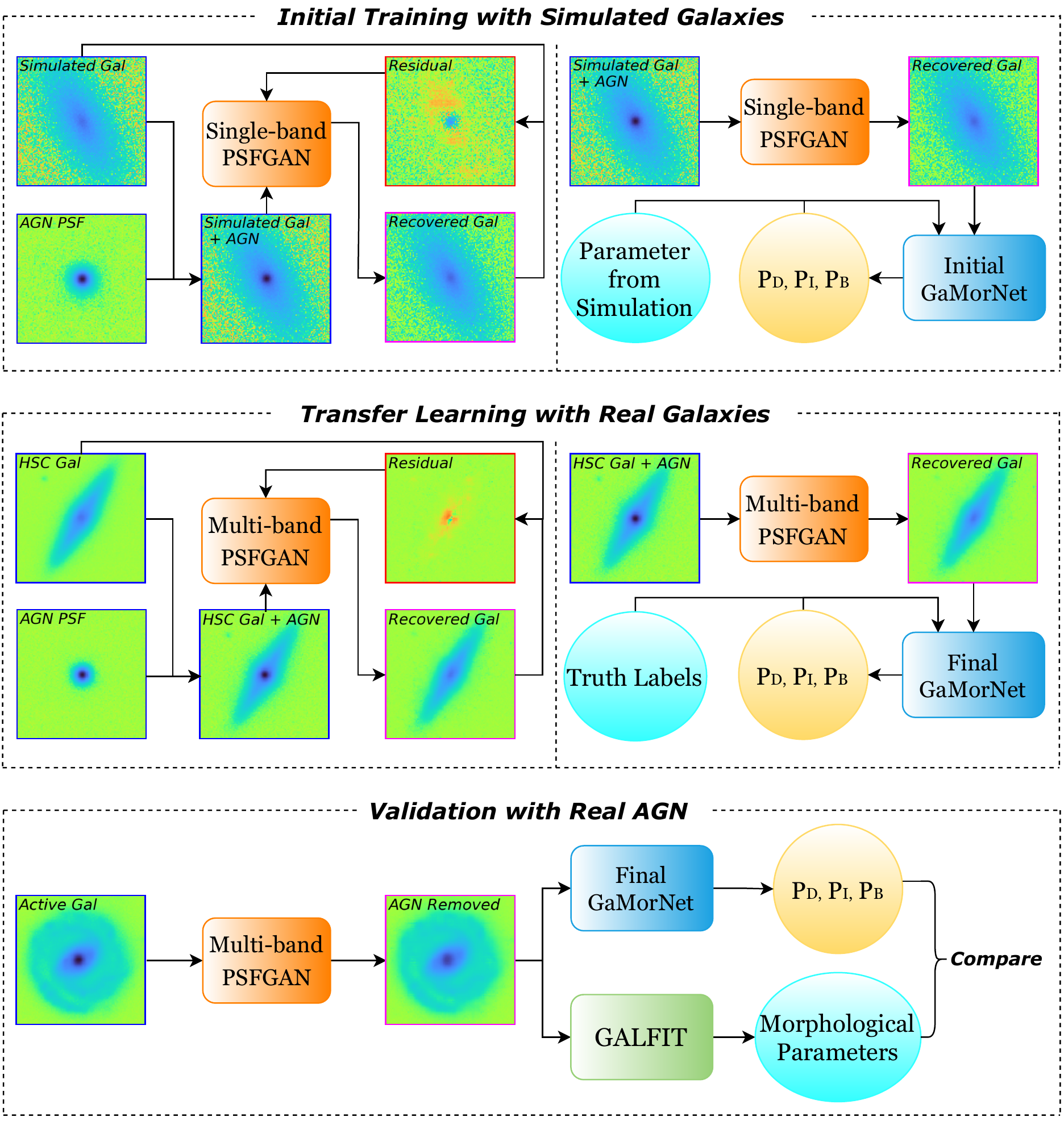}
\caption{
Diagram showing our use of PSFGAN and \gamornet\ to build a model for AGN morphology classification.\\
Top: initial training with simulated galaxies. We create simulated galaxies (convolved with real PSFs before adding noise) with known morphologies and reasonable parameters for their redshifts. 
Artificial AGN point sources are added and both the original galaxy and the original + AGN are fed to PSFGAN for training. 
In the training process, PSFGAN compares each recovered galaxy (i.e., its own guess) to the original galaxy and tries to adjust itself such that the difference between these two are minimal (by incorporating a residual term to the loss function of its generator).
The PSFGAN-recovered galaxies are then used by \gamornet\ for initial training to classify galaxy as either disk-dominated, indeterminate, or bulge-dominated.\\
Bottom: transfer learning and validation with real galaxies. We use real galaxies which were classified using traditional 2D fitting methods in place of simulated galaxies and repeat process in the top chart. We directly take the \gamornet\ trained with simulated galaxies and fine-tune its parameters using labelled real galaxies. Once both PSFGAN and \gamornet\ are trained, we feed active galaxy in HSC Wide in place of original + AGN (PSFGAN), and classify the morphology of the recovered host galaxy (\gamornet). \gamornet\ outputs are then compared against GALFIT fitting results.
\label{fig:method_diagram}}
\end{figure*}

Our approach includes the following steps, described in the three panels in Figure\,\ref{fig:method_diagram}:
\begin{enumerate}
    \item Initial training of PSFGAN + \gamornet\ with simulated galaxies;
    \item Transfer learning (described below) with labelled real galaxies (to which point sources were added, then removed with PSFGAN); 
    \item Validation against GALFIT 2D fitting results.
\end{enumerate}

\noindent
We first used GalSim \citep{2015A&C....10..121R} to create simulated galaxies with bulges and/or disks. Each galaxy was convolved with the PSF before noise was added. We created artificial AGNs by combining simulated galaxies and PSFs with a range of contrast ratios. We then trained a single-band version of PSFGAN to remove the added point source accurately, generating recovered galaxies with known morphological labels that are used to train \gamornet\ from scratch.
We can generate arbitrarily large sets of simulated AGNs with no cost, and more training data will make the ML tools more accurate.
However, the precision obtained for the final PSFGAN+\gamornet\ tool is limited by the computational resources available for the initial training, so in practice we used a few hundred thousand simulated galaxies.

The same training process was then repeated, in the transfer learning step, using real galaxies with known morphological parameters---a much smaller data set (see \S~\ref{subsection:data used}). The only difference is that instead of training another \gamornet\ from scratch, we start with the \gamornet\ trained on simulated galaxies, and fine-tune its parameters using the real galaxies. (Note that we first added artificial point sources to these real galaxies, then removed them with PSFGAN, as for the simulated galaxy training set, in order to teach \gamornet{} to ignore any PSFGAN-induced patterns.)
The approach of first training with a large amount of simulated data, and then re-training with a small amount of real, labelled data, allows us to drastically reduce the number of previously labelled real galaxies needed for training. 

Once the transfer learning was done, we fed real active galaxies to PSFGAN. The recovered host galaxies were then fed to both \gamornet\ and GALFIT and their results were compared against each other.

\section{Neural Network Training}
\label{section:training}
\subsection{Preparing Simulated AGN Host Galaxies for Initial Training}
\label{subsection:simgal}
Here we describe the simulation of galaxies and the addition of an artificial AGN point source, to be used in the initial training of PSFGAN and \gamornet\ described in \S\,3.2. 
We used GalSim to create 2D light profiles of $150,000$ galaxies in the $g$, $r$, and $i$ wavebands. 
In the low ($0<z<0.25$), medium ($0.25<z<0.5$), and high ($0.5<z<1.0$) redshift bins, we are focusing on galaxy morphology in the $g$, $r$, and $i$ bands, respectively, so that we are looking at roughly the same wavelength in the rest frame. 
We chose simulation parameters (e.g., ranges of galaxy flux and half-light radius) that match corresponding ranges observed in most real galaxies in each redshift bin and each band, according to \citet{binney_and_merrifield}, as summarized in Table\,\ref{tab:sim_gal}.
With these parameters, we created simulated galaxies in each redshift bin and each band. Furthermore, we convolved each simulated galaxy with real instrumental PSF and then we added noise. This step ensures our simulated galaxies closely resemble observed real galaxies, and are thus appropriate to be used in the training step for our initial models. Details are discussed below.

Three quarters of simulated galaxies ($112,500$) have both a disk and a bulge. 
To have a diverse range of light profiles in the training set, we used a range of \sersic\ index $n$ ($0.8<n<1.2$ for the disk component, $3.5<n<5.0$ for the bulge component) instead of single values ($n=1$ for a pure disk, $n=4$ for a pure bulge). 
The exact ranges of \sersic\ index are not important because training with simulated data is just the starting point; the final networks will be modified by training on real data (see \S\,\ref{subsection:tl}).
The remaining quarter of the galaxies ($37,500$) have a single component that is either a disk or a bulge (with the same fuzzy range of \sersic\ index; $18,750$ for each category).
For double component galaxies, we drew an axis ratio $b/a$ between $0.25$ and $1.0$ for the disk and the bulge components (mutually independent). 
For the position angle, we first chose the disk orientation randomly between $-90^\circ$ to $90^\circ$, then choose a random bulge orientation within
$\pm 15^\circ$ of the disk orientation.
The bulge-to-total flux ratio, $L_{B}/L_{T}$, is uniformly distributed between $0$ and $1$.
Single component galaxies were simulated using the same ranges of parameters used for the disk component in double component galaxies.
Each simulated galaxy cutout was chosen to have a size of 239 $\times$ 239 pixels (about 40 arcsec $\times$ 40 arcsec in HSC Wide); the same cutout size was used for all images discussed in this paper.

\begin{deluxetable*}{cccccc}[htbp]
\tablecaption{Parameter Ranges for Simulated Galaxies in Each Redshift Bin \label{tab:sim_gal}}
\tablecolumns{6}
\tablehead{
\colhead{Component} & \colhead{\sersic\ Index} & \colhead{Half-Light Radius} & \colhead{Flux} & \colhead{Axis Ratio} & \colhead{Position Angle} \\ 
\colhead{} & \colhead{} & \colhead{(Arcseconds)} &  \colhead{(ADUs)} & \colhead{} & \colhead{(Degrees)}
}
\startdata
    \hline
    \hline
    \multicolumn{6}{c}{37,500 Single-Component Galaxies\tablenotemark{a}, images in key band\tablenotemark{b}} \\
    \hline
    && 0.1 --- 5.0 (low-z)
    &30 --- 135000 (low-z)&&\\
    Single & 0.8 --- 1.2 or 3.5 --- 5.0 & 0.1 --- 2.0 (mid-z) & 30 --- 25000 (mid-z) & 0.25 --- 1.0 & $-$90.0 --- 90.0 \\
    && 0.2 --- 1.3 (high-z)
    &30 --- 40000 (high-z)&&\\
    \hline
    \hline
    \multicolumn{6}{c}{112,500 Double-Component Galaxies, images in key band} \\
    \hline
    Disk & 0.8 --- 1.2 & same as ``Single''\tablenotemark{c}
    & 0 --- 1\tablenotemark{d} & 0.25 --- 1.0 & $-$90.0 --- 90.0 \\
    && 0.1 --- 3.0 (low-z)
    &&&\\
    Bulge & 3.5 --- 5.0 & 0.1 --- 1.5 (mid-z)
    & 1 $-$ Disk Frac.  & 0.25 --- 1.0 & Disk Comp. $\pm\,(0,15)$\tablenotemark{e}\\
    && 0.2 --- 0.8 (high-z)
    &&&\\
    \hline
\enddata

\tablenotetext{a}{Divided equally into disks and bulges.}
\tablenotetext{b}{Band $g$, $r$, or $i$ for the low ($0<z<0.25$), medium ($0.25<z<0.5$), or high ($0.5<z<1.0$) redshift bin, respectively.}
\tablenotetext{c}{The range of half-light radius for the disk component in double-component galaxies is set equal to the range of the same parameter for single component galaxies, in each redshift bin respectively.}
\tablenotetext{d}{Only the fractional flux is mentioned here. Flux for the disk (bulge) component equals to the total flux multiplied by the disk (bulge) fractional flux. The total flux is chosen from the same range as for single component galaxies, in each redshift bin respectively.}
\tablenotetext{e}{The bulge position angle differs from the disk position angle by a randomly chosen value between $-15^\circ$ and $+15^\circ$.}
\tablecomments{The table shows details of parameter ranges of simulated galaxies in each of the three redshift bins. All parameters are drawn from uniform distributions except bulge position angle in double component galaxies. We choose parameter ranges such that given the size of our simulated data, they resemble real galaxies in corresponding redshifted bands. See \S~\ref{subsection:simgal} for more details.}
\end{deluxetable*}

\begin{figure}
\figurenum{3}
\epsscale{1.2}
\plotone{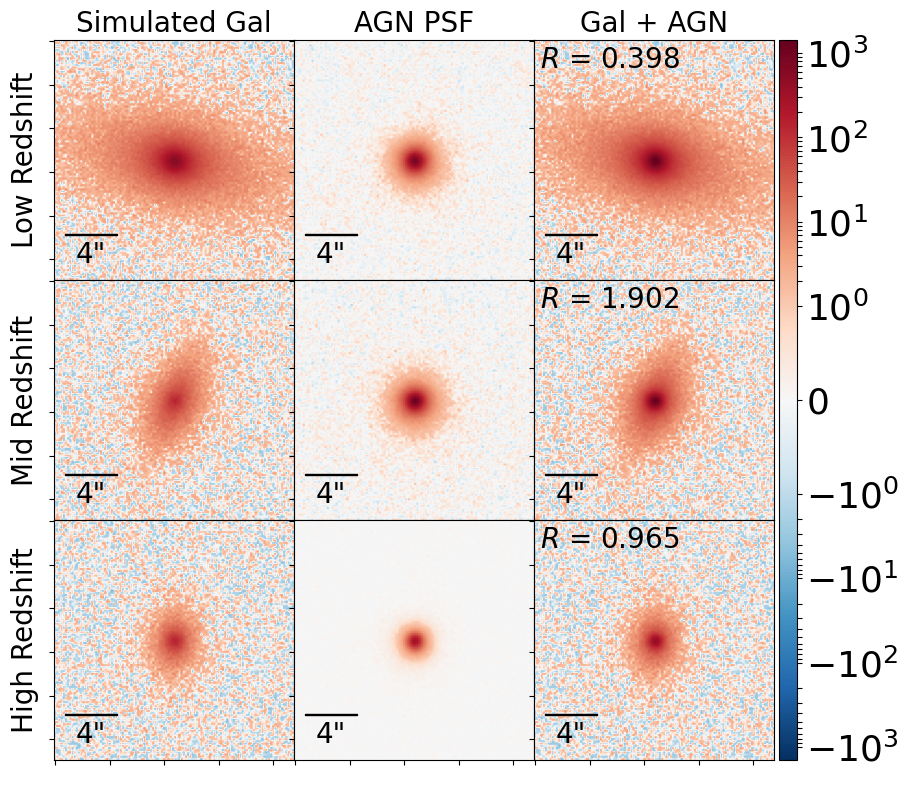}
\caption{Examples of AGN host galaxy simulations in three redshift bins. 
{\it Left to right:} Original galaxy, AGN point source, combined galaxy and AGN PSF. 
{\it Top to bottom:} Examples from low ($0<z<0.25$), medium ($0.25<z<0.5$), and high ($0.5<z<1$) redshift bins, using observed wavelength bands (g, r, and i, respectively) that correspond to roughly the same rest-frame wavelength.  
Simulated galaxies ({\it left column}) were convolved with the PSF and noise was added. AGNs were simulated with a wide range of contrast ratios, of which three randomly chosen values are shown ({\it right column}). The colorbar at far right shows the flux in Analog-Digital Units (ADUs). 
\label{fig:agn_addition}}
\end{figure}

To make the simulated galaxies closely resemble real ones, we convolve them with the instrumental PSF.
Using the HSC PSF Picker Tool\footnote{https://hsc-release.mtk.nao.ac.jp/psf/pdr2/}, we select 50 PSFs for each band from random locations in HSC Wide fields \citep{2019PASJ...71..114A}.
We then convolve each simulated galaxy with a PSF randomly selected from this group of 50. 
Then we add noise using GalSim's inherent noise module\footnote{https://galsim-developers.github.io/GalSim/\_build/html/noise.html} (to replicate the real HSC data\footnote{See table 1 in \citealp{2018PASJ...70S...4A}} we choose: type=CCD; gain=3.0; read\_noise=4.5). 
The first column of Figure\ \ref{fig:agn_addition} shows three examples of the final simulated galaxies, one for each redshift band.\footnote{For these and subsequent figures, we show the central part of the image, not the full cutout analyzed by the model.}

To represent the AGN PSF, we prepare 239-pixel square cutouts of 392 local, bright, stellar point sources ($\{filter\}\_extendedness\_value=0$ in all five bands) from HSC Wide, each of which:
\begin{enumerate}
    \itemsep0em 
    \item Is not blended with other sources and is in the inner region of a coadd patch and tract (\textit{isprimary = TRUE});
    \item Is not near the image edge (\textit{\{filter\}\_pixelflags\_edge = FALSE});
    \item SDSS centroid algorithm succeeded in all bands (\textit{\{filter\}\_sdsscentroid\_flag = FALSE});
    \item Flux well determined in all bands (i.e., final CModel fit succeeded; \textit{\{filter\}\_cmodel\_flag = FALSE});
    \item Is not contaminated by cosmic rays in any band
    (\textit{\{filter\}\_pixelflags\_cr = FALSE});
    \item Cutout contains no bad CCD pixels in any band (\textit{\{filter\}\_pixelflags\_bad = FALSE});
    \item HSC image is clean in all bands (\textit{\{filter\}\_cleanflags\_center = TRUE},\\
    \textit{\{filter\}\_cleanflags\_any = TRUE}).
\end{enumerate}

\noindent
We randomly select 50 of the 392 stars (combining them minimizes the impact of different seeing conditions). 
For a given simulated galaxy, we scale the flux of each star to the desired AGN-to-host-galaxy contrast ratio, $R$, which is randomly selected from a logarithmic uniform distribution between $-1 < \log R < 0.6$, appropriate to realistic AGN host galaxies (e.g., \citealp{Gabor_2009}) and optimal for PSFGAN subtraction \citep{2018MNRAS.477.2513S}. 
We average all 50 stars to create the artificial AGN point source (PS) and add it to this galaxy to obtain a simulated AGN. We repeat this process 150,000 times for all the simulated galaxies.

The second and third column of Figure\,\ref{fig:agn_addition} depict examples of a simulated AGN, one for each band.
We are now ready to train PSFGAN to remove these point sources, then train \gamornet\ to classify the PSFGAN-recovered host galaxies. 
The value of using large numbers of simulated galaxies to train these networks is that fewer real galaxies are then needed in the final transfer-learning step.

\subsection{Initial Training with Single-band PSFGAN and GaMorNet}
\label{subsection:train}
Using simulated AGN host galaxies described in the previous section, we train three separate, single-band PSFGAN networks, one for each band (as in the original version; \citealp{2018MNRAS.477.2513S}). Likewise, we train \gamornet\ separately for each band. The $g$ band is used for the lowest redshift bin, the $r$ band for medium redshifts, and the $i$ band for high redshifts.

We first split each simulated dataset---$150,000$ galaxies for each band---into five smaller subsets: a training set for PSFGAN ($9,000$ galaxies), a validation set for PSFGAN ($1,000$ galaxies), a training set for \gamornet\ ($90,000$ galaxies), a validation set for \gamornet\ ($10,000$ galaxies) and a test set for the combined framework ($40,000$ galaxies). Since the data split is completely random, each subset has roughly the same ratio of double to single component galaxies, and the same ratio of bulges to disks. We intentionally assign the majority of galaxies to train \gamornet\ from scratch, as that is generally desired in training convolutional neural networks for image classification (e.g., \citealp{2015arXiv151106348C}); in general, networks that require labels, like \gamornet, require more training data, whereas PSFGAN compares images internally.

We create morphology labels for each galaxy that describe the ground truth in \gamornet\ training. 
Galaxies are defined as disk-dominated, indeterminate\footnote{As we will show in later sections, we introduced the indeterminate type for galaxies which either have similar amount of light from disk and bulge or have sizes too small or signal-to-noise ratios too low. See \S~\ref{subsection:dependency} for details.}, or bulge-dominated, following \citet{2020ApJ...895..112G}.
For double component galaxies ($\sim 75\%$), we define the galaxy as:
\begin{enumerate}
        \item disk-dominated if $L_{B}/L_{T} < 0.45$,
        \item bulge-dominated if $L_{B}/L_{T} > 0.55$,
        \item indeterminate if $0.45  \leq L_{B}/L_{T}  \leq 0.55$.
    \end{enumerate}
\noindent
For single component galaxies ($\sim 25\%$), we assign these labels:
    \begin{enumerate}
        \item disk dominated if $n < 2.0$,
        \item bulge dominated if $n > 2.5$,
        \item indeterminate if $2.0 \leq n \leq 2.5$,
     \end{enumerate}
\noindent
Simulated single-component galaxies can only be classified as disks or bulges, and were simulated with limited values of the \sersic\ index (see Table\,\ref{tab:sim_gal}). 

A GAN consists of two sub-networks: a generator and a discriminator. While the generator tries to create fake images, the discriminator tries to differentiate generator-created images from real ones. Both of them are trained simultaneously in an adversarial way (i.e., performances of both sub-networks improve simultaneously). PSFGAN is a conditional GAN \citep{2016arXiv160505396R}: its generator takes an image of a host galaxy with an AGN point source (generator input), and it tries to create a fake image that contains only the host galaxy (generator output). At the same time, the discriminator in PSFGAN compares this fake image with the image of the actual host galaxy. Note that in our experiments, since the discriminator always has access to the image of the actual host galaxy, at the end of the training step, the generator has to create fake images that very closely resemble the real ones.

We train each single-band PSFGAN by requiring that a processed AGN image (i.e., with PSF removed by PSFGAN) resemble the original galaxy (i.e., the simulated galaxy to which we added a PSF to make the simulated AGN). 
Following \citet{2018MNRAS.477.2513S}, we use a generator loss function, $L_{gen}$, with an attention window to up-weight the central pixels, while the discriminator loss function, $L_{dsc}$, is unchanged:
\begin{equation}
\begin{split}
\label{eq:1}
    L_{gen} &= L_{adv\_gen} \\
    &+ \frac{\lambda P_{att}}{N^2}\sum_{i, j=0}^{N-1}|X^{ori}_{ij} - X^{gen}_{ij}| \\
    &+ \frac{\lambda}{{S_{att}}^2}\sum_{i, j=(N-1)/2-(S_{att}/2)+1}^{(N-1)/2+(S_{att}/2)}|X^{ori}_{ij} - X^{gen}_{ij}|, \\
    L_{dsc} &= L_{adv\_dsc}.
\end{split}
\end{equation}
where $L_{adv\_gen}$ and $L_{adv\_dsc}$ represent adversarial loss terms in the generator and the discriminator of the vanilla GAN, respectively\footnote{See Eq. 1 in \citet{goodfellow2014generative} for the mathematical forms of $L_{adv\_gen}$ and $L_{adv\_dsc}$.}; $\lambda=100$ and $P_{att}=0.05$ (attention parameter) are two parameters that control the relative importance of terms in the loss function; $N=239$ and $S_{att}$\footnote{See Table \ref{tab:psfgan_sim_details} for chosen values of the attention window size, $S_{att}$, in different redshift bins.} are image and attention window sizes in pixels, respectively; $X^{ori}_{ij}$ and $X^{gen}_{ij}$ denote the flux values for pixel $ij$ in the original and generated images. With the chosen values of $\lambda$ and $P_{att}$, the absolute difference between the entire original and generated images are $100*0.05 = 5$ times more important than the adversarial loss, while the absolute difference between the original and generated images within the central attention window is $100$ times more important than the  adversarial loss and 20 times more important than the absolute difference across the entire image. This ensures that while minimizing the residuals across the entire image\footnote{Residual is defined as recovered (generated) galaxy minus original galaxy}, PSFGAN puts much more focus on the central region.
Minimization of the loss function is done using Adam, a well-known first-order gradient-based optimizer \citep{2014arXiv1412.6980K}.
In addition to the attention parameter ($P_{att}$) and the attention window size ($S_{att}$), certain hyper-parameters are necessary to the training. 
These include the learning rate, which governs the parameter increment at each learning step and
the number of epochs over which the network is trained.

Moreover, three hyper-parameters are used in image normalization,
both in a normalization prior to PSFGAN, and the inverse operation prior to \gamornet. Every pixel value, $x$, is transformed using: 
\begin{equation}
\label{eq:2}
    x^\prime = \frac{\rm arcsinh (A*x)}
    {\rm arcsinh (A* x_{\rm max})}.
\end{equation}
where $x_{\rm max}$ is the maximum pixel value for the entire set of images to be normalized, and $A$ is a stretch factor. 
This normalization step has been found very helpful, especially in cases where the dynamic range of pixel values is high.
After trying a range of hyper-parameters, we determined the values that give the best results, for each band, as shown in Table\,\ref{tab:psfgan_sim_details}.

\begin{deluxetable}{cccc}[htbp]
\tablecaption{Hyper-Parameters for Single-band PSFGAN Training On Simulated Galaxies}
\label{tab:psfgan_sim_details}
\tablecolumns{4}
\tablehead{
\colhead{Redshift Bin} & \colhead{Low} & \colhead{Mid} & \colhead{High} \\
\colhead{} & \colhead{($g$ band)} & \colhead{($r$ band)} & \colhead{($i$ band)} 
} 
\startdata
    \hline
    Attention Window Size $(S_{att})$\tablenotemark{a} & $22$ & $12$ & $12$ \\
    Attention Parameter ($P_{att}$) & $0.05$ & $0.05$ & $0.05$ \\
    Learning Rate & $5\times10^{-5}$ & $1.5\times10^{-5}$ & $2\times10^{-5}$ \\
    Epoch & $20$ & $30$ & $20$ \\
    Stretch Function & arcsinh & arcsinh & arcsinh \\
    Stretch Factor & $50$ & $50$ & $50$ \\
    Maximum Pixel Value & $25,000$ & $5,000$ & $10,000$ \\
    \hline
    \hline
\enddata
\tablenotetext{a}{The length of each side, in pixels, of the square attention window at the image center.}
\end{deluxetable}

\begin{figure}
\figurenum{4}
\epsscale{1.3}
\gridline{\fig{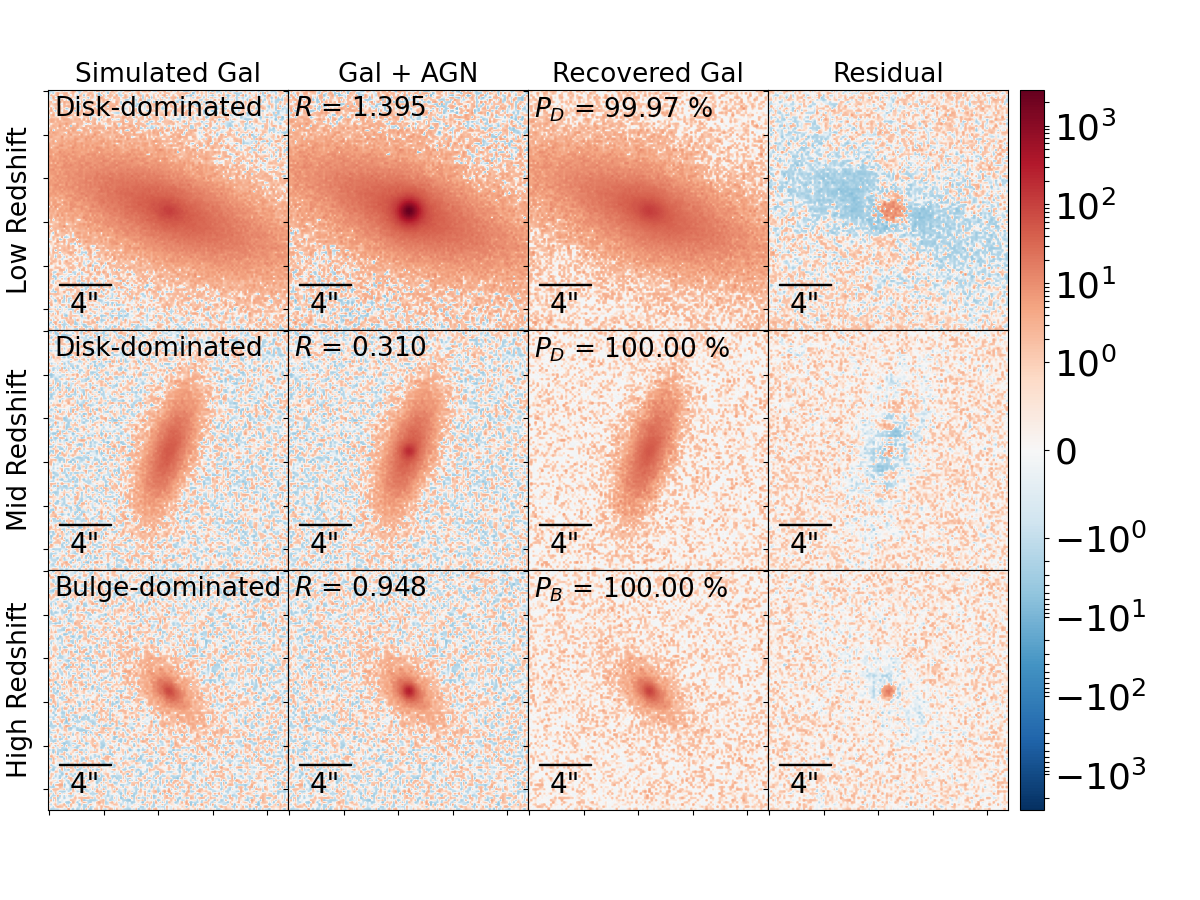}{0.55\textwidth}{}}
\vspace*{-0.4cm}
\caption{The effectiveness of PSFGAN trained on simulations is illustrated with examples from low, medium, and high redshift, for which we used $g$, $r$, and $i$ band images, respectively.
The first column shows the original galaxy simulated with GalSim; 
the second column adds a point source at the galaxy center; 
the third column shows the recovered galaxy after processing with PSFGAN; and the final column shows the residual image (recovered minus original).
Labels denote, from left to right, morphology of the original galaxy, contrast ratio for the simulated AGN, and probability that \gamornet\ correctly predicts the morphology of the recovered galaxy. 
The intensity is given by the color bar at the right, which is the same as in the previous figure. 
\label{fig:psf_gmn_sim_a}}
\end{figure}

The original PSFGAN used an input image size of $424 \times 424$ pixels \citep{2018MNRAS.477.2513S}. Since our cutouts are much smaller ($239 \times 239$ pixels), we removed one layer from the 8 layers in the encoding and decoding parts of the original generator\footnote{Our motivation of this modification is explained in Appendix \ref{section:appendix:mdf_psf_gnrt}.};
apart from this change, we used the original PSFGAN architecture to train PSFGAN from scratch.
We stopped the training once the PSFGAN loss converged, then we verified with the validation set that there was no sign of over-training.
Figure \ref{fig:psf_gmn_sim_a} shows three examples of the application of the trained PSFGAN to galaxies from the final test set, for each of the three redshift ranges.

We next apply PSFGAN to the training and validation sets for \gamornet, as well as the common test set, removing the artificial AGN point source and generating recovered images for each galaxy. 
Originally, we had tried running PSFGAN-ed images through a version of \gamornet\ trained on an independent set of galaxies; however, we found that the residuals, though very small, have a structure quite different from unprocessed (natural) galaxy images, which led to low precision for the \gamornet\ classifications. 
Therefore, we devised the present approach---training a new version of \gamornet\ only on PSFGAN-recovered galaxies---which results in much higher precision.

\begin{deluxetable}{cccc}[htbp]
\tablecaption{Hyper-Parameters\tablenotemark{a} for Initial Training of \gamornet\ On Simulated Galaxies}
\label{tab:gamornet_sim_details}
\tablecolumns{4}
\tablehead{
\colhead{Redshift Bin} & \colhead{Low} & \colhead{Mid} & \colhead{High} \\
\colhead{} & \colhead{($g$ band)} & \colhead{($r$ band)} & \colhead{($i$ band)} 
} 
\startdata
    \hline
    Learning Rate & $5\times10^{-5}$ & $5\times10^{-5}$ & $5\times10^{-5}$ \\
    Epoch & $100$ & $100$ & $100$ \\
    Batch Size & $128$ & $128$ & $128$ \\
    Momentum & $0.9$ & $0.9$ & $0.9$ \\
    Learning Rate Decay & $0$ & $0$ & $0$ \\
    Stretch Factor & $50$ & $50$ & $50$ \\
    Loss Function & cross entropy & cross entropy & cross entropy \\
    \hline
    \hline
\enddata
\tablenotetext{a}{
For a description of these hyper-parameters,
see \href{https://gamornet.readthedocs.io/en/latest/tutorials.html}{\gamornet\ Tutorials}.
}
\end{deluxetable}

CNN is a type of feed-forward neural network which specializes in image processing and recognition. A typical CNN contains three types of layers: convolutional layers, pooling layers, and fully connected layers. An input image is first sent to a convolutional layer which usually has multiple channels. Within each channel, each artificial neuron only responses to a limited region of the input image (``receptive field''). Specifically, two neighboring neurons separated by a certain distance in a certain dimension respond to two receptive fields that are displaced by a proportional distance in the same dimension.
For a certain channel, the connection between each receptive field and each neuron is translation invariant (i.e. they share the same weight matrix). Thus, it can be interpreted that neurons in each channel are trying to extract a particular feature at different locations of the input image. Feature maps of a convolutional layer, or equivalently the output of neurons of its channels, are then sent to a pooling layer, which usually preserves only the maximum values of its input, making the network insensitive to minor differences between images. A CNN usually contains multiple pairs of convolutional and pooling layers. The output of the last pooling layer is sent to a few fully connected layers, and the output of the last fully connected layer is therefore the output of the network, which contains information of the classified labels. During the training step, these output labels are compared against ground truth labels via a loss function (``supervised learning''). The losses are back-propagated to each layer of the network, and the network itself will try to adjust its parameters to making better classifications next time. As a CNN, our \gamornet{} shares all these features. Readers are encouraged to refer to Section 3.2, Figure 4, and Table 2 in \citet{2020ApJ...895..112G} for details of its structure.

We train \gamornet\ from scratch with its training set and determine the best hyper-parameters with its validation set, using PSFGAN-processed images (i.e. the AGN point source is removed from each simulated AGN in the training and validation sets for \gamornet{}).  Our choices of hyper-parameters for each band are shown in Table \ref{tab:gamornet_sim_details}. Using these hyper-parameters, we next test \gamornet\ on the common test set. 
We map each \gamornet{} output, which has a form of a 3-tuple ($P_{\rm D}$, $P_{\rm I}$, $P_{\rm B}$), into a predicted morphological type by choosing the maximum value of the three (i.e. if $P_{\rm D}$ is the highest, we map the input galaxy as a predicted disk). We then compare these predicted morphological types against the ground truth morphological labels (which we know from the simulation), and we compute the precision on disks (bulges) as the ratio between true disks (bulges), which are actual disks (bulges) that are also predicted as disks (bulges), to all galaxies predicted as disks (bulges).\footnote{Readers are encouraged to refer to \S~\ref{subsection:tradeoff}, in which we discussed precision and other metrics in detail.}
\gamornet\ achieves a precision $\gtrsim 90$\%.
Precision values for each redshift bin are shown in Table \ref{tab:sim_summary}.

\begin{figure}
\figurenum{5}
\epsscale{1.3}
\gridline{\fig{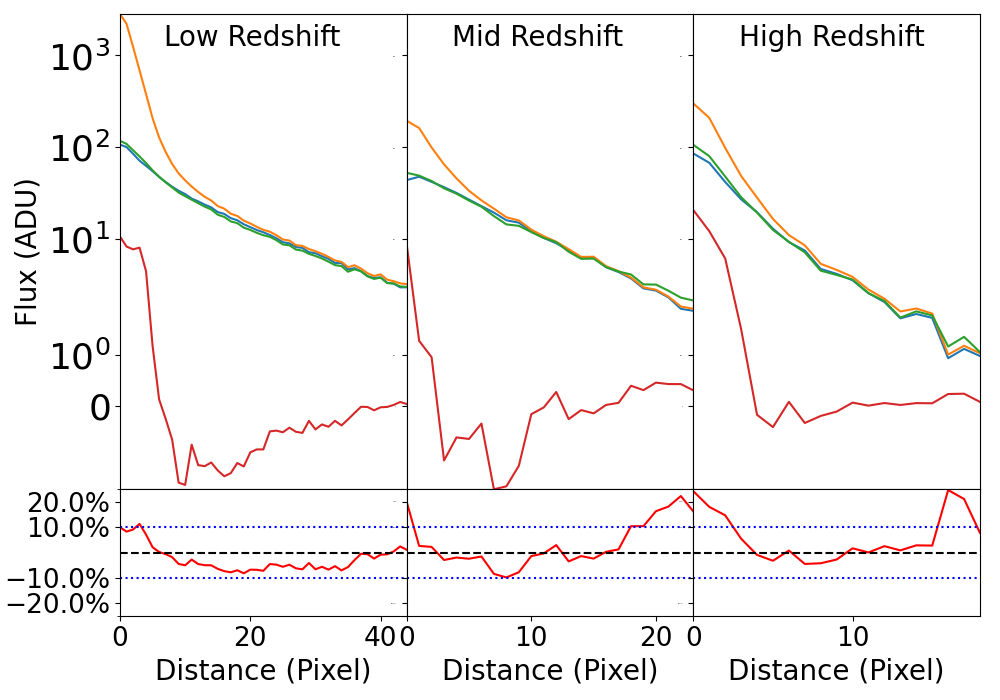}{0.5\textwidth}{}}
\vspace*{-0.4cm}
\caption{
(Top) Radial profiles of simulated AGN ({\it orange}), original galaxy ({\it blue}), recovered galaxy ({\it green}), and residuals (i.e., recovered minus original galaxy; {\it red}), for the three example galaxies shown in Figure \ref{fig:psf_gmn_sim_a}. 
(Bottom) Fractal residuals ($residual / original$) of the same example galaxies.
Over most of the galaxy, the residuals are a small percentage of the original flux. 
Even at image center where residual becomes larger, \gamornet{} is not confused, as shown evidently by its output probabilities in the third column of Figure \ref{fig:psf_gmn_sim_a}.
\label{fig:psf_gmn_sim_b}}
\end{figure}
\begin{deluxetable}{cccc}[htbp]
\tablecaption{Summary for \gamornet\ Classification Results on Test Sets of Simulated Galaxies}
\label{tab:sim_summary}
\tablecolumns{4}
\tablehead{
\colhead{Redshift Bin} & \colhead{Low} & \colhead{Mid} & \colhead{High} \\
\colhead{} & \colhead{($g$ band)} & \colhead{($r$ band)} & \colhead{($i$ band)} 
} 
\startdata
    \hline
    Precision on Disks & $91.15\%$ & $90.33\%$ & $89.75\%$ \\
    Precision on Bulges & $91.83\%$ & $90.63\%$ & $88.92\%$ \\
    \hline
    \hline
\enddata
\end{deluxetable}

Figure\,\ref{fig:psf_gmn_sim_b} shows radial profiles of the original simulated galaxy, AGN, and residuals, for the three example galaxies from Figure~\ref{fig:psf_gmn_sim_a}, obtained from application of the initial trained PSFGAN+\gamornet\ to galaxies in the test set. (These are still all simulated galaxies.) 
These models are used as the starting point for training on real galaxies, as described in the next two sections.
\subsection{Combining Point Sources with Real Galaxies in Preparation for Transfer Learning}
\label{subsection:realgal}

After preparing simulated datasets and training and testing our networks with them, we next made more realistic simulated AGNs out of real galaxies. Details are described below. Use of these datasets in refining PSFGAN and \gamornet\ is described in \S~\ref{subsection:tl}.

We used the three sets of morphologically classified real galaxies described in \S~\ref{subsection:data used} (also Fig.\,\ref{fig:training_photoz}, \ref{fig:training_gmag}, and \ref{fig:training_gradius_kron}). The low-, medium-, and high-redshift bins contain 50,873, 13,192, and 13,528 real HSC galaxies, respectively. We prepared 239$\times$239-pixel cutouts for each galaxy in these sets, in all five HSC Wide bands: $g$, $r$, $i$, $z$, and $y$. We again used the set of 392 local stars discussed in \S~\ref{subsection:data used} and \ref{subsection:simgal}, preparing 239$\times$239-pixel cutouts in all five bands.

For each real galaxy, we randomly selected 50 of the 392 stars to create the artificial AGN PSF (as before). 
For real AGNs, the contrast ratios in the five HSC bands are not independent, so we used results from actual AGN SEDs in order to make band-specific contrast ratios that are physically realistic. In detail, we used the 41 AGN SEDs fitted by \citet{2019MNRAS.489.3351B}, which represent a wide range of AGN types (see Table 1 in \citealp{2019MNRAS.489.3351B}).
Randomly selecting one of 41 AGN SEDs, we redshifted it to match the real galaxy. We then chose an overall AGN-to-host-galaxy contrast ratio, $R$, from a uniform logarithmic distribution between $-1 < \log R < 0.6$. Using band-specific and bolometric (all five HSC bands combined) fluxes of the real galaxy and the chosen AGN SED, we converted this overall contrast ratio into band-specific contrast ratios, as follows.

First, the contrast ratio is defined as the ratio of the desired bolometric flux of the AGN point source, $F_{\rm AGN\_desired}$, to the bolometric flux of the real galaxy, $F_{\rm galaxy}$; that is,  
$R = {F_{\rm AGN\_desired}}/{F_{\rm galaxy}}$.
To convert the actual bolometric flux of the randomly chosen AGN SED, $F_{\rm AGN\_actual}$, to $F_{\rm AGN\_desired}$, we defined a scale factor $s = {F_{\rm AGN\_desired}}/{F_{\rm AGN\_actual}}$.
Combining the expressions for $s$ and $R$ gives $s = R * {F_{\rm galaxy}}/{F_{\rm AGN\_actual}}$.
In practice, $R$ must be calculated separately for each band:

\begin{equation}
    \label{eq:3}
    R^{\rm band} = \frac{F^{\rm band}_{\rm AGN\_desired}}{F^{\rm band}_{\rm galaxy}}.
\end{equation}
while $s$ is independent of wavelength, so that
\begin{equation}
    \label{eq:4}
    R * \frac{F_{\rm galaxy}}{F_{\rm AGN\_actual}} = R^{\rm band} * \frac{F^{\rm band}_{\rm galaxy}}{F^{\rm band}_{\rm AGN\_actual}} .
\end{equation}
Rearranging, we obtained an expression for band-specific contrast ratio $R^{band}$:
\begin{equation}
    \label{eq:5}
    R^{\rm band} = R * \frac{F_{\rm galaxy}}{F_{\rm AGN\_actual}} * \frac{F^{\rm band}_{\rm AGN\_actual}}{F^{\rm band}_{\rm galaxy}} .
\end{equation}
\noindent
Randomly selecting from the $R$ distribution,
using the randomly selected AGN SED and real galaxy, provides all the quantities needed to create the realistic simulated AGN (terms 
on the right side of Equation\,\ref{eq:5}), following the steps described in \S\,\ref{subsection:simgal}.

\begin{figure}
\figurenum{6a}
\gridline{\fig{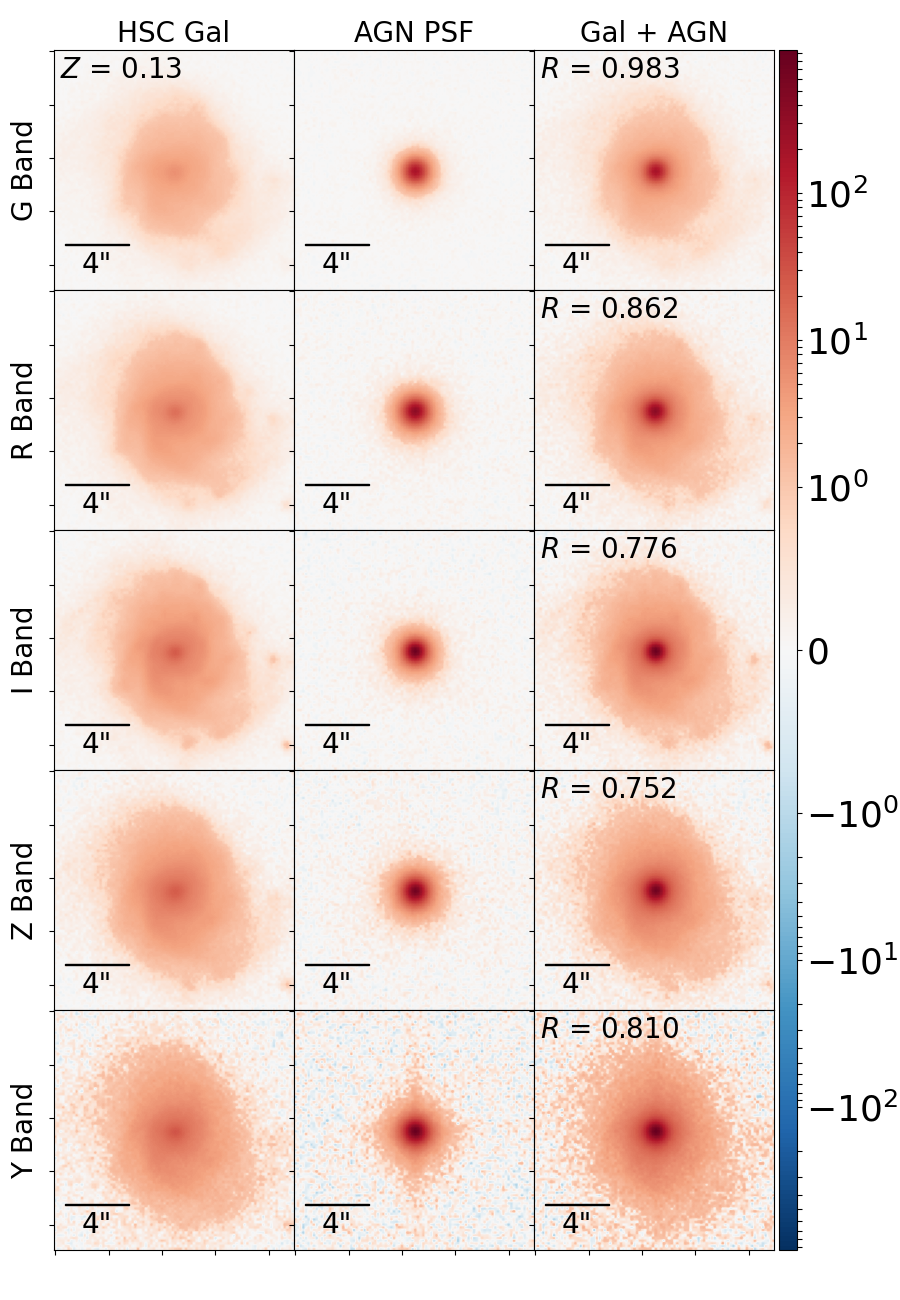}{0.5\textwidth}{}}
\vspace*{-0.4cm}
\caption{Example of adding an artificial AGN point source to a real galaxy in the low redshift bin.
{\it Top to bottom:} 
Images in the $g$, $r$, $i$, $z$, and $y$ bands.
{\it Left to right:} Original galaxy, AGN point source, galaxy plus AGN point source (what we call a ``realistic simulated AGN'' in the text). 
Contrast ratios were determined using real AGN SEDs, as described in the text.
All 15 sub-figures use the same colorbar limits.
\label{fig:agn_addition_low}}
\end{figure}

\begin{figure}
\figurenum{6b}
\gridline{\fig{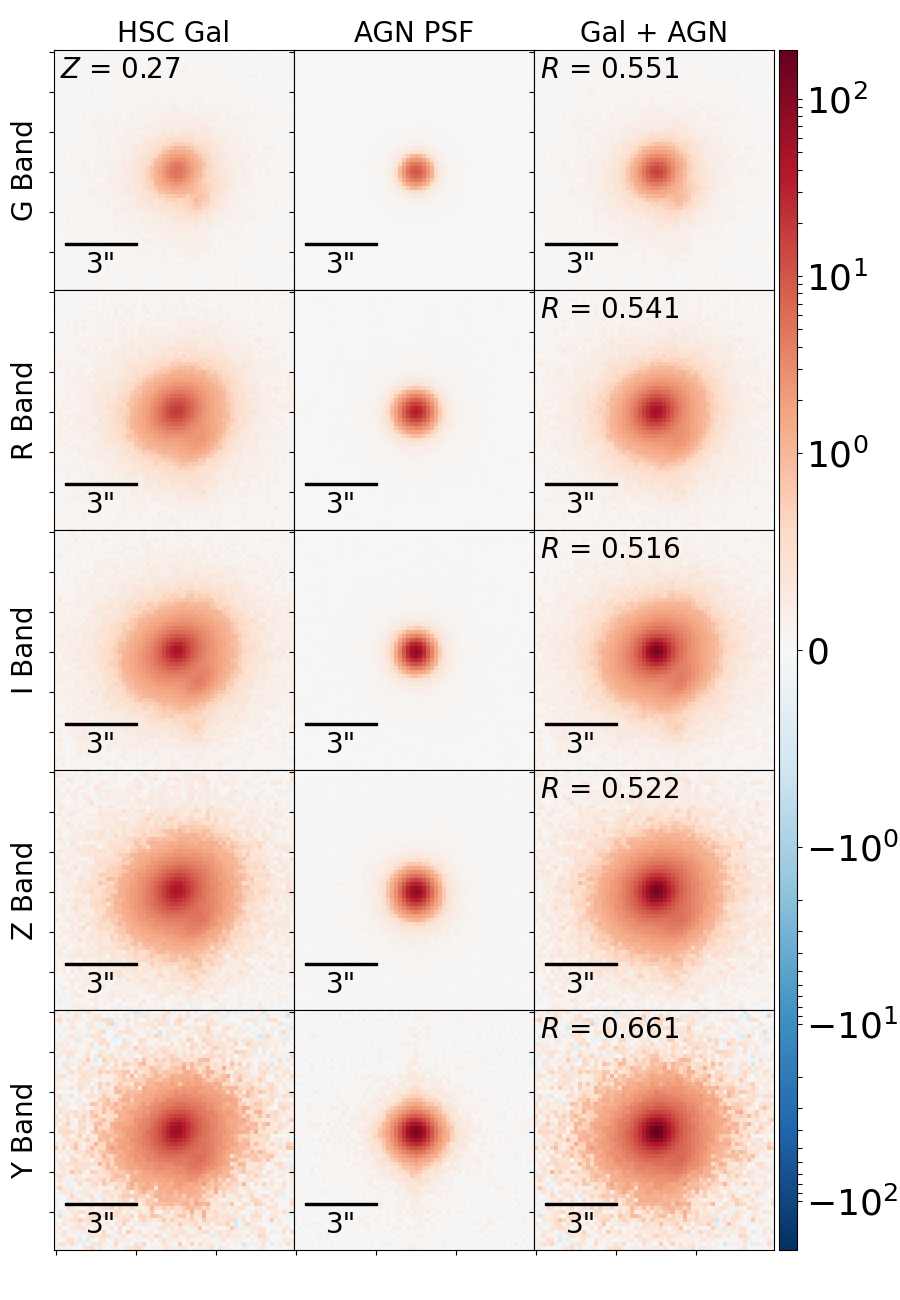}{0.5\textwidth}{}}
\vspace*{-0.4cm}
\caption{Example of a realistic simulated AGN in the medium redshift bin. All details are as in Figure \ref{fig:agn_addition_low}.
\label{fig:agn_addition_mid}}
\end{figure}

\begin{figure}
\figurenum{6c}
\gridline{\fig{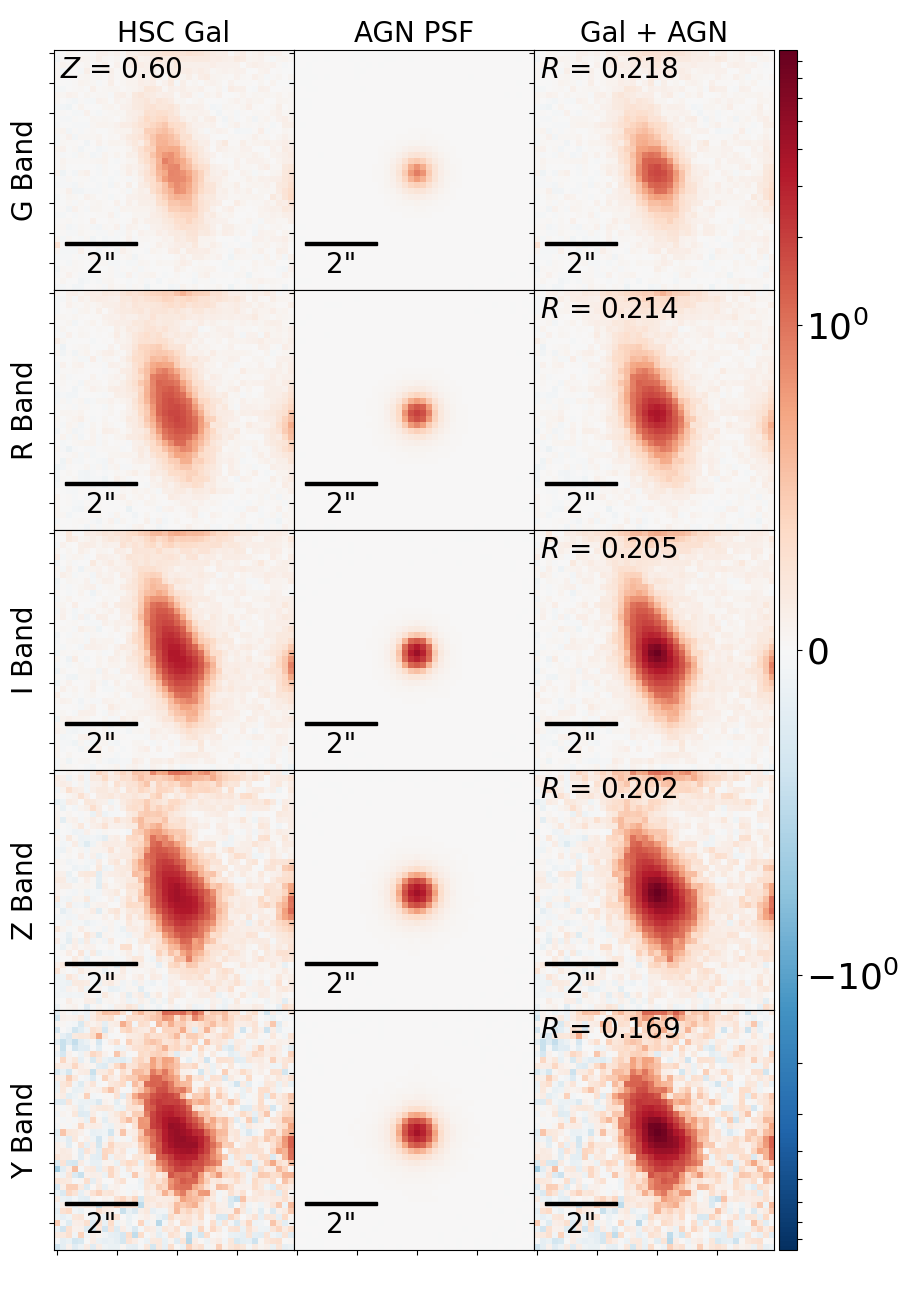}{0.5\textwidth}{}}
\vspace*{-0.4cm}
\caption{Example of a realistic simulated AGN for the high redshift bin. All details are as in Figure \ref{fig:agn_addition_low}.
\label{fig:agn_addition_high}}
\end{figure}

Figures \ref{fig:agn_addition_low}, \ref{fig:agn_addition_mid}, and \ref{fig:agn_addition_high} show three randomly selected examples of our multi-band simulated AGNs, one figure for each redshift bin. 
We are now ready to replicate the steps of \S~\ref{subsection:train}, modifying the version of \gamornet\ previously trained on simulated galaxies---specifically, fine-tuning its parameters using realistic simulated AGNs in a transfer learning step rather than re-training from scratch.

\subsection{Transfer Learning with Multi-band PSFGAN and GaMorNet}
\label{subsection:tl}

Having assembled all the pieces needed to train our final models, the next step was to convert the original PSFGAN, which processed images from a single band only, to a final version that processes all five bands at the same time (i.e., changing the number of input channels from one to five). Hereafter, we refer to this final version as multi-band PSFGAN\footnote{Note that compared to the original \citet{2018MNRAS.477.2513S} PSFGAN, our version of the encoding and decoding parts of the generator has only 7 instead of 8 layers, as explained in \S~\ref{subsection:train}.}. For each redshift bin, we trained a multi-band PSFGAN from scratch, such that it removed AGN point sources in all five bands simultaneously. 

We then took the \gamornet\ trained previously on simulated galaxies (\S~\ref{subsection:train}), made five copies of it (one for each band), and fine-tuned the hyper-parameters of each copy using PSFGAN-processed images of realistic simulated AGNs (\S~\ref{subsection:realgal}) in each of the five bands. Details of these steps are described below.

First, we created ground-truth morphology labels using bulge-to-total flux ratios and single \sersic\ indices, for galaxies from \citet{2011ApJS..196...11S} and galaxies from \citet{2018MNRAS.478.5410D}, respectively.  Specifically, following conventions introduced in \citet{2020ApJ...895..112G}, for \citet{2011ApJS..196...11S} data, we defined galaxies with $(B/T)<0.45$ ($(B/T)>0.55$) as disk-dominated (bulge-dominated), while galaxies with $0.45<(B/T)<0.55$ are indeterminate. On another hand, for \citet{2018MNRAS.478.5410D} data, as indicated by \citet{2020ApJ...895..112G}, there does not exist a perfect nor unique way to map \sersic\ indices to bulge-to-total flux ratios. Rather, they illustrated in their Figure 6 using simulation results in the redshift bin $z=1.075$ from \citet{2008ApJ...683..644S} that a choice of $n<2$ gives purely disk-dominated ($(B/T)<0.45$) galaxies, while galaxies with $n>2.5$ are mostly bulge-dominated ($(B/T)>0.55$)\footnote{We rechecked this correspondence between single \sersic\ indices and bulge-to-total flux ratios using our 50,873 low-redshift (mostly $z<0.25$) SDSS galaxies \citep{2011ApJS..196...11S} by comparing $(B/T)_{g}$ from Table 1 (SDSS Structural Parameters from $n_b=4$ bulge + disk Decompositions) and $n_g$ from Table 3 (SDSS Structural Parameters from Pure \sersic\ Decompositions), and confirmed that the $n_g<2$ bin is overwhelmingly dominated by galaxies with $(B/T)_{g}<0.45$, while the majority of galaxies with $n_g>2.5$ satisfy $(B/T)_{g}>0.55$.}. 

To summarize, we used the following redshift bin-dependent rules:
\begin{enumerate}
    \item For the low-redshift bin \citep{2011ApJS..196...11S}:
    \begin{enumerate}
        \item Galaxy is disk-dominated if the $g$-band bulge fraction, $(B/T)_{g}$, is less than 0.45.
        \item Galaxy is bulge-dominated if $(B/T)_{g} > 0.55$.
        \item Galaxy is indeterminate otherwise.
    \end{enumerate}
    \item For the medium-redshift bin \citep{2018MNRAS.478.5410D}:
    \begin{enumerate}
        \item Galaxy is disk-dominated if the F606-band single \sersic\ index, $n\_F606$, is less than 2.0.
        \item Galaxy is bulge-dominated if $n\_F606$ is greater than 2.5.
        \item Galaxy is indeterminate otherwise.
    \end{enumerate}
    \item For the high-redshift bin \citep{2018MNRAS.478.5410D}:
    \begin{enumerate}
        \item Galaxy is disk-dominated if the F814-band single \sersic\ index, $n\_F814$, is less than 2.0.
        \item Galaxy is bulge-dominated if $n\_F814$ is greater than 2.5.
        \item Galaxy is indeterminate otherwise.
    \end{enumerate}
\end{enumerate}
The point of using different observed bands for each redshift bin is to characterize galaxy morphology in the same approximate rest frame.
That said, a separate version of \gamornet{} was trained for each of the five HSC bands and we have made all five available to the public.

\begin{deluxetable*}{ccccccc}[htbp]
\tablecaption{Number of Real Galaxies Used to Train and Test PSFGAN \& \gamornet\ by Morphology and Redshift}
\label{tab:real_datasplit}
\tablecolumns{7}
\tablehead{
\colhead{Morphology} & \colhead{PSFGAN Train} & \colhead{PSFGAN Eval} & \colhead{\gamornet\ Train} & \colhead{\gamornet\ Eval} & \colhead{Test} & \colhead{Total} 
}
\startdata
    \hline
    \hline
    \multicolumn{7}{c}{$0<z<0.25$} \\
    \hline
    Disk-dominated & 5304 & 572 & 18849 & 2107 & 2795 & 29627 \\
    Indeterminate & 1012 & 113 & 3549 & 398 & 534 & 5606 \\
    Bulge-dominated & 2684 & 315 & 10002 & 1095 & 1544 & 15640 \\
    Total & 9000 & 1000 & 32400 & 3600 & 4873 & 50873\\
    \hline
    \hline
    \multicolumn{7}{c}{$0.25<z<0.5$} \\
    \hline
    Disk-dominated & 4113 & 443 & 3859 & 450 & 767 & 9632 \\
    Indeterminate & 387 & 57 & 432 & 46 & 78 & 1000 \\
    Bulge-dominated & 0 & 0 & 2009 & 204 & 347 & 2560 \\
    Total & 4500 & 500 & 6300 & 700 & 1192 & 13192\\
    \hline
    \hline
    \multicolumn{7}{c}{$0.5<z<1.0$} \\
    \hline
    Disk-dominated & 4059 & 444 & 3372 & 389 & 796 & 9060 \\
    Indeterminate & 441 & 56 & 359 & 37 & 111 & 1004 \\
    Bulge-dominated & 0 & 0 & 2569 & 274 & 617 & 3460 \\
    Total & 4500 & 500 & 6300 & 700 & 1524 & 13524\\
    \hline
    \hline
\enddata
\tablecomments{This table shows the number of disk-dominated, indeterminate, and bulge-dominated galaxies in each of the five training sets, for each of three redshift bins. Morphology labels were generated according to the rules in \S\,\ref{subsection:tl}. No bulges were included in the training and validation sets for PSFGAN in the mid- and high-redshift bins, in order to better balance the numbers of disks and bulges for \gamornet\ training and validation and in the common test set.}
\end{deluxetable*}

Next, we randomly split each sample of realistic simulated AGNs, in each redshift bin, into five subsets, analogous to what was done in \S~\ref{subsection:train}: a training set for PSFGAN, a validation set for PSFGAN, a training set for \gamornet, a validation set for \gamornet, and a common test set. The number of galaxies of each morphology in each subset and each redshift bin is shown in Table \ref{tab:real_datasplit}. 

For real galaxies, there are generally more disks than bulges. As we empirically tested, if a certain type (either disk or bulge) has a disproportionately small number of examples in the training set, \gamornet\ will be unable to correctly identify that type. This is exactly the case in the medium- and high-redshift bins (there are too few bulges, see \S\,\ref{subsection:data used}).
Moreover, even after creating rotated and reflected copies of bulge-dominated systems for these redshift bins, the number of real spheroidal galaxies is still too limited. 
Rather than discard some disks in order to make the numbers more even, 
we put all the bulge-dominated galaxies into the \gamornet{} training, validation and common test sets, allocating none to the PSFGAN training and validation.
This works because removal of the bright AGN point source is basically independent of the host galaxy morphology. 
We checked this by comparing two PSFGANs: one trained with only disk-dominated and indeterminate galaxies, and a second trained with all three morphological types (fluxes, sizes, and all other parameters were the same). We found no difference between the two; the quality of the host galaxy reconstruction, as measured by the residuals is the same.
Apart from this exclusion, data sets were evenly distributed with respect to any other parameter (e.g., half light radius, galaxy flux, etc.).

We trained each multi-band PSFGAN by requiring the processed AGN image closely resembles the original galaxy, simultaneously in all five HSC bands.
We trained a separate multi-band PSFGAN for each redshift bin (i.e., there are three versions of multi-band PSFGAN). 
The best sets of hyper-parameters for each redshift bin are summarized in Table \ref{tab:psfgan_real_details}.
These PSFGANs were applied to the training and validation sets of \gamornet, as well as to the common test set. 

We then used the PSFGAN-processed images in the transfer learning step for \gamornet. 
For each redshift bin, we took the \gamornet\ previously trained on simulated galaxies (\S~\ref{subsection:train}), and made five copies of it. Each copy was re-trained on the training set of \gamornet\ in each of the five HSC bands. In total, there are 15 versions of \gamornet\, corresponding to $3$ redshift bins and $5$ bands.
For each version, we tested various combinations of hyper-parameters using the validation set of \gamornet, as summarized in Table \ref{tab:gamornet_search_space}.
The best sets of hyper-parameters\footnote{We empirically found that, although in the low-redshift bin the difference of fixing different layers is usually minor or moderate (i.e. a less than $5\%$ increase in precision), in the mid- or high-redshift bin this difference could be major (i.e. an increase in precision close to $10\%$). We therefore choose to fix different layers in each redshift bin and each band, in order to maximize the performance of each version of \gamornet{}.}, shown in Table \ref{tab:gamornet_real_details}, were used to train the final versions of \gamornet.

However, choosing which layer(s) to fix for different bands and/or redshift bins is purely an empirical decision. As our goal is to maximize model performance for each individual version of \gamornet{}, we let the choice of layers with fixed parameters vary independently in each redshift bin and each band. The performance difference between cases of fixing different layers in the low-redshift bin is usually minor to moderate (e.g. a precision increase from $84\%$ to $88\%$), but in the mid- and high-redshift bins it could be major (e.g. a precision increase from $71\%$ to $82\%$). We therefore decided to use redshift bin and band dependent choices of fixed layers. 

\begin{deluxetable}{cccc}[htbp]
\tablecaption{Hyper-Parameters for Multi-band PSFGAN Training On Real Galaxies}
\label{tab:psfgan_real_details}
\tablecolumns{4}
\tablehead{
\colhead{Redshift Bin} & \colhead{Low} & \colhead{Mid} & \colhead{High} 
} 
\startdata
    \hline
    Learning Rate & $9\times10^{-5}$ & $2\times10^{-5}$ & $5\times10^{-6}$ \\
    Epoch & $20$ & $40$ & $40$ \\
    Attention Window\tablenotemark{a} & $22$ & $16$ & $12$ \\
    Attention Parameter & $0.05$ & $0.05$ & $0.05$ \\
    Stretch Function & arcsinh & arcsinh & arcsinh \\
    Stretch Factor & $50$ & $50$ & $50$ \\
    Maximum Pixel Value & $45,000$ & $10,000$ & $1,000$ \\
    \hline
    \hline
\enddata
\tablenotetext{a}{The length of each side, in pixels, of the square attention window at the image center.}
\end{deluxetable}

\begin{deluxetable*}{ccccc}[htbp]
\tablecaption{Hyper-Parameters Tested for \gamornet\ Transfer Learning On Recovered Real Galaxies
}
\label{tab:gamornet_search_space}
\tablecolumns{5}
\tablehead{
\colhead{Hyper-Parameter (Values Tested on the Right)} & \colhead{~~~} & \colhead{~~~} & \colhead{~~~} & \colhead{~~~}
} 
\startdata
    \hline
    Epochs & $50$ & $100$ & $200$ & $500$ \\
    Batch Size & $64$ & $128$ & $256$ & $512$ \\
    Learning Rate & $1\times 10^{-5}$ & $2\times 10^{-5}$ & $5\times 10^{-5}$ & $1\times 10^{-4}$ \\
    Fixed Layers\tablenotemark{a} (Loaded from Previous Model)
    & None & (2) & (2, 5) & (2, 5, 8) \\
    Layers Trained from Scratch & None & (17) & (15, 17) & (13, 15, 17)\\
    Trainable Layers Loaded from Previous Model & ~~~ & All remaining layers\tablenotemark{b} & ~~~ & ~~~ \\
    \hline
    \hline
\enddata
\tablenotetext{a}{There are 17 layers in \gamornet\ (see Figure\,4 and Table 2 in \citealp{2020ApJ...895..112G}), of which five are convolutional (layers 2, 5, 8, 9, 10), three are fully connected (13, 15, and 17), and the rest are functional (specifically, max pooling, dropout, and local response normalization layers, which contain no trainable parameters). Further details are given in \S\,\ref{subsection:tl}.}
\tablenotetext{b}{All layers that are not fixed or trained from scratch.}
\end{deluxetable*}

\begin{deluxetable*}{ccccccc}[htbp]
\tablecaption{Hyper-Parameters Used for Transfer Learning of \gamornet\ On Recovered Real Galaxies
\label{tab:gamornet_real_details}}
\tablecolumns{7}
\tablehead{
\colhead{Band} & \colhead{Epoch} & \colhead{Batch Size} & \colhead{Learning Rate} & \colhead{Layers\tablenotemark{a} Loaded, Fixed} & \colhead{Layers Loaded, Trainable} & \colhead{Layers Trained from Scratch}
}
\startdata
    \hline
    \hline
    \multicolumn{7}{c}{$0<z<0.25$} \\
    \hline
    G & 100 & 128 & $5\times 10^{-5}$ & (2) & (5, 8, 9, 10) & (13, 15, 17)  \\
    R & 100 & 128 & $5\times 10^{-5}$ & (2, 5, 8) & (9, 10, 13, 15, 17) & None  \\
    I & 100 & 128 & $5\times 10^{-5}$ & None & All layers & None  \\
    Z & 100 & 128 & $5\times 10^{-5}$ & (2, 5) & (8, 9, 10, 13, 15, 17) & None \\
    Y & 100 & 128 & $5\times 10^{-5}$ & (2) & (5, 8, 9, 10, 13, 15, 17) & None \\
    \hline
    \hline
    \multicolumn{7}{c}{$0.25<z<0.5$} \\
    \hline
    G & 200 & 128 & $5\times 10^{-5}$ & (2, 5, 8) & (9, 10, 13, 15, 17) & None  \\
    R & 200 & 128 & $5\times 10^{-5}$ & (2, 5, 8) & (9, 10) & (13, 15, 17)  \\
    I & 200 & 128 & $5\times 10^{-5}$ & (2, 5, 8) & (9, 10, 13, 15) & (17)  \\
    Z & 200 & 128 & $5\times 10^{-5}$ & (2, 5, 8) & (9, 10, 13, 15) & (17) \\
    Y & 200 & 128 & $5\times 10^{-5}$ & (2, 5, 8) & (9, 10) & (13, 15, 17) \\
    \hline
    \hline
    \multicolumn{7}{c}{$0.5<z<1.0$} \\
    \hline
    G & 200 & 128 & $5\times 10^{-5}$ & (2) & (5, 8, 9, 10, 13, 15, 17) & None  \\
    R & 200 & 128 & $5\times 10^{-5}$ & None & All layers & None  \\
    I & 200 & 128 & $5\times 10^{-5}$ & (2, 5) & (8, 9, 10) & (13, 15, 17)  \\
    Z & 200 & 128 & $5\times 10^{-5}$ & (2) & (5, 8, 9, 10, 13, 15) & (17) \\
    Y & 200 & 128 & $5\times 10^{-5}$ & (2, 5, 8) & (9, 10, 13) & (15, 17) \\
    \hline
    \hline
\enddata
\tablenotetext{a}{There are 17 layers in \gamornet\ (see Figure\,4 and Table 2 in \citealp{2020ApJ...895..112G}), of which five are convolutional (layers 2, 5, 8, 9, 10), three are fully connected (13, 15, and 17), and the rest are functional (specifically, max pooling, dropout, and local response normalization layers, which contain no trainable parameters). Further details are given in \S\,\ref{subsection:tl}.}
\tablecomments{This table shows hyper-parameters for 15 versions of \gamornet\ (5 bands $\times$ 3 redshift bins) which are trained separately in the transfer learning step. 
Other (common) hyper-parameters include momentum (0.9), learning rate decay (0), loss function (cross entropy).}
\end{deluxetable*}

\begin{figure}
\figurenum{7a}
\gridline{\fig{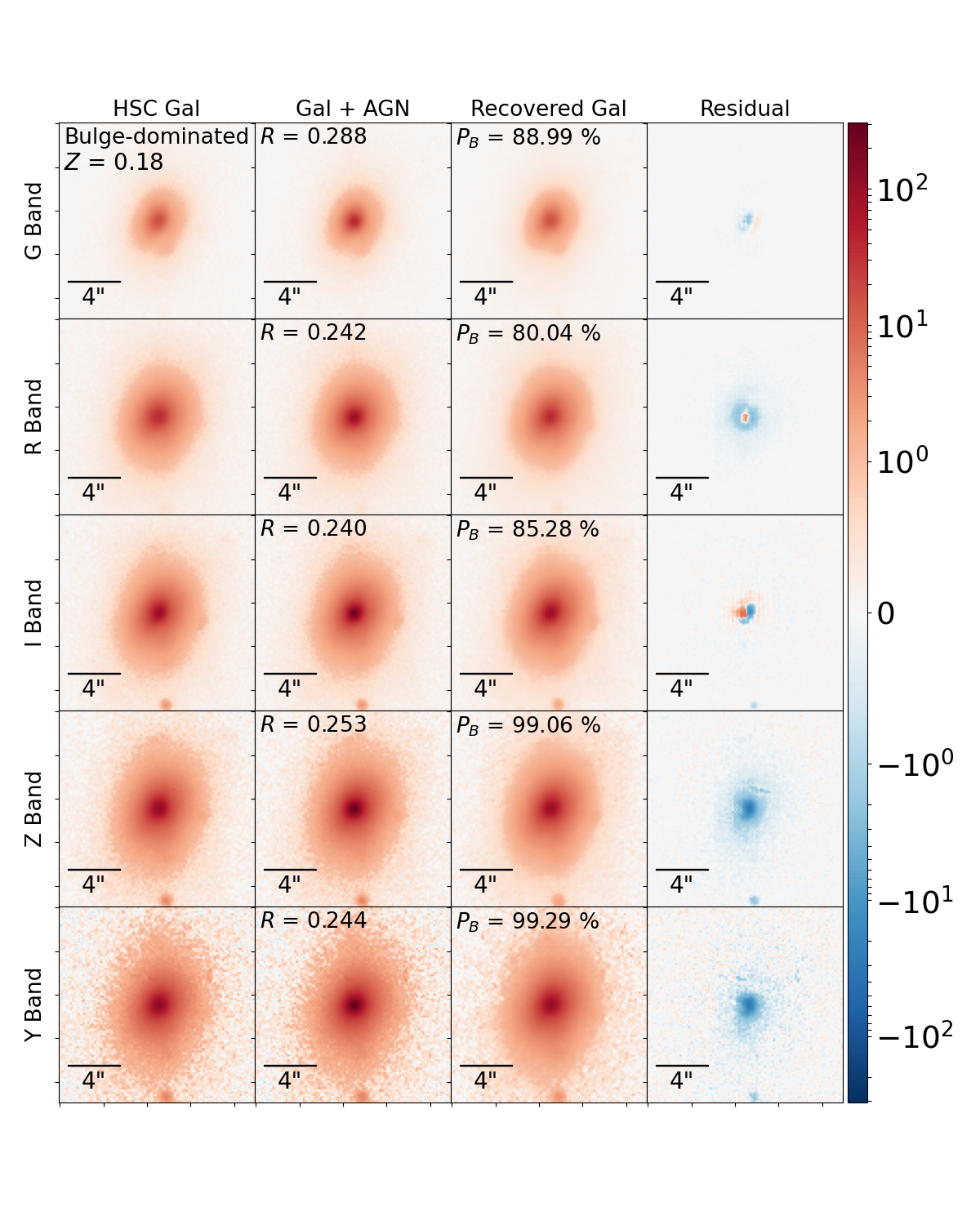}{0.54\textwidth}{}}
\vspace*{-0.4cm}
\caption{
Examples of the performance of multi-band PSFGAN and \gamornet{} on a realistic simulated AGN in the low-redshift bin ($0<z<0.25$). 
Each row corresponds to a different HSC band 
({\it top to bottom}): 
$g$, $r$, $i$, $z$, and $y$.
{\it Left to right:}
original galaxy (from the HSC Wide survey), 
realistic simulated AGN (i.e., original galaxy with an AGN point source added), 
galaxy recovered by multi-band PSFGAN, 
and the residual image (recovered galaxy minus original galaxy). 
Labels in the boxes represent ({\it left to right}) the morphology labels from \citet{2011ApJS..196...11S} or \citet{2018MNRAS.478.5410D}, contrast ratios (AGN flux to galaxy flux in that band), and probabilities that \gamornet\ correctly identifies the morphology of the recovered galaxy.
The small residuals and accurate morphologies shows that PSFGAN+\gamornet{} works well to characterize AGN host galaxies. 
}
\label{fig:psf_gmn_real_low_a}
\end{figure}

\begin{figure}
\figurenum{7b}
\gridline{\fig{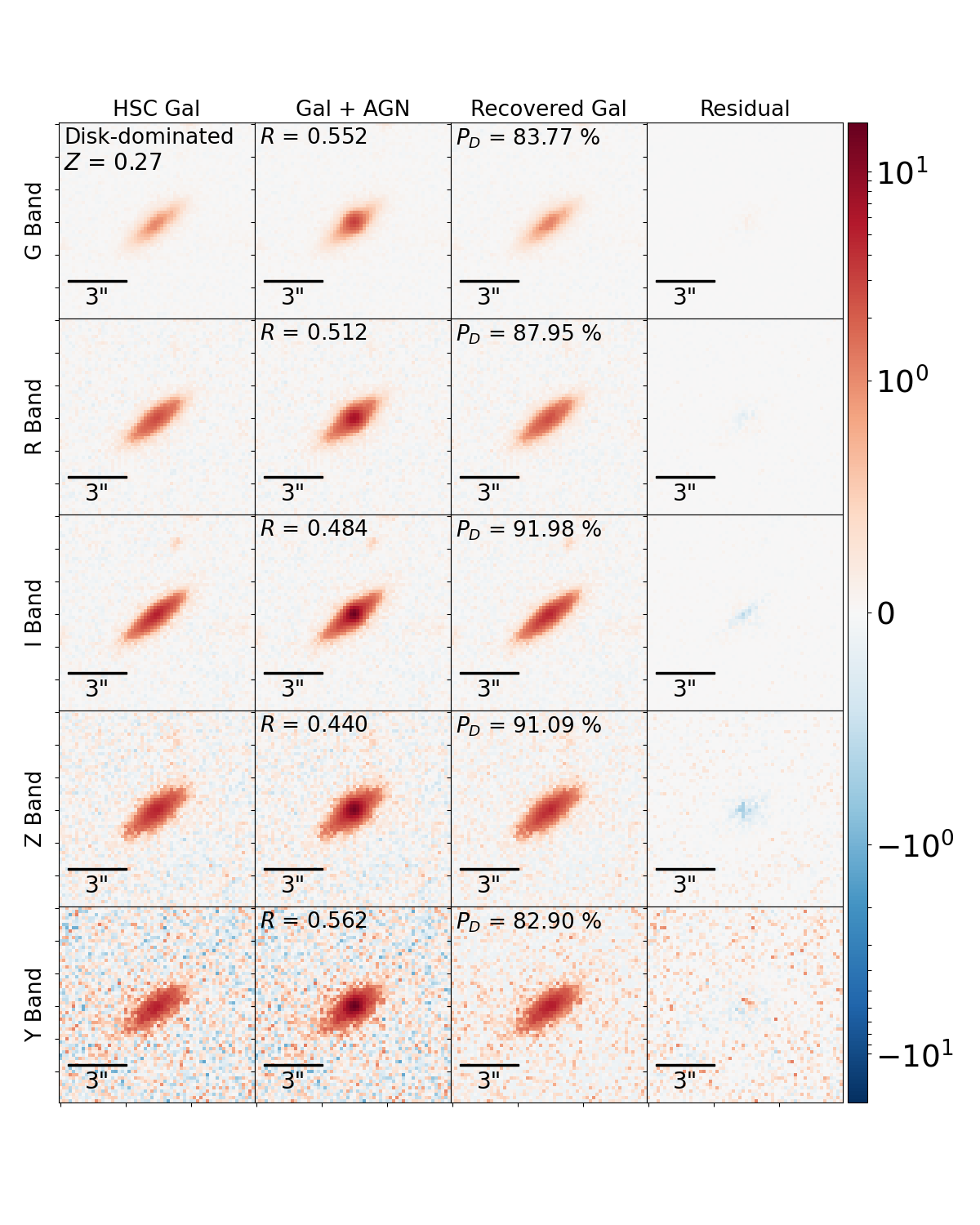}{0.54\textwidth}{}}
\caption{Performance of multi-band PSFGAN and \gamornet{} on a realistic simulated AGN in the medium-redshift bin ($0.25<z<0.5$).
Format and details as in Fig.\,\ref{fig:psf_gmn_real_low_a}.}
\label{fig:psf_gmn_real_mid_a}
\end{figure}

\begin{figure}
\figurenum{7c}
\gridline{\fig{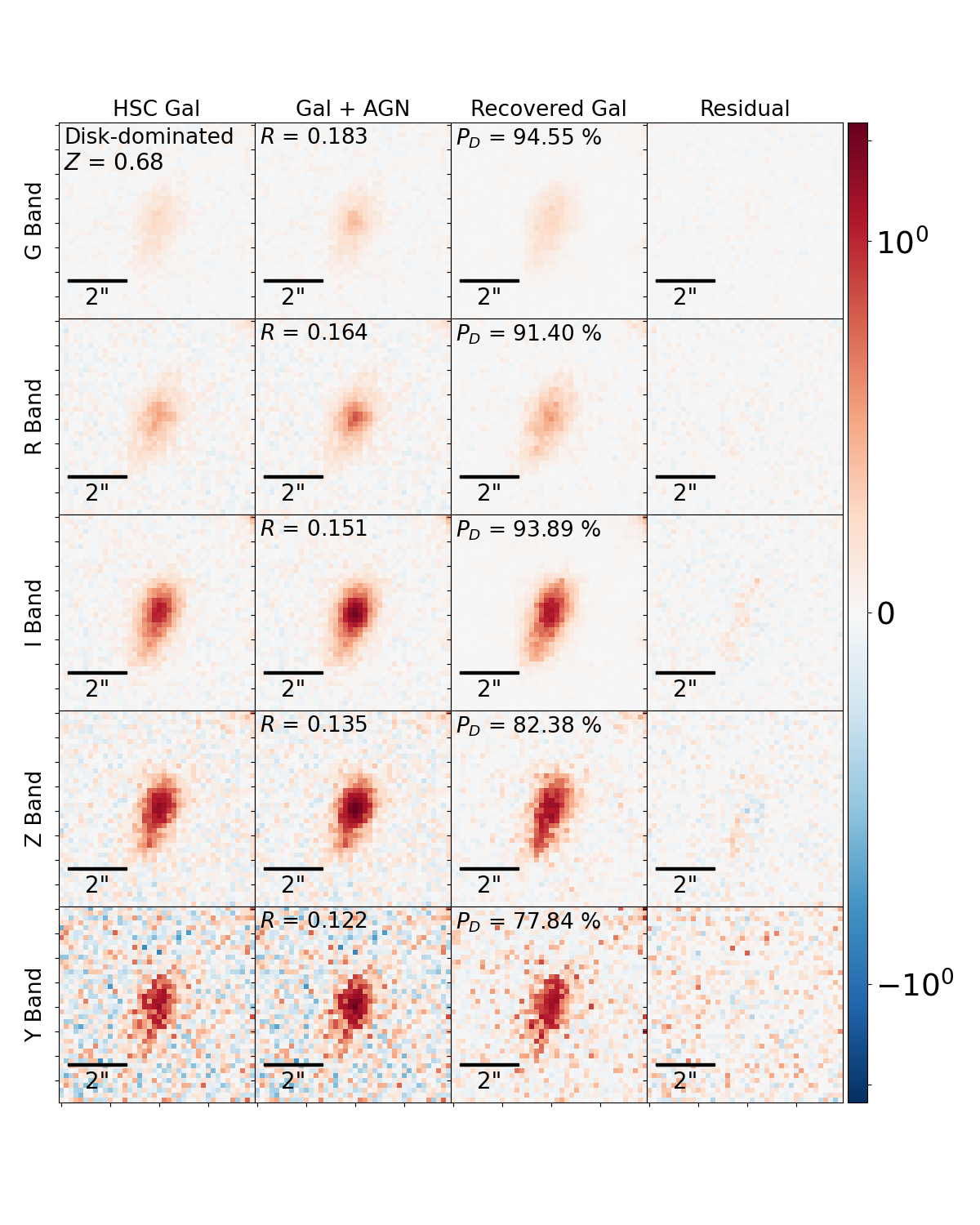}{0.54\textwidth}{}}
\caption{Performance of multi-band PSFGAN and \gamornet{} on a realistic simulated AGN in the high-redshift bin ($0.5<z<1$).
Format and details as in Fig.\,\ref{fig:psf_gmn_real_low_a}.}
\label{fig:psf_gmn_real_high_a}
\end{figure}

\begin{figure}
\figurenum{8a}
\gridline{\fig{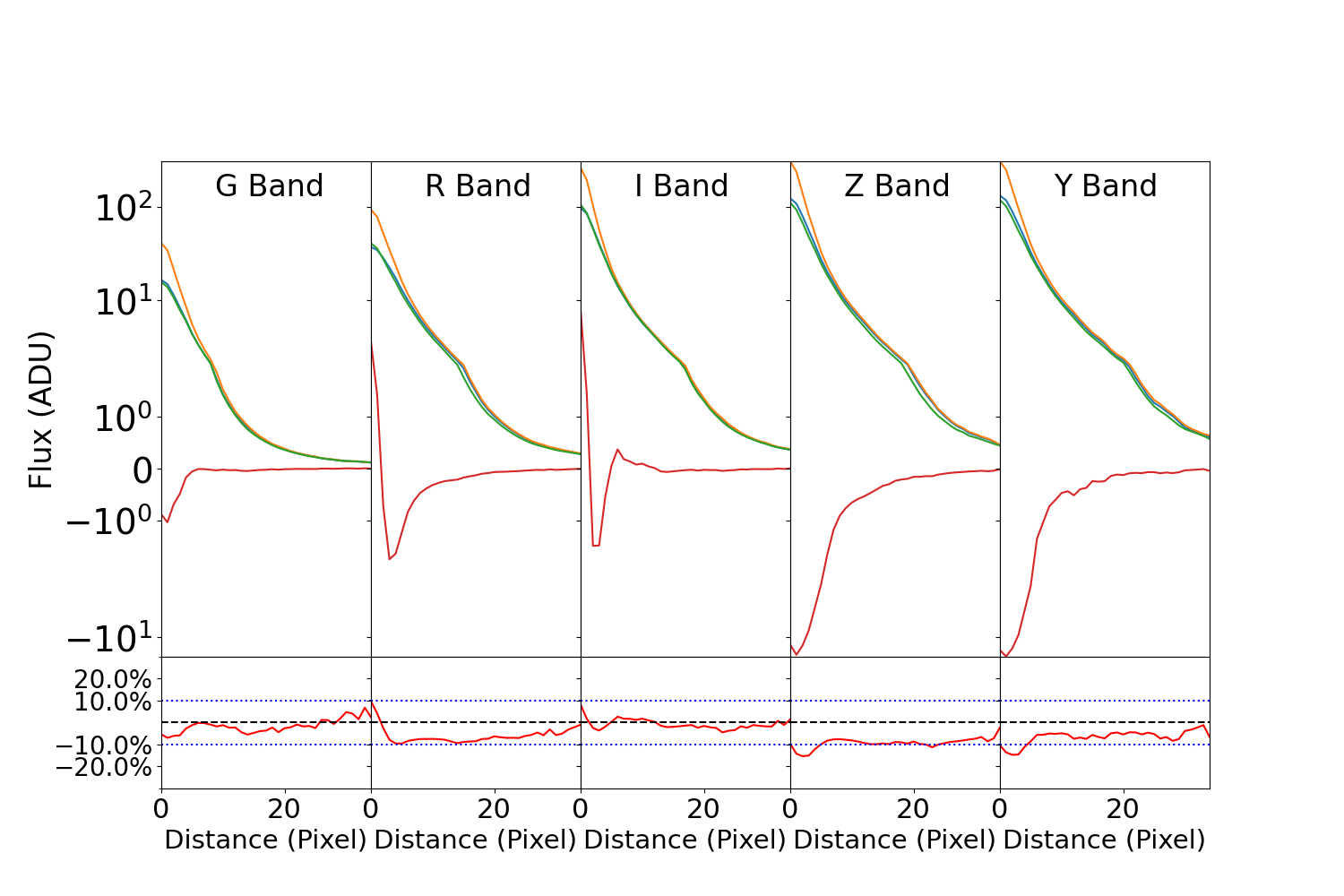}{0.56\textwidth}{}}
\vspace*{-0.4cm}
\caption{Radial profiles ({\it top}) of the example images from Fig.\,\ref{fig:psf_gmn_real_low_a}, from the low-redshift bin, for the realistic simulated AGN ({\it orange curve}), original galaxy ({\it blue curve}), recovered galaxy ({\it green curve}), and residual ({\it red curve}, defined as original minus recovered flux). 
Bottom panels show the absolute value of the residuals ({\it red curve}), expressed as the percentage of the flux of the original galaxy. 
In most cases, the residuals are a few percent of the original flux across most of the profile. 
}
\label{fig:psf_gmn_real_low_b}
\end{figure}

\begin{figure}
\figurenum{8b}
\gridline{\fig{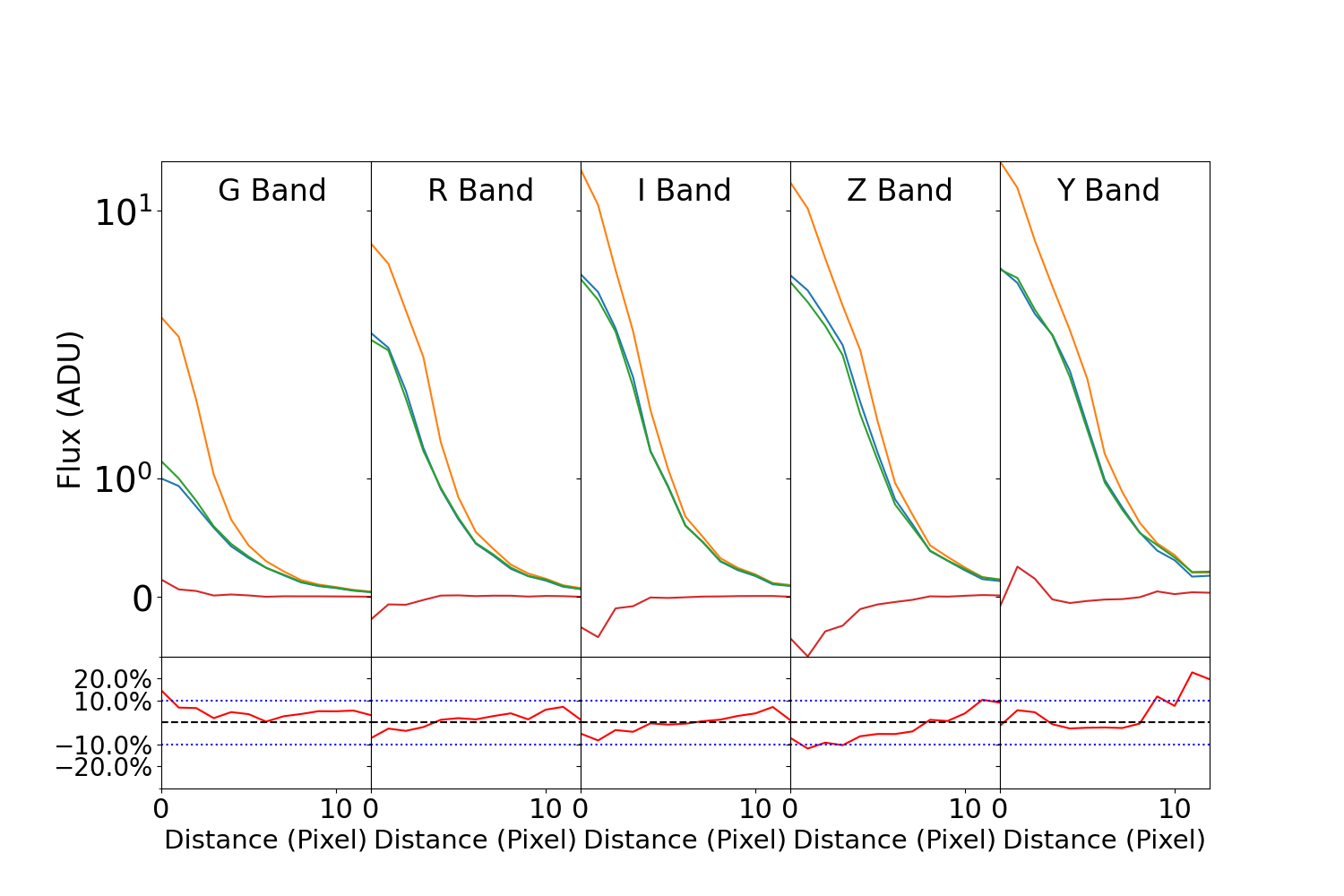}{0.56\textwidth}{}}
\caption{Radial profiles of the example images from Fig.\,\ref{fig:psf_gmn_real_mid_a}, from the medium-redshift bin. 
Format and colors as in Fig.\,\ref{fig:psf_gmn_real_low_b}. Again, residuals are a few percent of the original flux across most of the profile.
\label{fig:psf_gmn_real_mid_b}}
\end{figure}

\begin{figure}
\figurenum{8c}
\gridline{\fig{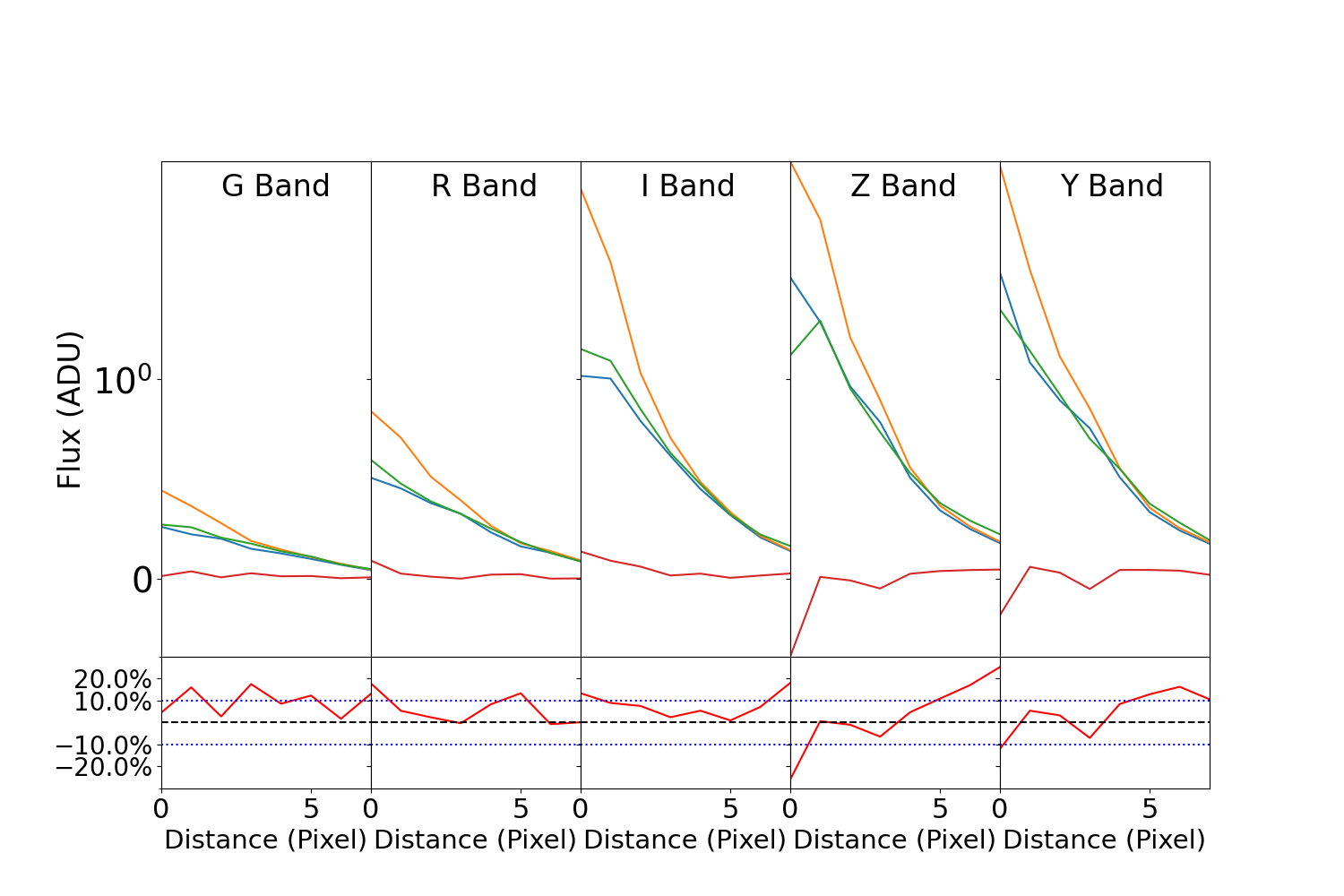}{0.56\textwidth}{}}
\caption{Radial profiles of the example images from Fig.\,\ref{fig:psf_gmn_real_high_a}, from the high-redshift bin. 
Format and colors as in Fig.\,\ref{fig:psf_gmn_real_low_b}. Again, residuals are a few percent of the original flux across most of the profile.
\label{fig:psf_gmn_real_high_b}}
\end{figure}

The final PSFGAN + \gamornet\ model consists of, for each redshift bin, 1 multi-band PSFGAN and 5 versions of \gamornet, one for each band. 
We applied this final model to realistic simulated AGNs in the common test set---specifically, applying the multi-band PSFGAN to remove the AGN point source in five bands simultaneously, then using the appropriate \gamornet\ to characterize the morphology of the PSFGAN-processed image in each band. 
Figures \ref{fig:psf_gmn_real_low_a}, \ref{fig:psf_gmn_real_mid_a}, and \ref{fig:psf_gmn_real_high_a} shows a representative example for each of the three redshift bins, respectively, using the same format as Figure \ref{fig:psf_gmn_sim_a}. \gamornet{} predictions (given in the third column of each row of images) are generally pretty accurate, falling off slightly at higher redshift as the signal-to-noise ratio decreases. For galaxies similar in size to the AGN point source, \gamornet\ becomes less accurate but still has $\gtrsim80$\% precision in most bands.
Corresponding radial profiles are shown in Figures \ref{fig:psf_gmn_real_low_b}, \ref{fig:psf_gmn_real_mid_b}, and \ref{fig:psf_gmn_real_high_b}, in the same format as Figure \ref{fig:psf_gmn_sim_b}. 
Most of the residuals are a few percent of the flux of the original galaxy over most of the profile.

\begin{deluxetable}{cccc}[htbp]
\tablecaption{
Summary of Terminology for True and False Identifications
\label{tab:true_vs_pred}}
\tablecolumns{4}
\tablehead{
\colhead{} & \colhead{Pred Disk} & \colhead{Pred Indet} & \colhead{Pred Bulge}
}
\startdata
    \hline
    \hline
    Actual Disk&
    $TD$ &
    $FI_D$ &
    $FB_D$  \\
    Actual Indet&
    $FD_I$ &
    $TI$ &
    $FB_I$  \\
    Actual Bulge &
    $FD_B$ &
    $FI_B$ &
    $TB$  \\
    \hline
    \hline
\enddata
\tablecomments{Each PSFGAN-recovered galaxy from the common test set has an ``actual'' morphology (given by \citealp{2011ApJS..196...11S} or \citealp{2018MNRAS.478.5410D}) and a ``predicted'' morphology (from \gamornet{}). For example, ``$TD$'' (True Disk) stands for actual disks that are correctly predicted as disks, 
while ``$FD_B$'' (False Disk [Actual Bulge])  stands for actual bulges that are incorrectly predicted as disks.
These quantities are used in \S~\ref{subsection:tradeoff} to evaluate performance of the \gamornet\ models. 
}
\end{deluxetable}

\begin{figure}
\figurenum{9}
\gridline{\fig{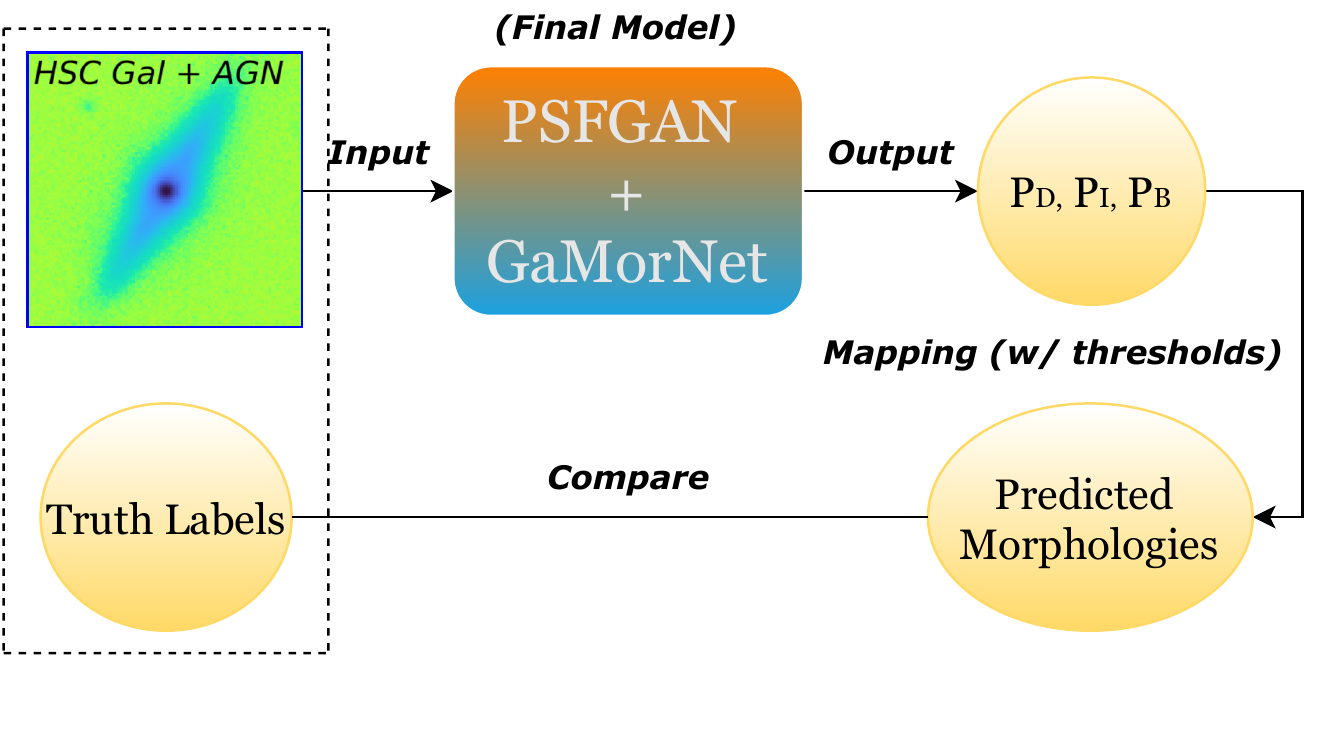}{0.5\textwidth}{}}
\vspace*{-0.6cm}
\caption{
How output from PSFGAN+\gamornet{} is mapped to a predicted morphology.
A realistic simulated AGN ({\it top left}) is input to the final PSFGAN+\gamornet, which outputs 3 raw probabilities that the host galaxy is a disk, indeterminate, or bulge ({\it top right}). These are mapped to a predicted morphology ({\it bottom right})
by comparing with the actual morphology (\citealp{2011ApJS..196...11S} or \citealp{2018MNRAS.478.5410D}; {\it bottom left}), 
as described in \S~\ref{subsection:tradeoff}.
\label{fig:true_vs_pred_diagram}}
\end{figure}

\begin{figure*}
\figurenum{10}
\gridline{\fig{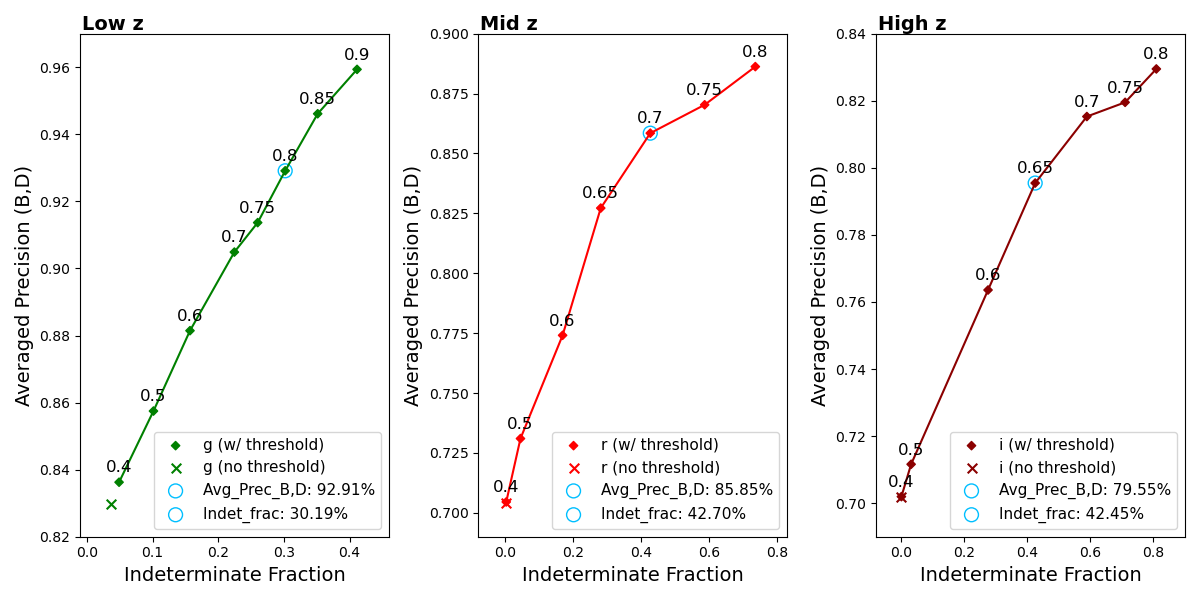}{1.0\textwidth}{}}
\vspace*{-0.4cm}
\caption{
Precision versus indeterminate fraction for galaxies in the common test set.
\textit{Left to right}:
low-, mid- and high-redshift bins, in the same rest-frame band ($g$, $r$, $i$, respectively).
Numbered points ({\it diamonds}) indicate a possible threshold for mapping the \gamornet{} probability to a galaxy classification (see Eq. \ref{eq:12}), for disks or bulges only.
The unnumbered point ({\it cross}) corresponds to not using any threshold. 
In this case, we use the maximum probability for each galaxy (i.e., the maximum of $P_D$, $P_B$, $P_I$).
Our chosen trade-off between $Indet\_Frac$ and $Ave\_Precision$ ({\it cyan circle}) ensures high precision for at least half the galaxies; depending on science goals, other users may prefer a different threshold. 
}
\label{fig:confid_thold_valid}
\end{figure*}

\begin{figure*}
\figurenum{11}
\gridline{\fig{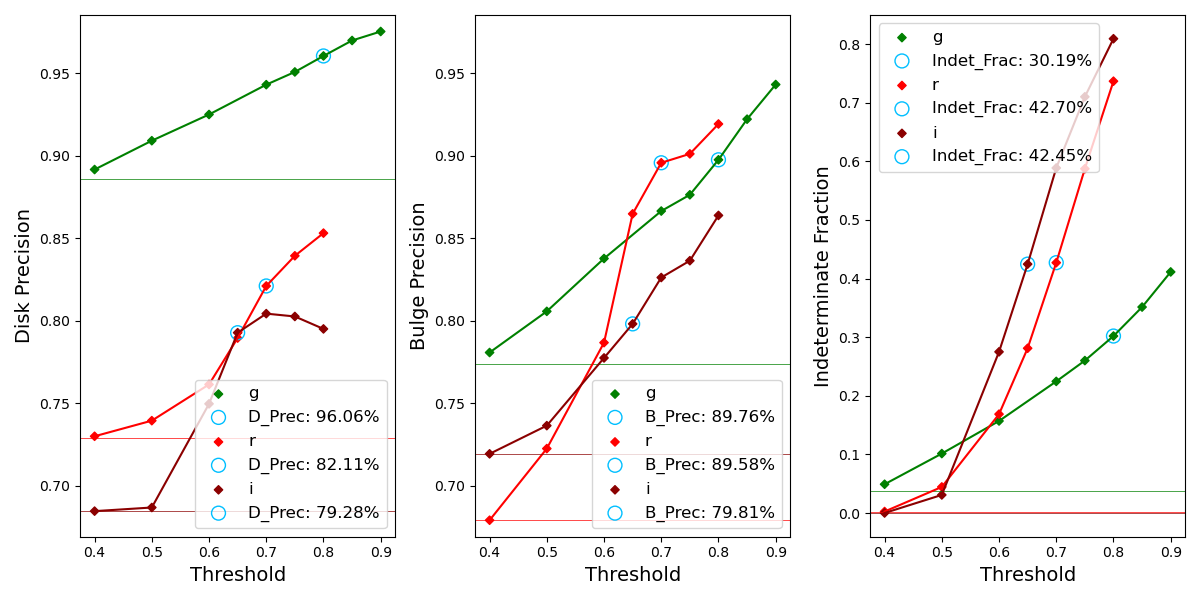}{1.0\textwidth}{}}
\vspace*{-0.4cm}
\caption{
Disk precision, bulge precision and indeterminate fraction (\textit{left to right}) versus threshold, for galaxies in the common test set.
Each panel shows low-redshift ($g$-band, {\it green}), medium-redshift ($r$-band, {\it red}), and high-redshift ($i$-band, {\it brown}) galaxies, for comparison.
$Diamonds$ were generated using the threshold mapping in \S~\ref{subsection:tradeoff}.
$Disk\_Precision$, $Bulge\_Precision$ and $Indet\_Frac$ obtained without using a threshold (i.e., using the highest \gamornet{} probability) are shown in horizontal lines.
Our chosen thresholds are marked by \textit{light blue open circles}. 
In general, precision is best at lower redshifts, and better for disks than for bulges.
}
\label{fig:confid_thold_all}
\end{figure*}

\begin{figure*}
\figurenum{12}
\gridline{\fig{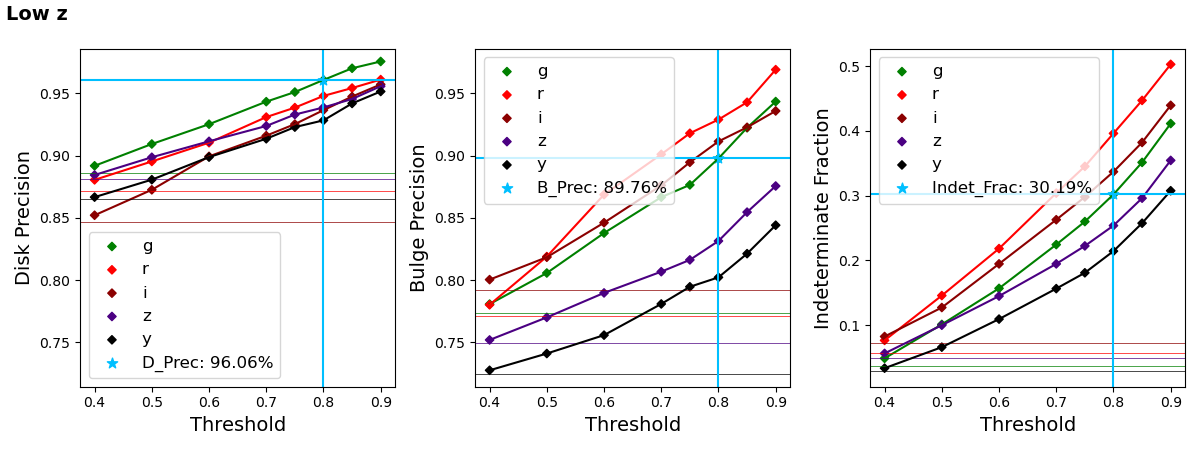}{1.0\textwidth}{}}
\vspace*{-0.6cm}
\gridline{\fig{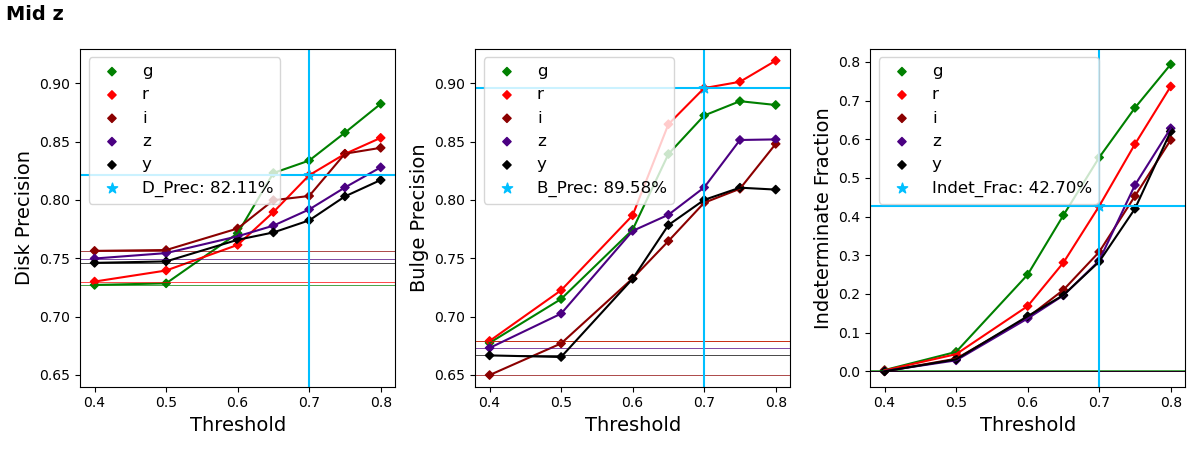}{1.0\textwidth}{}}
\vspace*{-0.6cm}
\gridline{\fig{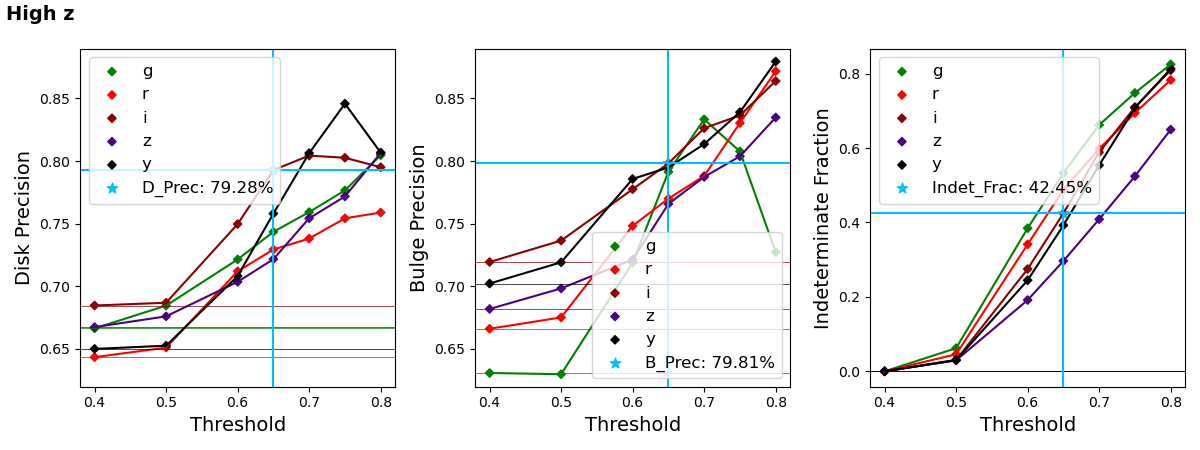}{1.0\textwidth}{}}
\vspace*{-0.6cm}
\caption{
Disk precision, bulge precision and indeterminate fraction (\textit{left to right}) versus confidence threshold, for galaxies in the common test set, for all five bands (colors shown in the key).
Low-, mid- and high-redshift bins are in the {\it top}, {\it middle}, and {\it bottom} rows, respectively. 
$Disk\_Precision$, $Bulge\_Precision$ and $Indet\_Frac$ obtained without using a threshold (i.e., using the highest \gamornet{} probability) are shown in horizontal lines.
Chosen thresholds are shown by \textit{cyan lines}.
Overall, both the precision and the indeterminate fraction increase with increasing thresholds (in most bands/bins monotonically). 
\label{fig:confid_thold_each}}
\end{figure*}

\begin{figure*}
\figurenum{13}
\gridline{\fig{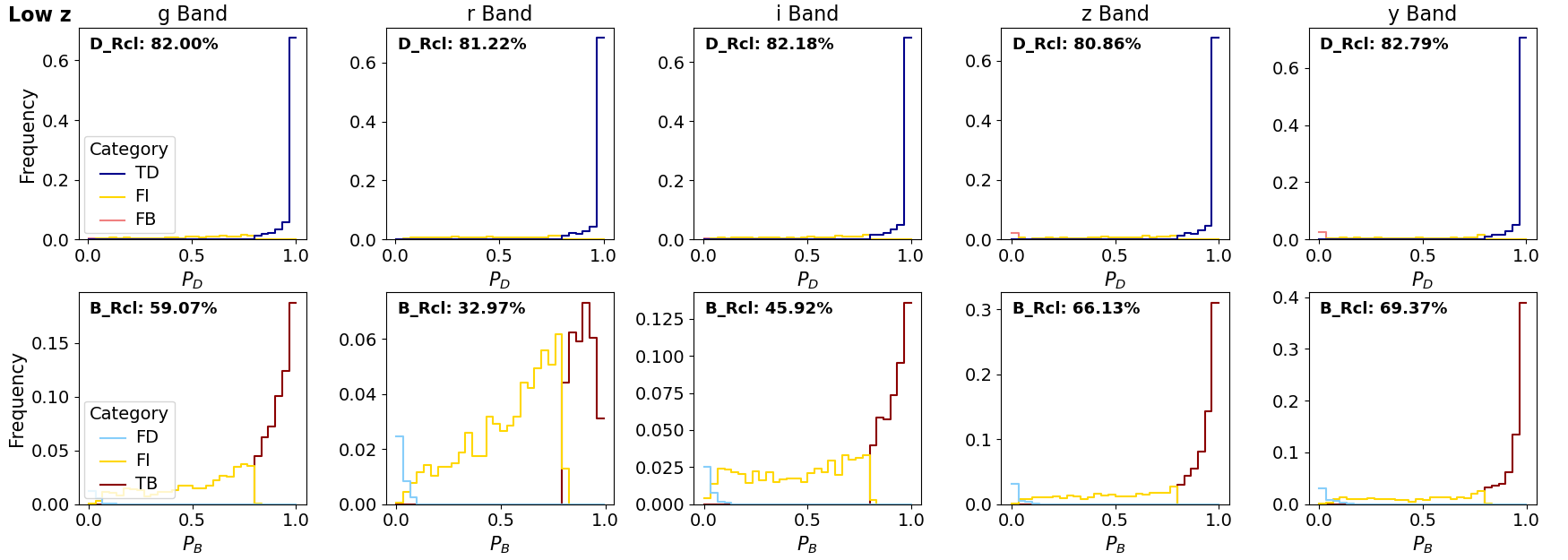}{1.05\textwidth}{}}
\vspace*{-0.8cm}
\gridline{\fig{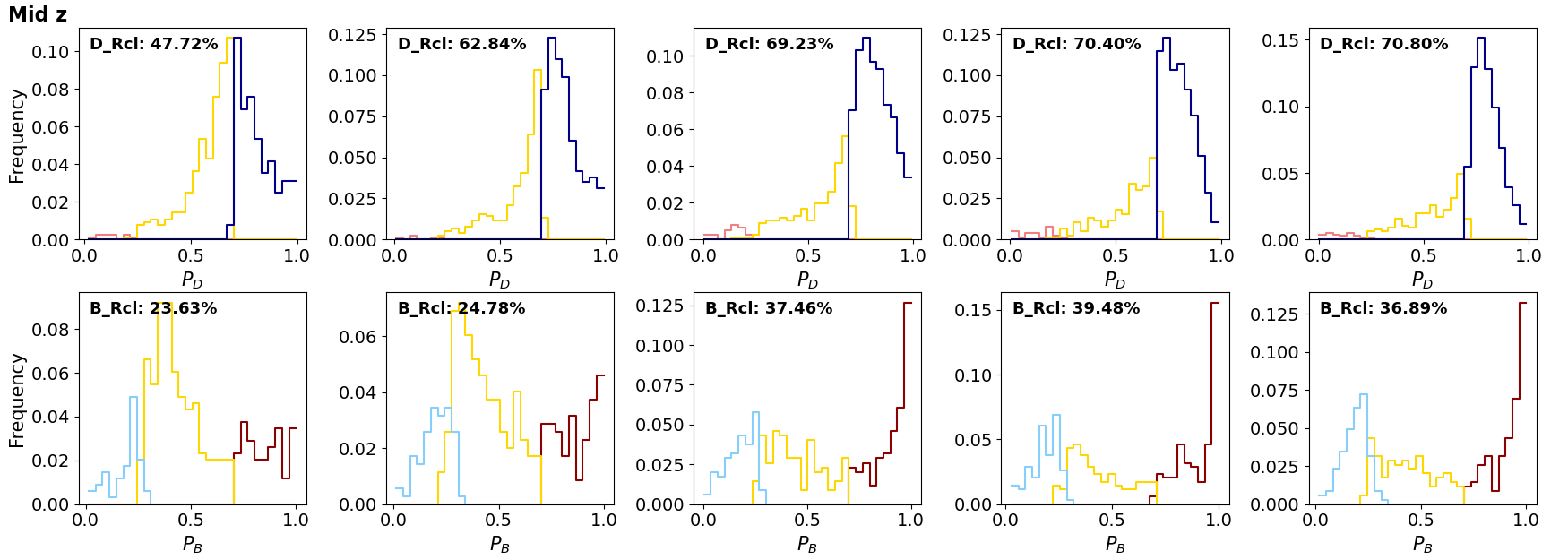}{1.05\textwidth}{}}
\vspace*{-0.8cm}
\gridline{\fig{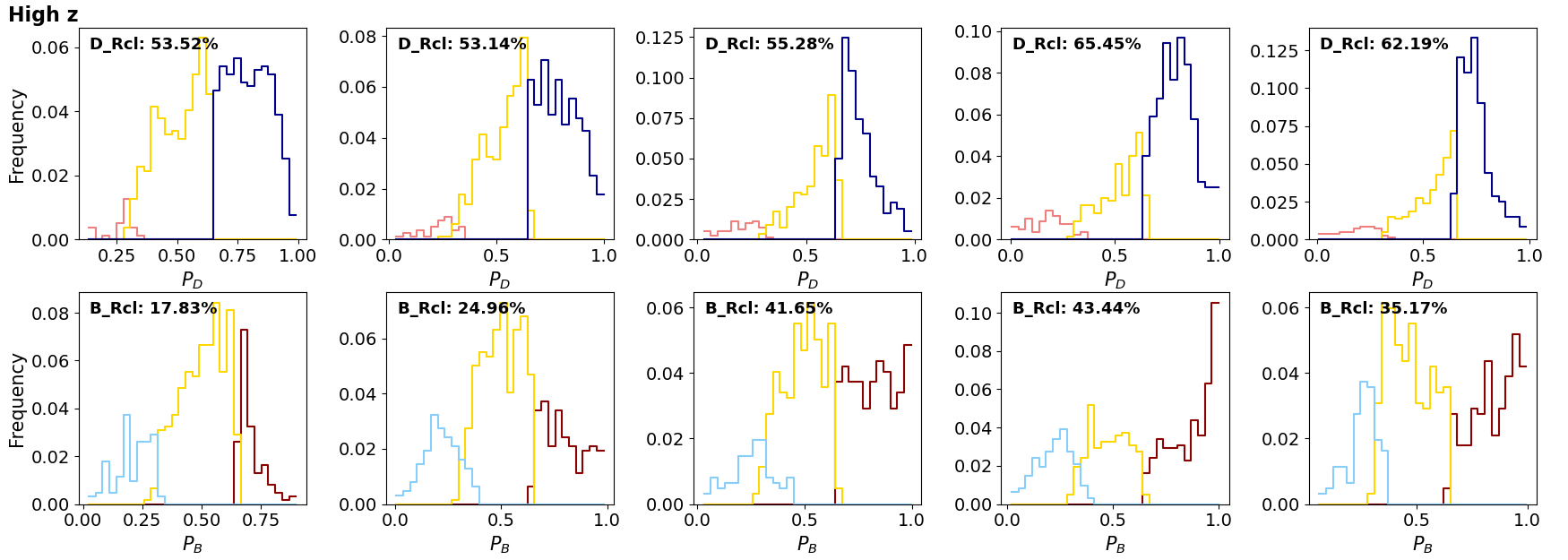}{1.05\textwidth}{}}
\vspace*{-0.4cm}
\caption{
{\it First row}: Normalized histogram of the \gamornet{} probability that the recovered galaxy predicted as a disk, $P_D$, for all actual disks in the low-redshift common test set. Disk recall is annotated at the top left of each histogram.
{\it Second row}: Same for the probability that the recovered galaxy is a bulge, $P_B$, for all actual bulges in the low-redshift common test set. Bulge recall is similarly annotated.
Color code: True Disk (TD): {\it dark blue}; False Disk (FD): {\it light blue}; True Bulge (TB): {\it dark red}; False Bulge (FB): {\it light red}; False Indeterminate (FI): {\it yellow}. Note True Indeterminate (TI) are not included, for we only selected actual disks and bulges.
{\it Third and fourth rows}: Same histograms for the medium-redshift bin.
{\it Fifth and sixth rows}: Same histograms for the high-redshift bin.
{\it Left to right}: band $g$, $r$, $i$, $z$, $y$.
}
\label{fig:ptrue_hist_each}
\end{figure*}

\section{Performance of Final PSFGAN+\gamornet{} Model}
\label{section:performance}
\subsection{Mapping Three Prediction Probabilities to One Morphological Type}
\label{subsection:tradeoff}

In this section, we map the final PSFGAN+ \gamornet{} predicted probabilities---for disks, bulges, and indeterminate types---to a single morphological determination, by incorporating thresholds.
Below we describe the choice of threshold and apply the best threshold to the common test set for each band and redshift bin.

Specifically, we apply PSFGAN+\gamornet{} to each realistic simulated AGN from the common test set in each redshift bin. Figure \ref{fig:true_vs_pred_diagram} shows the flowchart for our approach: (1) a realistic simulated AGN is fed as an input to the final model; (2) \gamornet{} produces three probabilities ($P_{\rm D}$, $P_{\rm I}$, $P_{\rm B}$) that a given galaxy is a disk, indeterminate, or a bulge, respectively (see \citealp{2020ApJ...895..112G}); (3) these probabilities are mapped (using a set of rules described below) to a predicted morphology; (4) the predicted and actual morphologies of the host galaxy are compared. 
We now describe exactly how these last two steps are done.

Note that every galaxy in the common test set has an ``actual'' morphology, given by \citet{2011ApJS..196...11S} or \citet{2018MNRAS.478.5410D}, and a ``predicted'' morphology from \gamornet{}.
Since we have three morphological types, we assess the precision of predicted morphologies using the following metrics (see summary in Table \ref{tab:true_vs_pred}), analogous to the usual binary classifiers\footnote{The usual definitions for a binary classifier are ``$TP$'' (True Positive), ``$TN$'' (True Negative), ``$FP$'' (False Positive), and ``$FN$'' (False Negative), with precision defined as $Precision=TP/(TP+FP)$.}:
\begin{itemize}
    \item ``$TD$'' represents actual disks that are correctly predicted as disks.
    \item ``$FD_B$'' represents actual bulges that are incorrectly predicted as disks.
    \item ``$FD_I$'' represents actual indeterminate types that are incorrectly predicted as disks. 
    \item ``$FD$'' represents the sum of actual bulges and indeterminate types that are incorrectly predicted as disks ($FD=FD_I+FD_B$). 
    \item Analogous definitions apply to the other morphological labels; for instance, $FB=FB_I+FB_D$ represents the sum of actual indeterminate types and disks that are incorrectly predicted as bulges.
\end{itemize}

These new metrics, appropriate to the 3-tuple outputs of \gamornet{}, are directly related to the binary metrics. 
When ``Positive'' corresponds to ``Disk'' and ``Negative'' to ``Non-disk'' (i.e., actual bulges or indeterminate types), the following relations hold:

\begin{equation}
\begin{split} 
    \label{eq:6}
    TP &= TD \\
    TN &= TI + FB_{I} + FI_{B} + TB \\
    FP &= FD_{I} + FD_{B} \\
    FN &= FI_{D} + FB_{D} 
\end{split}
\end{equation}
Analogous equations apply when ``Positive'' means correctly identifying a galaxy as ``Indeterminate'' or as a ``Bulge''.

We now define metrics to evaluate \gamornet{}'s performance, with the goal of maximizing the fraction of actual disks (bulges) in galaxies that are predicted as disks (bulges). 
Therefore, we define $Disk\_Precision$ and $Bulge\_Precision$ as follows: 

\begin{equation}
\begin{split}
    \label{eq:7}
    Disk\_Precision = \frac{TD}{TD+FD} ,\\
    Bulge\_Precision = \frac{TB}{TB+FB} .
\end{split}
\end{equation}
Note that we do not seek to maximize the precision of the indeterminate types. 
Some galaxies are difficult to classify as disk- or bulge-dominated, either because they have similar amounts of light in the disk and bulge, or they are too small or have too low single-to-noise ratios.
Therefore, we avoid classifying such galaxies as disks or bulges (which would decrease $Disk\_Precision$ and $Bulge\_Precision$) by identifying them as indeterminate.
That is, indeterminate types include both galaxies with bulge-to-total light ratios between 0.45 and 0.55 and galaxies about which \gamornet{} cannot make accurate determinations.
We define the fraction of galaxies being predicted as indeterminate types as:

\begin{equation}
\begin{split} 
    \label{eq:9}
    Indet\_Frac = \frac{TI+FI}{N} . 
\end{split}
\end{equation}
where $FI=FI_D + FI_B$ and $N=TD+FD+TI+FI+TB+FB$.

Ideally, we would like to classify every galaxy correctly; however, there is an obvious trade-off between the size of the indeterminate fraction and the precision of disk and bulge classifications.
Each \gamornet{} user must decide this balance according to their scientific goals. 
Here, we have enormous numbers of galaxies, so correctly identifying 100\% of the sample is not essential, while achieving high precision for the sample is important to our science goals.

To find this balance, we maximize an averaged precision for disks and bulges (rather than individual $Disk\_Precision$ or $Bulge\_Precision$):

\begin{equation}
\begin{split} 
    \label{eq:10}
    Ave\_Precision_{B,D} = \frac{1}{2}(\frac{TD}{TD+FD} + \frac{TB}{TB+FB}) .
\end{split}
\end{equation}
We then decide on a mapping rule from \gamornet{} outputs (3-tuples) to predicted morphological types that strikes an optimal balance between (1) averaged precision and (2) the indeterminate fraction.

For our scientific goals, we care less about the fraction of actual disks or bulges that are correctly predicted, which we call $Disk\_Recall$ and $Bulge\_Recall$, but these quantities may be important for other applications:

\begin{equation}
\begin{split}  
    \label{eq:8}
    Disk\_Recall = \frac{TD}{TD+FI_{D}+FB_{D}} ,\\
    Bulge\_Recall = \frac{TB}{TB+FD_{B}+FI_{B}} .
\end{split}
\end{equation}
For completeness, we also define accuracy, in the usual way:
\begin{equation}
\begin{split} 
    \label{eq:11}
    Accuracy = \frac{TD+TI+TB}{N},
\end{split}
\end{equation}
where $N=TD+FD+TI+FI+TB+FB$.

\gamornet{} produces three outputs, $P_{\rm D}$, $P_{\rm I}$, $P_{\rm B}$, which are the probabilities that the galaxy is a disk, bulge, or indeterminate, respectively. (By design, $P_{\rm D} + P_{\rm I} + P_{\rm B} = 1$.) 
One way to deduce the morphology would be to simply choose the highest probability; that is, the galaxy is $\{type\}$ if $P_{\{type\}} = max(P_{D}, P_{I}, P_{B})$.
However, this does not result in the highest possible precision.
Instead, we use the following threshold mapping suggested by \citet{2020ApJ...895..112G}:

\begin{equation}
\begin{split}  
    \label{eq:12}
    Galaxy\:is\:\textit{disk-dominated}\;\textit{if}\;\;\;\;\;\;\;\;\;\;\;\;\;\;\;\;\;\;\;\;\;\;\\
    P_{D} = max(P_{D}, P_{I}, P_{B})\:and\:P_{D} > Th_{D};\\
    Galaxy\:is\:\textit{bulge-dominated}\;\textit{if}\;\;\;\;\;\;\;\;\;\;\;\;\;\;\;\;\;\;\;\;\\
    P_{B} = max(P_{D}, P_{I}, P_{B})\:and\:P_{B} > Th_{B};\\
    Galaxy\:is\:\textit{indeterminate}\;otherwise.\;\;\;\;\;\;\;\;\;\;
\end{split}
\end{equation}

\noindent
Here $Th_{D}$ and $Th_{B}$ are confidence thresholds (between 0 and 1) chosen by the user.
The higher the threshold, the higher the precision but also, the greater the fraction of galaxies classified as indeterminate (as discussed further below).
The threshold mapping results in a higher precision compared to the simpler most-probable mapping, at a cost of having a larger fraction of galaxies being classified as indeterminate.

For each redshift bin, we searched for the sweet spot between the predicted indeterminate fraction (defined by Equation \ref{eq:9}) and averaged precision (between disks and bulges only, defined by Equation \ref{eq:10}), by varying the common threshold, $Th$\footnote{We use the same threshold for both disks and bulges: $Th_{D}$ = $Th_{B}$ = $Th$.}. 
Figure\,\ref{fig:confid_thold_valid} shows the threshold to be used by our model in each redshift bin, which was chosen by user's preference, based on the principle of maximizing averaged precision while minimizing indeterminate fraction.
Namely, $Th_{Low}=0.8$, $Th_{Mid}=0.7$, and $Th_{High}=0.65$.
It is evident that our choices of thresholds are \textit{not} unique: (1) depending on how favorable of a higher averaged precision to a lower indeterminate fraction (based on the scientific question at hand), one may use a different threshold
(2) there could be a case of multiple candidates with negligible differences (i.e. in low-redshift bin, Figure\,\ref{fig:confid_thold_valid})
(3) we have tested only a finite number of thresholds --- one could surely find another threshold that has better performance than any one we have tested. 
Such exploration can be done by heuristically testing other values.
Performance based on our chosen threshold only represents a lower bound of this approach.

Figure\,\ref{fig:confid_thold_all} shows disk precision, bulge precision and indeterminate fraction vs. confidence threshold, using results from all three redshift bins (key bands) for a quantitative comparison. Generally speaking, precision values in the low-redshift bin are better than the ones in the mid- and high-redshift bins. In addition, the indeterminate fraction in the low-redshift bin is also much lower than cases in the other two bins. For completeness, Figure\,\ref{fig:confid_thold_each} shows a similar comparison in each redshift bin using results from all five bands. In the majority of cases, disk and bulge precision values increase monotonically with increasing threshold, as does the indeterminate fraction. In both figures, the chosen thresholds are highlighted with light blue open circles (lines). At these thresholds, our PSFGAN+\gamornet\ model achieves a precision of, in the low-redshift bin,  $96\%$ for disks and $90\%$ for bulges, with fewer than one third of galaxies in the test set being indeterminate. For the mid- and high-redshift bins, our model is still robust enough to achieve precision values between $79\%$ and $90\%$ for at least three fifths of galaxies.

In Figure \ref{fig:ptrue_hist_each}, we show histograms of the \gamornet-predicted probabilities, for actual disks and actual bulges only, in two rows per redshift bin. 
Actual disks are colored by their predicted morphologies ($TD$, $FB$, and $FI$) and similarly for actual bulges ($FD$, $FI$, and $TB$).
Note we used our final models (with chosen thresholds) on recovered galaxies in the common test set when plotting these histograms.
Since we chose to optimize disk and bulge precision (instead of recall), recall is lower.
Specifically, disk recall ranges from $50\%$ to $80\%$, while bulge recall is lower, $30\%$-$70\%$ in the low-redshift bin and $20\%$-$40\%$ in the mid- and high-redshift bins.
Essentially, this is a consequence of choosing fairly high thresholds, which effectively moves some actual disks (bulges) to the predicted indeterminate category (thus reducing disk and bulge recall), in order to guarantee a high disk (bulge) precision in the category of predicted disks (bulges).
Finally, it is worth mentioning that the balance between precision and recall depends heavily on the chosen thresholds. Different thresholds can be selected to achieve (for instance) higher recall and lower precision, based on the scientific question at hand.

\subsection{Relation Between GaMorNet Precision and Key Variables}

\label{subsection:dependency}
Here we discuss how disk and bulge precision relate to three key parameters of the host galaxies, namely, g-band Kron radius, g-band magnitude, and contrast ratio.
We first sort galaxies into True Disk (TD), False Disk (FD), True Bulge (TB), False Bulge (FB), True Indeterminate (TI), and False Indeterminate (FI) categories (see Table \ref{tab:true_vs_pred}). 
The indeterminate category includes galaxies that have actual bulge-to-total flux ratios between $0.45$ and $0.55$ (i.e., actual indeterminate types) or that \gamornet{} incorrectly identifies as indeterminate; 
this category increases the fraction of actual disks (bulges) left in the predicted disk (predicted bulge) category.
Here, we do not differentiate between TI and FI, nor do we particularly care about indeterminate precision or recall. 
We instead focus on how disk and bulge precision depends on size, magnitude, and contrast ratio.

\begin{figure*}
\figurenum{14a}
\gridline{\fig{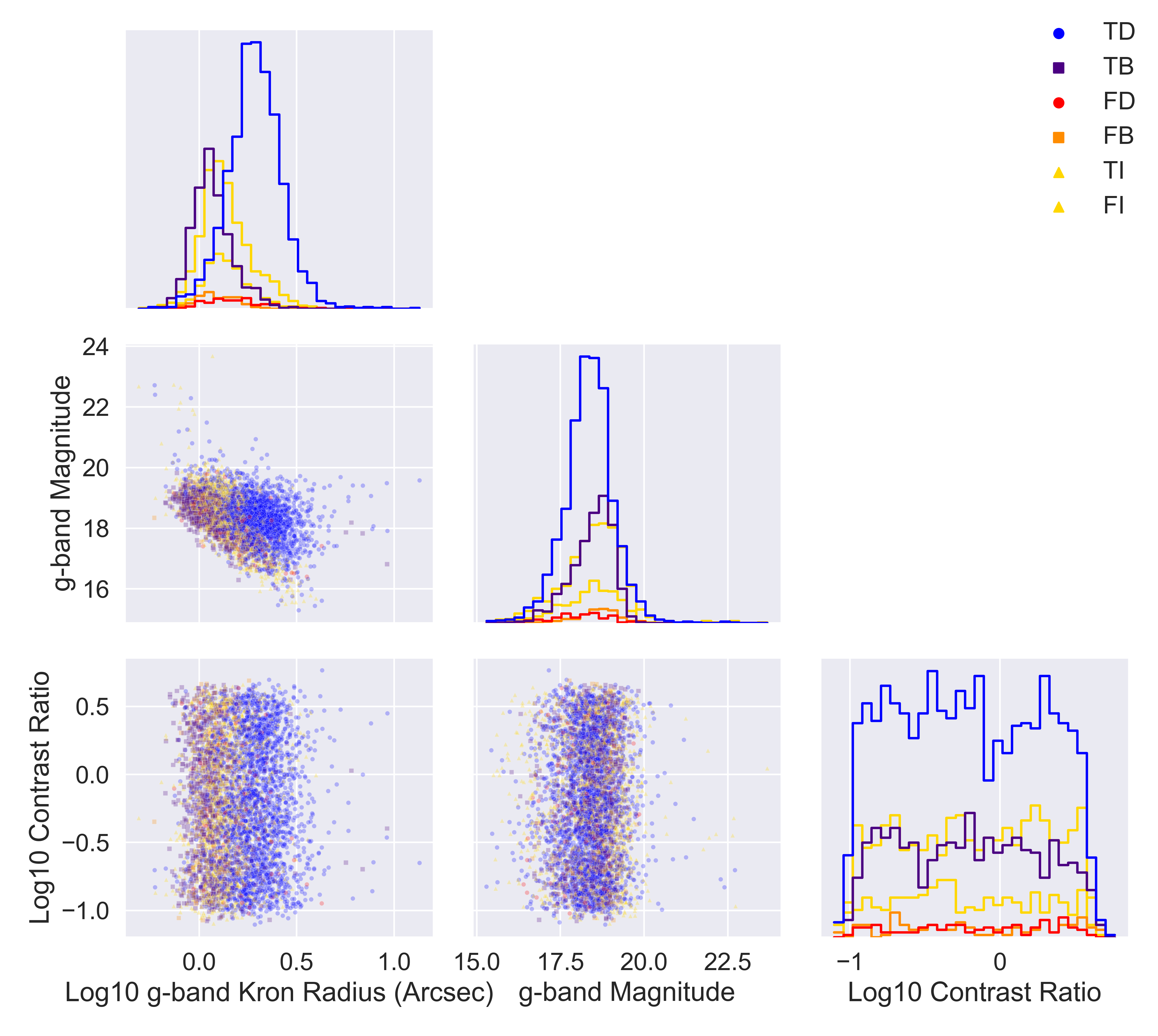}{1.0\textwidth}{}}
\caption{
Distributions of galaxies in the common test set of the low-redshift bin ($0<z<0.25$), colored by their \gamornet{} predicted morphologies in g-band and the correctness of this prediction, on log10 g-band Kron radius (arcsec), g-band magnitude, and log10 contrast ratio (in g-band).
This corner plot shows True Disks ({\it blue filled circles}), True Bulges ({\it purple filled squares}), False Disks ({\it red filled circles}), False Bulges ({\it orange filled squares}), and True and False Indeterminate cases ({\it yellow filled triangles}).
Specifically, histograms ({\it diagonal}) show the dependency of these categories on each key variable as an univariate function; scatter plots ({\it off-diagonal}) show the joint dependence on two variables. See \S~\ref{subsection:dependency} for analysis in detail.
\label{fig:parm_anlys_low}}
\end{figure*}

\begin{figure*}
\figurenum{14b}
\gridline{\fig{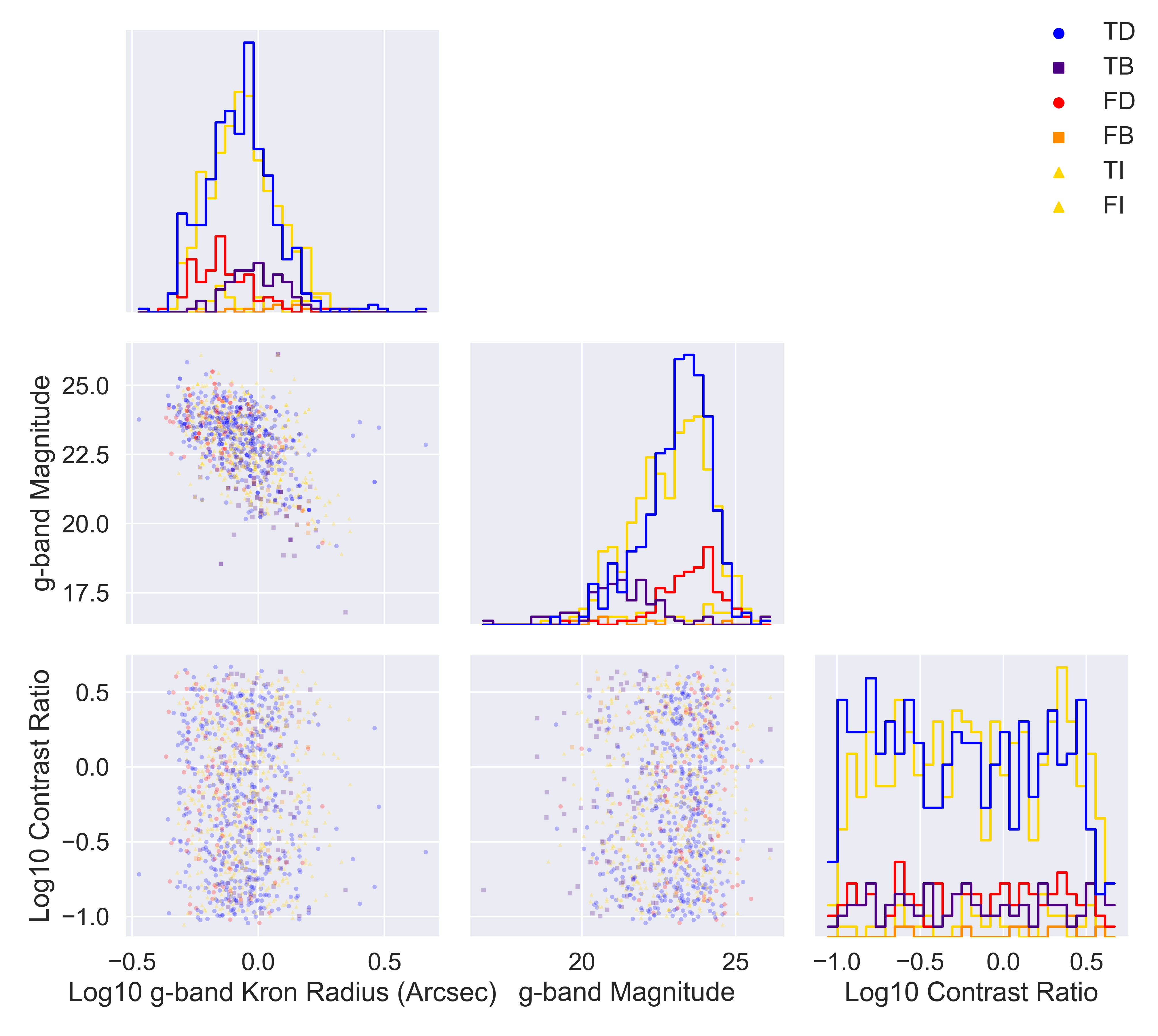}{1.0\textwidth}{}}
\caption{Distributions of galaxies in the common test set of the mid-redshift bin ($0.25<z<0.5$), colored by their \gamornet{} predicted morphologies in r-band and the correctness of this prediction, on log10 g-band Kron radius (arcsec), g-band magnitude, and log10 contrast ratio (in r-band).
Symbols and colors as in Fig.\,\ref{fig:parm_anlys_low}.
See \S~\ref{subsection:dependency} for analysis in detail.
\label{fig:parm_anlys_mid}}
\end{figure*}

\begin{figure*}
\figurenum{14c}
\gridline{\fig{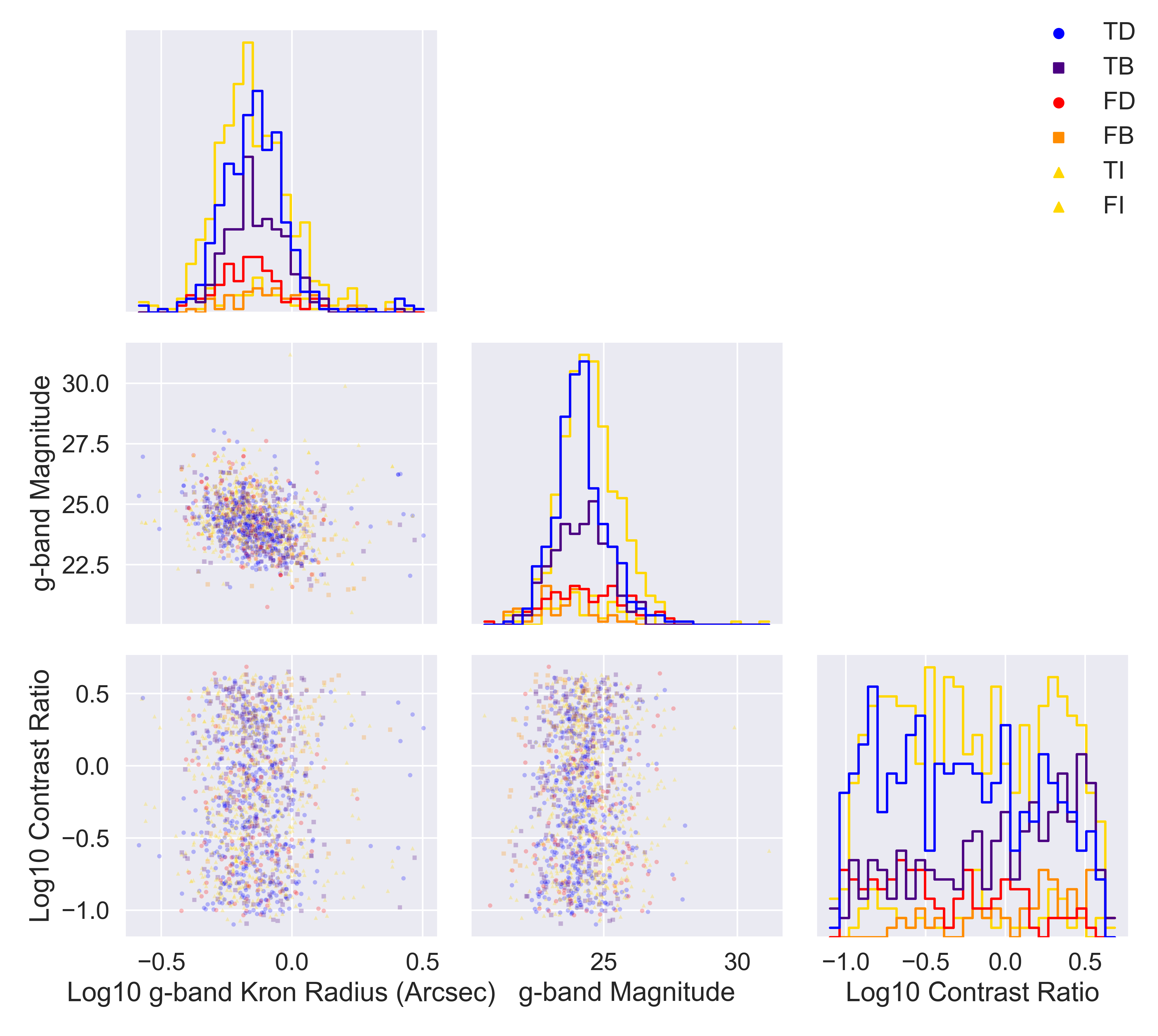}{1.0\textwidth}{}}
\caption{Distributions of galaxies in the common test set of the high-redshift bin ($0.5<z<1.0$), colored by their \gamornet{} predicted morphologies in i-band and the correctness of this prediction, on log10 g-band Kron radius (arcsec), g-band magnitude, and log10 contrast ratio (in i-band).
Symbols and colors as in Fig.\,\ref{fig:parm_anlys_low}.
See \S~\ref{subsection:dependency} for analysis in detail.
\label{fig:parm_anlys_high}}
\end{figure*}

Figures\,\ref{fig:parm_anlys_low}, \ref{fig:parm_anlys_mid}, and \ref{fig:parm_anlys_high} show the distributions of these six subgroups, as a function of the three chosen parameters, in each of the three redshift bins. 
To assess the similarity of distributions, we did Kolmogorov-Smirnov (K-S) tests \citep{10.2307/2280095}---one for each parameter in each redshift bin---between the parameter distributions for true and false disks (Table \ref{tab:pvalue_disk}). We did the same for bulges (Table \ref{tab:pvalue_bulge}). 

These tests and figures suggest that true disks and falsely identified disks have different size distributions, at least for low and mid redshifts. (The high-redshift data might be noisier and thus less sensitive to this difference.) 
This phenomenon is obvious in the lower redshift bins, where false disks tend to have smaller radii than true disks; that is, smaller galaxies are clearly harder to classify correctly, and thus disk precision depends on galaxy size. 
Likewise, bulge precision shows some dependency on galaxy radius, although the K-S tests are less significant.
In some cases, disk and bulge precision depend slightly on magnitude, in the sense that fainter galaxies are harder to classify correctly. 
Finally, neither precision depends strongly on contrast ratio.

\begin{deluxetable}{cccc}[htbp]
\tablecaption{P-values from K-S Tests of Each Galaxy Parameter Distribution between True Disks and False Disks\tablenotemark{a} 
}
\label{tab:pvalue_disk}
\tablecolumns{4}
\tablehead{
\colhead{~~~~Redshift~~~~} & \colhead{Low} & \colhead{Mid} & \colhead{High} \\
\colhead{~~~~~~~~~~~~~~~~} & \colhead{($g$ band)} & \colhead{($r$ band)} & \colhead{($i$ band)} \\
\colhead{Galaxy Parameter~~~~~} & \colhead{~~~~~~~~~~} & \colhead{~~~~~~~~~~} & \colhead{~~~~~~~~~~} 
} 
\startdata
    \hline
    $log_{10}(R_{Kron})$\tablenotemark{b} & $<10^{-4}$ & $<10^{-4}$ & $0.3761$ \\
    $m_{g}$\tablenotemark{c} & $0.0017$ & $0.0052$ & $0.0004$ \\
    $log_{10}(R_{\{key band\}})$\tablenotemark{d} & $0.6103$ & $0.5538$ & $0.0578$ \\
    \hline
    \hline
\enddata
\tablenotetext{a} {K-S probability that True Disks and False Disks have the same underlying distribution of a given parameter in a given redshift bin.}
\tablenotetext{b}{Log10 $g$-band Kron radius (arcsec).}
\tablenotetext{c}{$g$-band magnitude.}
\tablenotetext{d}{Log10 contrast ratio between point source and host galaxy in $g$, $r$, and $i$ bands for the low-, mid-, and high-redshift bins, respectively.}
\end{deluxetable}

\begin{deluxetable}{cccc}[htbp]
\tablecaption{
P-values from K-S Tests of Each Galaxy Parameter Distribution between True Bulges and False Bulges\tablenotemark{a}
}
\label{tab:pvalue_bulge}
\tablecolumns{4}
\tablehead{
\colhead{~~~~Redshift~~~~} & \colhead{Low} & \colhead{Mid} & \colhead{High} \\
\colhead{~~~~~~~~~~~~~~~~} & \colhead{($g$ band)} & \colhead{($r$ band)} & \colhead{($i$ band)} \\
\colhead{Galaxy Parameter~~~~~} & \colhead{~~~~~~~~~~} & \colhead{~~~~~~~~~~} & \colhead{~~~~~~~~~~} 
} 
\startdata
    \hline
    $log_{10}(R_{Kron})$\tablenotemark{b} & $0.0044$ & $0.0667$ & $0.0011$ \\
    $m_{g}$\tablenotemark{c} & $0.1234$ & $0.2076$ & $0.0004$ \\
    $log_{10}(R_{\{key band\}})$\tablenotemark{d} & $0.0355$ & $0.232$ & $0.0195$ \\
    \hline
    \hline
\enddata
\tablenotetext{a} {K-S probability that True Bulges and False Bulges have the same underlying distribution of a given parameter in a given redshift bin.}
\tablenotetext{b}{Log10 $g$-band Kron radius (arcsec).}
\tablenotetext{c}{$g$-band magnitude.}
\tablenotetext{d}{Log10 contrast ratio between point source and host galaxy in $g$, $r$, and $i$ bands for the low-, mid-, and high-redshift bins, respectively.}
\end{deluxetable}

\subsection{GaMorNet Predictions Compared to GALFIT Results for Previously Unclassified Real AGNs}
\label{subsection:galfitvalid}

To examine the robustness and precision of our \gamornet\ model, we compared its predictions to GALFIT results for PSFGAN-processed AGN images.
That is, we compared \gamornet{} (not PSFGAN+\gamornet{}) to GALFIT. 
Specifically, we fitted light profiles of AGNs from the HSC Wide survey, for three samples: 1,330 low-redshift broad H-alpha selected AGNs (mostly $0<z<0.25$) from \citealp{2019ApJS..243...21L}, 227 medium-redshift X-ray and/or IR selected AGNs (mostly $0.25<z<0.5$) from \citealp{2020ApJ...888...78S}, and 644 high-redshift X-ray and/or IR selected AGNs ($0.5<z<1.0$) from \citealp{2020ApJ...888...78S} again. 
(Cross matching of these catalogs with HSC was done as in \S~\ref{subsection:data used}.) 
For each AGN, we first removed the central point source using multi-band PSFGAN from our final model, in all five HSC bands.
We then used GALFIT to fit a single \sersic\ profile plus sky background to each recovered host galaxy, in the band of interest---i.e., $g$, $r$, and $i$ for the low-, medium- and high-redshift bins, respectively.  
We mapped each \sersic\ index into a morphological type using the same rules for single-component galaxies described in \S\,\ref{subsection:train} (i.e., $n<2$ is a disk, $n>2.5$ is a bulge).

For comparison, we used \gamornet\ to classify each recovered host galaxy in the band of interest. 
We mapped the three \gamornet\ probabilities into morphological types, as discussed in \S~\ref{subsection:tradeoff}, choosing the following thresholds (for both $P_{D}$ and $P_{B}$):
$\textit{Th}=0.8$, $0.7$, and $0.65$ for the low-, medium- and high-redshift bins, respectively.

\begin{figure}
\figurenum{15}
\gridline{\fig{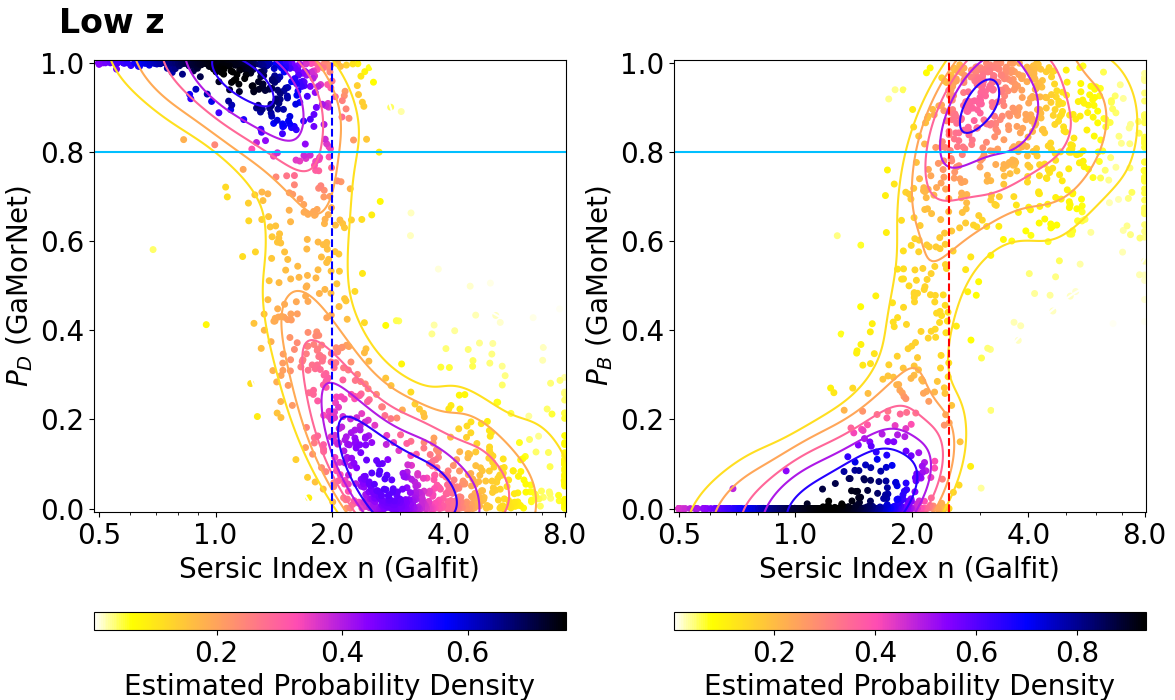}{0.5\textwidth}{}}
\vspace*{-0.7cm}
\gridline{\fig{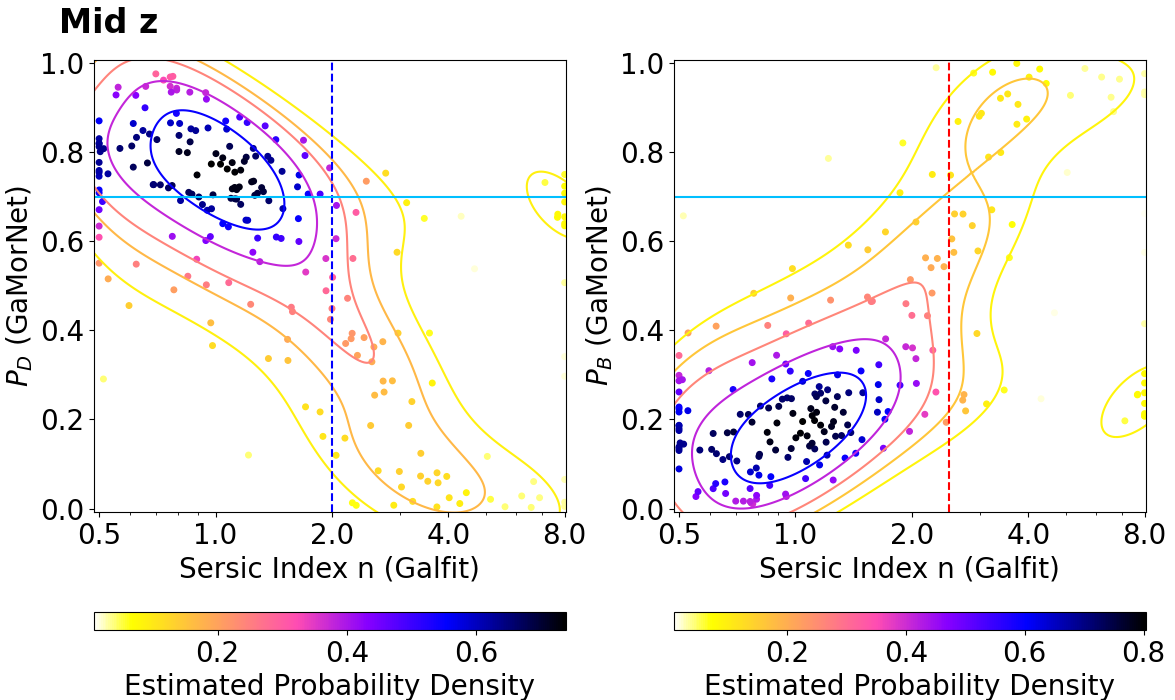}{0.5\textwidth}{}}
\vspace*{-0.7cm}
\gridline{\fig{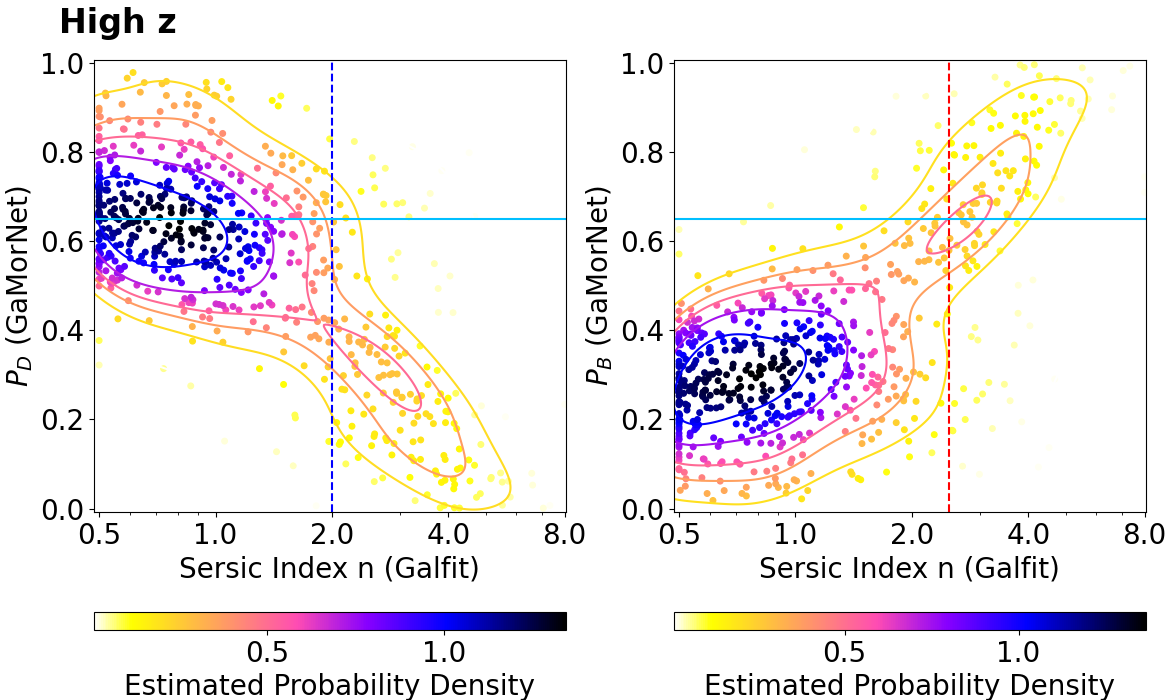}{0.5\textwidth}{}}
\vspace*{-0.7cm}
\caption{
\gamornet\ predictions agree well with GALFIT results.
(Specifically, here we compare PSFGAN+\gamornet{} predictions to GALFIT analysis of PSFGAN-processed AGN images.)
{\it Top row:} Scatter plot with contours superimposed, for galaxy morphologies in the low redshift sample ($0<z<0.25$).
{\it Middle row:} Same for the medium redshift ($0.25<z<0.5$).
{\it Bottom row:} Same for the high redshift ($0.5<z<1.0$).
{\it Left:} Probability that the galaxy is a disk ($P_{D}$ from \gamornet) versus single \sersic\ index ($n$ from GALFIT).
{\it Right:} Probability that the galaxy is a bulge ($P_{B}$ from \gamornet) versus \sersic\ index.
The horizontal {\it long-dash cyan line} indicates the chosen threshold for \gamornet{}; above this line, \gamornet\ classifies galaxies as disks ({\it left}) or bulges ({\it right}).
The vertical {\it dark blue dashed line} ($n=2$, on the left) and {\it red dashed line} ($n=2.5$, on the right) mark the boundaries for GALFIT-ed \sersic{} indices indicating disks ($n<2$) and bulges ($n>2.5$).
Each subplot uses a unique \textit{color bar} to show the estimated probability density function of its data (generated using a Kernel Density Estimator).
\label{fig:validation}}
\end{figure}

We then compared the morphological labels from GALFIT and \gamornet{}.
Results are shown in Figure\,\ref{fig:validation}, for each of the three redshift bins.
In the low-redshift bin, the vast majority of galaxies predicted as disks by \gamornet\ ($P_{D}>T$, above the cyan solid line) are also predicted as disks by GALFIT ($n<2$, to the left of the blue dashed line).
Similarly, the majority of galaxies predicted as bulges by \gamornet\ ($P_{B}>T$, above the cyan solid line) are also predicted as bulges by GALFIT ($n>2.5$, to the right of red dashed line).
Results for the mid- and high-redshift bins are similar (medium and bottom rows, Figure \ref{fig:validation}).
In very few cases are galaxies characterized as bulges by GALFIT and as disks by \gamornet{}, or vice versa.

Table \ref{tab:validation} shows the exact numbers for these morphology comparisons, as well as the associated precision values. 
For instance, in the low-redshift bin, of the 452 galaxies predicted as disks by \gamornet{}, 426 (94\%) are also determined to be disks by GALFIT. 
Overall, classification results between \gamornet\ and GALFIT agree well ($>90$\% for disks, $>80$\% for bulges).
This means \gamornet\ (preceded by PSFGAN) determines the morphologies of AGN host galaxies as well as GALFIT, while running much faster and not needing customized input parameters.

\begin{deluxetable}{ccc}[htbp]
\tablecaption{Comparison of \gamornet\ and GALFIT Classifications
\label{tab:validation}}
\tablecolumns{3}
\tablehead{
\colhead{} & \colhead{Disks (\gamornet)} & \colhead{Bulges (\gamornet)}}
\startdata
    \hline
    \hline
    \multicolumn{3}{c}{~~~~~~~~~~~~~~~~~~~~~~~~~Low redshift bin} \\
    \hline
    Disks (GALFIT) & $426$($94\%$) & $3$($1\%$)  \\
    Indets (GALFIT) & $23$($5\%$) & $25$($7\%$)  \\
    Bulges (GALFIT) & $3$($1\%$) & $355$($93\%$) \\
    Total (GALFIT) & $452$ & $383$ \\
    \hline
    \hline
    \multicolumn{3}{c}{~~~~~~~~~~~~~~~~~~~~~~~~~Medium redshift bin} \\
    \hline
    Disks (GALFIT) & $97$($93\%$) & $3$($9\%$)  \\
    Indets (GALFIT) & $1$($1\%$) & $2$($6\%$)  \\
    Bulges (GALFIT) & $6$($6\%$) & $27$($84\%$) \\
    Total (GALFIT) & $104$ & $32$ \\
    \hline
    \hline
    \multicolumn{3}{c}{~~~~~~~~~~~~~~~~~~~~~~~~~High redshift bin} \\
    \hline
    Disks (GALFIT) & $225$($91\%$) & $8$($8\%$)  \\
    Indets (GALFIT) & $4$($2\%$) & $12$($11\%$)  \\
    Bulges (GALFIT) & $17$($7\%$) & $86$($81\%$) \\
    Total (GALFIT) & $246$ & $106$ \\
    \hline
    \hline
\enddata
\tablecomments{Percentages refer to precision, i.e., the number of correctly classified $\{type\}$ divided by the number of total galaxies predicted as this $\{type\}$.
}
\end{deluxetable}

\section{Summary and Discussion}
\label{section:summary} 

We developed a comprehensive machine learning framework to morphologically classify AGN host galaxies, tailored specifically to HSC Wide survey images.
Because the signal-to-noise ratio depends strongly on redshift, we divided data into, and separately built models for, three redshift bins: low ($0<z<0.25$), medium ($0.25<z<0.5$), and high ($0.5<z<1.0$).
We focused on specific HSC bands for each---namely $g$, $r$, and $i$ for the low-, medium-, and high-redshift bins, respectively. This ensures we are classifying within approximately the same rest-frame wavelength range.
In each redshift bin, we modified the original version of PSFGAN \citep{2018MNRAS.477.2513S} to remove the AGN point source in five HSC bands simultaneously.
We then trained a version of \gamornet\ \citep{2020ApJ...895..112G} to morphologically classify PSFGAN-processed images, for each of the five bands.
We used a large number of simulated galaxies for the initial training, for which we know the correct morphological parameters perfectly. This initial training was followed by a transfer-learning step, in which we used a small amount of real data to fine-tune model parameters.
The final PSFGAN+\gamornet\ model was then tested on either simulated realistic AGNs (i.e., inactive galaxies that were already morphological classified, to which point sources were added) or unclassified real AGNs (which we fitted with GALFIT, for comparison). 
The precision is highest for disks and slightly lower for bulges, because of the much higher fraction of disks in the training samples. 
The precision is also highest for the lowest redshift bin ($\sim95$\%), where the signal-to-noise ratio is highest, but is still very good ($80-90$\%) for the higher redshift bins. 

Specifically, by first training our models from scratch using 150,000 simulated galaxies with artificial AGN point sources (\S~\ref{subsection:simgal}), we achieved $\sim 90\%$ classification precision values and mis-classification rates between $2\%$ and $5\%$ on test sets (\S~\ref{subsection:train}). We then used transfer learning to fine-tune parameters of trained models with respect to $\sim$50,000 (low-redshift bin) or $\sim$13,000 (medium- and high-redshift bins) real galaxies with artificial AGN point sources, which were previously classified by traditional methods (\S~\ref{subsection:realgal} and \S~\ref{subsection:tl}). We showed explicitly that in the low-, medium- and high-redshift bins, using thresholds of $T=0.8$, 0.7, and 0.65, respectively, for both disks and bulges, our models achieved precision values of 96\%, 82\%, and 79\% on disks, and 90\%, 90\%, and 80\% on bulges, with corresponding incompleteness (indeterminate fractions) of 30\%, 43\%, and 42\% (\S~\ref{subsection:tradeoff}). 
We explored how this precision depends on host galaxy magnitude, radius, and the contrast ratio between AGN point source and host galaxy. We verified the parameter space in which PSFGAN+\gamornet{} works well (\S~\ref{subsection:dependency}). 
Finally, we GALFIT-ed 2,201 galaxies, after processing them with PSFGAN, and compared those classifications to results from \gamornet, demonstrating that our models agree $80\% - 95\%$ of time (\S~\ref{subsection:galfitvalid}).

We reached two main conclusions.
First, our experiment confirms the feasibility of combining PSFGAN (which removes the AGN point source) and \gamornet\ (which classifies the galaxy morphology).  
Our PSFGAN+\gamornet\ model achieves good precision for AGNs for redshifts $z<1$.
Although our model was tailored specifically to the HSC Wide survey, the same approach can be easily generalized to other surveys of interest, and it is particularly helpful that only  a limited number of morphology labels are needed for the transfer learning step.

Second, our final \gamornet{} model predicts morphologies that agree well with those from a traditional two-dimensional fitting method, GALFIT.
Moreover, it is a non-parametric method, that does not require specific fitting functions and galaxy-wise input parameters during inference. In addition, it runs two to three orders of magnitude faster at least. PSFGAN+\gamornet{} is an ideal tool for determining morphologies of AGNs in upcoming data-intensive surveys. 

In near future, we are planning to apply our PSFGAN+\gamornet{} framework on real AGNs to study their host galaxy morphologies. We are also interested in using a similar approach with other ML frameworks to study host galaxy properties beyond morphology.

\begin{acknowledgments}

The authors are very grateful to the anonymous referees for their insightful, encouraging, and pertinent comments, which were extremely helpful in improving the manuscript, clarifying discussions, and making the results more accessible to readers.

This material is based upon work supported by the National Science Foundation under Grant No. 1715512.

We acknowledge support from National Aeronautics and Space Administration via ADAP grant 80NSSC18K0418.

We thank the Yale Center for Research Computing (YCRC) for guidance and use of the research computing infrastructure.

The Hyper Suprime-Cam (HSC) collaboration includes the astronomical communities of Japan and Taiwan, and Princeton University. The HSC instrumentation and software were developed by the National Astronomical Observatory of Japan (NAOJ), the Kavli Institute for the Physics and Mathematics of the Universe (Kavli IPMU), the University of Tokyo, the High Energy Accelerator Research Organization (KEK), the Academia Sinica Institute for Astronomy and Astrophysics in Taiwan (ASIAA), and Princeton University. Funding was contributed by the FIRST program from the Japanese Cabinet Office, the Ministry of Education, Culture, Sports, Science and Technology (MEXT), the Japan Society for the Promotion of Science (JSPS), Japan Science and Technology Agency (JST), the Toray Science Foundation, NAOJ, Kavli IPMU, KEK, ASIAA, and Princeton University. 

This paper makes use of software developed for the Large Synoptic Survey Telescope. We thank the LSST Project for making their code available as free software at  http://dm.lsst.org

This paper is based (in part) on data collected at the Subaru Telescope and retrieved from the HSC data archive system, which is operated by the Subaru Telescope and Astronomy Data Center (ADC) at National Astronomical Observatory of Japan. Data analysis was in part carried out with the cooperation of Center for Computational Astrophysics (CfCA), National Astronomical Observatory of Japan. The Subaru Telescope is honored and grateful for the opportunity of observing the Universe from Maunakea, which has the cultural, historical and natural significance in Hawaii. 

The Pan-STARRS1 Surveys (PS1) and the PS1 public science archive have been made possible through contributions by the Institute for Astronomy, the University of Hawaii, the Pan-STARRS Project Office, the Max Planck Society and its participating institutes, the Max Planck Institute for Astronomy, Heidelberg, and the Max Planck Institute for Extraterrestrial Physics, Garching, The Johns Hopkins University, Durham University, the University of Edinburgh, the Queen’s University Belfast, the Harvard-Smithsonian Center for Astrophysics, the Las Cumbres Observatory Global Telescope Network Incorporated, the National Central University of Taiwan, the Space Telescope Science Institute, the National Aeronautics and Space Administration under grant No. NNX08AR22G issued through the Planetary Science Division of the NASA Science Mission Directorate, the National Science Foundation grant No. AST-1238877, the University of Maryland, Eotvos Lorand University (ELTE), the Los Alamos National Laboratory, and the Gordon and Betty Moore Foundation.

\end{acknowledgments}

\software{PSFGAN \citep{2018MNRAS.477.2513S},
          \gamornet{} \citep{2020ApJ...895..112G},
          TOPCAT \citep{2005ASPC..347...29T},
          GALFIT \citep{2002AJ....124..266P},
          NumPy \citep{2020Natur.585..357H},
          SciPy \citep{2020SciPy-NMeth},
          Astropy \citep{2018AJ....156..123A},
          Pandas \citep{pandas_2010},
          Matplotlib \citep{Hunter:2007}
          }


\appendix

\section{Online Data Access}
\label{section:appendix:online_data_access}
Our PSFGAN+\gamornet{} framework, along with trained models and detailed instructions, are made accessible to the public. Interested readers can access them at the following GitHub Repository:

\begin{itemize}
    \item \href{https://github.com/ufsccosmos/PSFGAN-GaMorNet}{https://github.com/ufsccosmos/PSFGAN-GaMorNet}. 
\end{itemize}

\section{Motivation of modifying the generator in PSFGAN}
\label{section:appendix:mdf_psf_gnrt}
The generator in PSFGAN, by design, has a structure of an encoder-decoder (i.e. image side length first decreases monotonically then increases monotonically). Image side length at the ``bottleneck'' (where side length is at its minimum) is 4 in the default structure of PSFGAN. (Side lengths of encoder layers, by default: [424, 212, 106, 53, 27, 14, 7, 4]; order reversed for decoder layers) In our experiments, since our image has a side length of 239, we removed one layer from the encoder-decoder, so that side length at the bottleneck remains the same. (Side lengths of encoder layers: [239, 120, 60, 30, 15, 8, 4]; order reversed for decoder layers) By design, (using the encoder as example) side lengths of two adjacent layers, $s_n$ and $s_{n+1}$, satisfy a relation such that $s_{n+1}$ equals to the integer division (i.e. division without remainder) of $(s_n + 1)$ by 2.

Recall that in both the original PSFGAN paper and our work, the PSFGAN is used for the same task: to remove the (usually bright) AGN point source at image center. Thus, a latent space (bottleneck) with similar size should suffice our need as well. We have not explicitly tested the case of having a different side length at the bottleneck and how this would affect our results. Such experiments would be very computationally expensive, for we trained versions of PSFGAN and \gamornet{} as linked pairs (i.e. for each unique structure of PSFGAN, one needs to train and test a set of versions of \gamornet{}).

\section{Prediction Continuity Analysis for Adjacent Models} \label{section:appendix:aja_cnty}
We divided our datasets into three redshift bins: low ($0<z<0.25$), medium ($0.25<z<0.5$), and high ($0.5<z<1.0$), before training PSFGAN and \gamornet{} models independently in each bin. Although this approach benefits our final models (i.e. each of them becomes more precise by focusing on narrower ranges of galaxy magnitude and radius), it is natural to ask whether two adjacent models (e.g. model A trained on low-redshift bin vs. model B trained on mid-redshift bin, both for the same band) will give one identical answer for an input galaxy. We answer this question specifically in this section.

Recall that for each redshift bin, we trained one version of multi-band PSFGAN and five versions (one for each band) of \gamornet{} (Section \ref{subsection:tl}). We then determined mapping rules (and common thresholds) to project \gamornet{} output probabilities into predicted morphological types (Section \ref{subsection:tradeoff}). Hereafter, for convenience, we denote the final model we trained for band $\{filter\}$ in $\{bin\_name\}$ as $model-\{filter\}-\{bin\_name\}$ (e.g. $model-g-low$ stands for our final model trained for band $g$ in the low-redshift bin). Note that we already applied the mapping rules ($Th_{Low}=0.8$, $Th_{Mid}=0.7$, and $Th_{High}=0.65$, regardless of the band) for each model (i.e. outputs of our final models are predicted morphological types, instead of probabilities).

For two models A and B and a certain bin X, we can define the Agreed Predictions Fraction as:

\begin{equation}
Agrd\_Pred\_Frac_{X}^{AB} = \frac{N(agrd\_disks)_{X}^{AB}+N(agrd\_indets)_{X}^{AB}+N(agrd\_bulges)_{X}^{AB}}{N(all\_gals)_{X}^{AB}}
\label{eq:appendix:agt_pred_frac}
\end{equation}

Where $N(agrd\_disks)_{X}^{AB}$ is the number of galaxies in bin X that are predicted as disks by both model A and B (similarly for $N(agrd\_indets)_{X}^{AB}$ and $N(agrd\_bulges)_{X}^{AB}$).

For completeness, these definitions can be related to metrics introduced in Table \ref{tab:true_vs_pred}. We first denote galaxies predicted as disks as $PD$, which can be related to $TD$ and $FD$ as simply $PD = TD + FD$ (see Table \ref{tab:true_vs_pred}). Then we denote galaxies predicted as disks by a certain model A as $PD^{A}$ (similarly for other metrics). At last, we denote galaxies predicted as disks by both model A and B as $PD^{A} \cap PD^{B}$ (that is, one may look at $PD^{A}$ as a set of all galaxies predicted by model A, so the $\cap$ symbol makes sense). Definitions for other morphological types are similar. The following relations holds:

\begin{equation}
N(agrd\_disks)_{X}^{AB} = PD^{A} \cap PD^{B}
\label{eq:appendix:agrd_disks}
\end{equation}

\begin{equation}
N(agrd\_indets)_{X}^{AB} = PI^{A} \cap PI^{B}
\label{eq:appendix:agrd_indets}
\end{equation}

\begin{equation}
N(agrd\_bulges)_{X}^{AB} = PB^{A} \cap PB^{B}
\label{eq:appendix:agrd_bulges}
\end{equation}

We compared predictions from each of the three final models, $model-g-low$, $model-r-mid$ and $model-i-high$ (note each of them was trained in the key band for corresponding redshift bin), against predictions from each model's adjacent model(s), in appropriate redshift bins, respectively. The following combinations of model A (main model), model B (adjacent model) and redshift bin X are chosen for analysis:

\begin{enumerate}
    \item A: $model-g-low$; B: $model-g-mid$; X: low-redshift bin or mid-redshift bin. 
    \item A: $model-r-mid$; B: $model-r-low$ ($model-r-high$); X: mid-redshift bin or low-redshift bin (high-redshift bin). 
    \item A: $model-i-high$; B: $model-i-mid$; X: high-redshift bin or mid-redshift bin. 
\end{enumerate}

It is worth to mention that in the high-redshift bin, all 1,524 galaxies from the common test set were used, while in the low-redshift (mid-redshift) bin, only 4,635 (958) galaxies with strictly $z<0.25$ ($0.25<z<0.5$) were used (see Figure \ref{fig:training_photoz} and Table \ref{tab:real_datasplit}). Moreover, in addition to analyzing the continuity of adjacent models in each redshift bin, we divide each redshift bin evenly into five smaller sub-bins, and repeat this analysis to capture how the agreed prediction fraction changes as a function of redshift.

Table \ref{tab:appendix:aja_cnty_g}, \ref{tab:appendix:aja_cnty_r}, and \ref{tab:appendix:aja_cnty_i} show our continuity analysis results for $model-g-low$, $model-r-mid$, and $model-i-high$ with each model's adjacent model(s), respectively. We observe that the agreed prediction fraction between two adjacent models is always highest at redshift bin boundaries, and it decreases monotonically as we go away. It is also noticeable that the agreed prediction fraction is lower at the low-mid boundary than at the mid-high boundary --- intuitively speaking, this could be a direct consequence of the fact that galaxies from mid- and high-redshift bins have relatively larger overlaps in magnitude and radius.

\begin{deluxetable*}{ccccccc}[htbp]
\tablecaption{Continuity Analysis Results for $model-g-low$ and Its Adjacent Model ($model-g-mid$)}
\label{tab:appendix:aja_cnty_g}
\tablecolumns{7}
\tablehead{
\colhead{Sub-redshift-bin} & \colhead{First bin} & \colhead{Second bin} & \colhead{Third bin} & \colhead{Fourth bin} & \colhead{Fifth bin} & \colhead{Five bins combined} 
}
\startdata
    \hline
    \hline
    \multicolumn{7}{c}{with $model-g-mid$ in low-redshift bin} \\
    \hline
    Redshift Range & $0<z<0.05$ & $0.05<z<0.1$ & $0.1<z<0.15$ & $0.15<z<0.2$ & $0.2<z<0.25$ & All ($0<z<0.25$) \\
    \hline
    Agreed Predictions & 40 (59.7\%) & 867 (69.1\%) & 1450 (76.2\%) & 815 (81.0\%) & 356 (87.7\%) & 3528 (76.1\%) \\
    All Galaxies & 67 (100\%) & 1254 (100\%) & 1902 (100\%) & 1006 (100\%) & 406 (100\%) & 4635 (100\%)\\
    \hline
    \hline
    \multicolumn{7}{c}{with $model-g-mid$ in mid-redshift bin} \\
    \hline
    Redshift Range & $0.25<z<0.3$ & $0.3<z<0.35$ & $0.35<z<0.4$ & $0.4<z<0.45$ & $0.45<z<0.5$ & All ($0.25<z<0.5$) \\
    \hline
    Agreed Predictions & 85 (85.9\%) & 142 (83.0\%) & 201 (79.1\%) & 126 (71.2\%) & 162 (63.0\%) & 716 (74.7\%) \\
    All Galaxies & 99 (100\%) & 171 (100\%) & 254 (100\%) & 177 (100\%) & 257 (100\%) & 958 (100\%)\\
    \hline
    \hline
\enddata
\tablecomments{We apply $model-g-low$ and its adjacent model ($model-g-mid$) in the low- and mid-redshift bins, and calculate their agreed prediction fractions. We then divide each redshift bin into five smaller sub-bins, and repeat this calculation in each of these sub-bins.}
\end{deluxetable*}

\begin{deluxetable*}{ccccccc}[htbp]
\tablecaption{Continuity Analysis Results for $model-r-mid$ and Its Adjacent Models ($model-r-low$ and $model-r-high$)}
\label{tab:appendix:aja_cnty_r}
\tablecolumns{7}
\tablehead{
\colhead{Sub-redshift-bin} & \colhead{First bin} & \colhead{Second bin} & \colhead{Third bin} & \colhead{Fourth bin} & \colhead{Fifth bin} & \colhead{Five bins combined} 
}
\startdata
    \hline
    \hline
    \multicolumn{7}{c}{with $model-r-low$ in low-redshift bin} \\
    \hline
    Redshift Range & $0<z<0.05$ & $0.05<z<0.1$ & $0.1<z<0.15$ & $0.15<z<0.2$ & $0.2<z<0.25$ & All ($0<z<0.25$) \\
    \hline
    Agreed Predictions & 37 (55.2\%) & 851 (67.9\%) & 1435 (75.4\%) & 846 (84.1\%) & 362 (89.2\%) & 3531 (76.2\%) \\
    All Galaxies & 67 (100\%) & 1254 (100\%) & 1902 (100\%) & 1006 (100\%) & 406 (100\%) & 4635 (100\%)\\
    \hline
    \hline
    \multicolumn{7}{c}{with $model-r-low$ in mid-redshift bin} \\
    \hline
    Redshift Range & $0.25<z<0.3$ & $0.3<z<0.35$ & $0.35<z<0.4$ & $0.4<z<0.45$ & $0.45<z<0.5$ & All ($0.25<z<0.5$) \\
    \hline
    Agreed Predictions & 87 (87.9\%) & 146 (85.4\%) & 197 (77.6\%) & 119 (67.2\%) & 153 (59.5\%) & 702 (73.3\%) \\
    All Galaxies & 99 (100\%) & 171 (100\%) & 254 (100\%) & 177 (100\%) & 257 (100\%) & 958 (100\%)\\
    \hline
    \hline
    \multicolumn{7}{c}{with $model-r-high$ in mid-redshift bin} \\
    \hline
    Redshift Range & $0.25<z<0.3$ & $0.3<z<0.35$ & $0.35<z<0.4$ & $0.4<z<0.45$ & $0.45<z<0.5$ & All ($0.25<z<0.5$) \\
    \hline
    Agreed Predictions & 78 (78.8\%) & 145 (84.8\%) & 226 (89.0\%) & 159 (90.0\%) & 236 (91.8\%) & 844 (88.1\%) \\
    All Galaxies & 99 (100\%) & 171 (100\%) & 254 (100\%) & 177 (100\%) & 257 (100\%) & 958 (100\%)\\
    \hline
    \hline
    \multicolumn{7}{c}{with $model-r-high$ in high-redshift bin} \\
    \hline
    Redshift Range & $0.5<z<0.6$ & $0.6<z<0.7$ & $0.7<z<0.8$ & $0.8<z<0.9$ & $0.9<z<1.0$ & All ($0.5<z<1.0$) \\
    \hline
    Agreed Predictions & 177 (92.7\%) & 372 (91.4\%) & 325 (86.2\%) & 269 (81.5\%) & 164 (74.9\%) & 1307 (85.8\%) \\
    All Galaxies & 191 (100\%) & 407 (100\%) & 377 (100\%) & 330 (100\%) & 219 (100\%) & 1524 (100\%)\\
    \hline
    \hline
\enddata
\tablecomments{We apply $model-r-mid$ and its adjacent models ($model-r-low$ and $model-r-high$) in the low-, mid- and high-redshift bins, and calculate their agreed prediction fractions. We then divide each redshift bin into five smaller sub-bins, and repeat this calculation in each of these sub-bins.}
\end{deluxetable*}

\begin{deluxetable*}{ccccccc}[htbp]
\tablecaption{Continuity Analysis Results for $model-i-high$ and Its Adjacent Model ($model-i-mid$)}
\label{tab:appendix:aja_cnty_i}
\tablecolumns{7}
\tablehead{
\colhead{Sub-redshift-bin} & \colhead{First bin} & \colhead{Second bin} & \colhead{Third bin} & \colhead{Fourth bin} & \colhead{Fifth bin} & \colhead{Five bins combined} 
}
\startdata
    \hline
    \hline
    \multicolumn{7}{c}{with $model-i-mid$ in mid-redshift bin} \\
    \hline
    Redshift Range & $0.25<z<0.3$ & $0.3<z<0.35$ & $0.35<z<0.4$ & $0.4<z<0.45$ & $0.45<z<0.5$ & All ($0.25<z<0.5$) \\
    \hline
    Agreed Predictions & 75 (75.8\%) & 142 (83.0\%) & 223 (87.8\%) & 157 (88.7\%) & 233 (90.7\%) & 830 (86.6\%) \\
    All Galaxies & 99 (100\%) & 171 (100\%) & 254 (100\%) & 177 (100\%) & 257 (100\%) & 958 (100\%)\\
    \hline
    \hline
    \multicolumn{7}{c}{with $model-i-mid$ in high-redshift bin} \\
    \hline
    Redshift Range & $0.5<z<0.6$ & $0.6<z<0.7$ & $0.7<z<0.8$ & $0.8<z<0.9$ & $0.9<z<1.0$ & All ($0.5<z<1.0$) \\
    \hline
    Agreed Predictions & 174 (91.1\%) & 369 (90.7\%) & 323 (85.7\%) & 266 (80.6\%) & 162 (74.0\%) & 1294 (84.9\%) \\
    All Galaxies & 191 (100\%) & 407 (100\%) & 377 (100\%) & 330 (100\%) & 219 (100\%) & 1524 (100\%)\\
    \hline
    \hline
\enddata
\tablecomments{We apply $model-i-high$ and its adjacent model ($model-i-mid$) in the mid- and high-redshift bins, and calculate their agreed prediction fractions. We then divide each redshift bin into five smaller sub-bins, and repeat this calculation in each of these sub-bins.}
\end{deluxetable*}

\clearpage

\bibliography{references}{}

\begin{thebibliography}{}
\expandafter\ifx\csname natexlab\endcsname\relax\def\natexlab#1{#1}\fi
\providecommand{\url}[1]{\href{#1}{#1}}
\providecommand{\dodoi}[1]{doi:~\href{http://doi.org/#1}{\nolinkurl{#1}}}
\providecommand{\doeprint}[1]{\href{http://ascl.net/#1}{\nolinkurl{http://ascl.net/#1}}}
\providecommand{\doarXiv}[1]{\href{https://arxiv.org/abs/#1}{\nolinkurl{https://arxiv.org/abs/#1}}}

\bibitem[{{Ackermann} {et~al.}(2018){Ackermann}, {Schawinski}, {Zhang},
  {Weigel}, \& {Turp}}]{2018MNRAS.479..415A}
{Ackermann}, S., {Schawinski}, K., {Zhang}, C., {Weigel}, A.~K., \& {Turp},
  M.~D. 2018, \mnras, 479, 415, \dodoi{10.1093/mnras/sty1398}

\bibitem[{{Aihara} {et~al.}(2018){Aihara}, {Arimoto}, {Armstrong}, {Arnouts},
  {Bahcall}, {Bickerton}, {Bosch}, {Bundy}, {Capak}, {Chan}, {Chiba}, {Coupon},
  {Egami}, {Enoki}, {Finet}, {Fujimori}, {Fujimoto}, {Furusawa}, {Furusawa},
  {Goto}, {Goulding}, {Greco}, {Greene}, {Gunn}, {Hamana}, {Harikane},
  {Hashimoto}, {Hattori}, {Hayashi}, {Hayashi}, {He{\l}miniak}, {Higuchi},
  {Hikage}, {Ho}, {Hsieh}, {Huang}, {Huang}, {Ikeda}, {Imanishi}, {Inoue},
  {Iwasawa}, {Iwata}, {Jaelani}, {Jian}, {Kamata}, {Karoji}, {Kashikawa},
  {Katayama}, {Kawanomoto}, {Kayo}, {Koda}, {Koike}, {Kojima}, {Komiyama},
  {Konno}, {Koshida}, {Koyama}, {Kusakabe}, {Leauthaud}, {Lee}, {Lin}, {Lin},
  {Lupton}, {Mandelbaum}, {Matsuoka}, {Medezinski}, {Mineo}, {Miyama},
  {Miyatake}, {Miyazaki}, {Momose}, {More}, {More}, {Moritani}, {Moriya},
  {Morokuma}, {Mukae}, {Murata}, {Murayama}, {Nagao}, {Nakata}, {Niida},
  {Niikura}, {Nishizawa}, {Obuchi}, {Oguri}, {Oishi}, {Okabe}, {Okamoto},
  {Okura}, {Ono}, {Onodera}, {Onoue}, {Osato}, {Ouchi}, {Price}, {Pyo}, {Sako},
  {Sawicki}, {Shibuya}, {Shimasaku}, {Shimono}, {Shirasaki}, {Silverman},
  {Simet}, {Speagle}, {Spergel}, {Strauss}, {Sugahara}, {Sugiyama}, {Suto},
  {Suyu}, {Suzuki}, {Tait}, {Takada}, {Takata}, {Tamura}, {Tanaka}, {Tanaka},
  {Tanaka}, {Tanaka}, {Terai}, {Terashima}, {Toba}, {Tominaga}, {Toshikawa},
  {Turner}, {Uchida}, {Uchiyama}, {Umetsu}, {Uraguchi}, {Urata}, {Usuda},
  {Utsumi}, {Wang}, {Wang}, {Wong}, {Yabe}, {Yamada}, {Yamanoi}, {Yasuda},
  {Yeh}, {Yonehara}, \& {Yuma}}]{2018PASJ...70S...4A}
{Aihara}, H., {Arimoto}, N., {Armstrong}, R., {et~al.} 2018, \pasj, 70, S4,
  \dodoi{10.1093/pasj/psx066}

\bibitem[{{Aihara} {et~al.}(2019){Aihara}, {AlSayyad}, {Ando}, {Armstrong},
  {Bosch}, {Egami}, {Furusawa}, {Furusawa}, {Goulding}, {Harikane}, {Hikage},
  {Ho}, {Hsieh}, {Huang}, {Ikeda}, {Imanishi}, {Ito}, {Iwata}, {Jaelani},
  {Kakuma}, {Kawana}, {Kikuta}, {Kobayashi}, {Koike}, {Komiyama}, {Li},
  {Liang}, {Lin}, {Luo}, {Lupton}, {Lust}, {MacArthur}, {Matsuoka}, {Mineo},
  {Miyatake}, {Miyazaki}, {More}, {Murata}, {Namiki}, {Nishizawa}, {Oguri},
  {Okabe}, {Okamoto}, {Okura}, {Ono}, {Onodera}, {Onoue}, {Osato}, {Ouchi},
  {Shibuya}, {Strauss}, {Sugiyama}, {Suto}, {Takada}, {Takagi}, {Takata},
  {Takita}, {Tanaka}, {Terai}, {Toba}, {Uchiyama}, {Utsumi}, {Wang}, {Wang}, \&
  {Yamada}}]{2019PASJ...71..114A}
{Aihara}, H., {AlSayyad}, Y., {Ando}, M., {et~al.} 2019, \pasj, 71, 114,
  \dodoi{10.1093/pasj/psz103}

\bibitem[{{Aihara} {et~al.}(2021){Aihara}, {AlSayyad}, {Ando}, {Armstrong},
  {Bosch}, {Egami}, {Furusawa}, {Furusawa}, {Harasawa}, {Harikane}, {Hsieh},
  {Ikeda}, {Ito}, {Iwata}, {Kodama}, {Koike}, {Kokubo}, {Komiyama}, {Li},
  {Liang}, {Lin}, {Lupton}, {Lust}, {MacArthur}, {Mawatari}, {Mineo},
  {Miyatake}, {Miyazaki}, {More}, {Morishima}, {Murayama}, {Nakajima},
  {Nakata}, {Nishizawa}, {Oguri}, {Okabe}, {Okura}, {Ono}, {Osato}, {Ouchi},
  {Pan}, {Plazas Malag{\'o}n}, {Price}, {Reed}, {Rykoff}, {Shibuya},
  {Simunovic}, {Strauss}, {Sugimori}, {Suto}, {Suzuki}, {Takada}, {Takagi},
  {Takata}, {Takita}, {Tanaka}, {Tang}, {Taranu}, {Terai}, {Toba}, {Turner},
  {Uchiyama}, {Vijarnwannaluk}, {Waters}, {Yamada}, {Yamamoto}, \&
  {Yamashita}}]{2021arXiv210813045A}
---. 2021, arXiv e-prints, arXiv:2108.13045.
\newblock \doarXiv{2108.13045}

\bibitem[{{Ananna} {et~al.}(2020){Ananna}, {Urry}, {Treister}, {Hickox},
  {Shankar}, {Ricci}, {Cappelluti}, {Marchesi}, \&
  {Turner}}]{2020ApJ...903...85A}
{Ananna}, T.~T., {Urry}, C.~M., {Treister}, E., {et~al.} 2020, \apj, 903, 85,
  \dodoi{10.3847/1538-4357/abb815}

\bibitem[{{Astropy Collaboration} {et~al.}(2018){Astropy Collaboration},
  {Price-Whelan}, {Sip{\H{o}}cz}, {G{\"u}nther}, {Lim}, {Crawford}, {Conseil},
  {Shupe}, {Craig}, {Dencheva}, {Ginsburg}, {VanderPlas}, {Bradley},
  {P{\'e}rez-Su{\'a}rez}, {de Val-Borro}, {Aldcroft}, {Cruz}, {Robitaille},
  {Tollerud}, {Ardelean}, {Babej}, {Bach}, {Bachetti}, {Bakanov}, {Bamford},
  {Barentsen}, {Barmby}, {Baumbach}, {Berry}, {Biscani}, {Boquien}, {Bostroem},
  {Bouma}, {Brammer}, {Bray}, {Breytenbach}, {Buddelmeijer}, {Burke},
  {Calderone}, {Cano Rodr{\'\i}guez}, {Cara}, {Cardoso}, {Cheedella}, {Copin},
  {Corrales}, {Crichton}, {D'Avella}, {Deil}, {Depagne}, {Dietrich}, {Donath},
  {Droettboom}, {Earl}, {Erben}, {Fabbro}, {Ferreira}, {Finethy}, {Fox},
  {Garrison}, {Gibbons}, {Goldstein}, {Gommers}, {Greco}, {Greenfield},
  {Groener}, {Grollier}, {Hagen}, {Hirst}, {Homeier}, {Horton}, {Hosseinzadeh},
  {Hu}, {Hunkeler}, {Ivezi{\'c}}, {Jain}, {Jenness}, {Kanarek}, {Kendrew},
  {Kern}, {Kerzendorf}, {Khvalko}, {King}, {Kirkby}, {Kulkarni}, {Kumar},
  {Lee}, {Lenz}, {Littlefair}, {Ma}, {Macleod}, {Mastropietro}, {McCully},
  {Montagnac}, {Morris}, {Mueller}, {Mumford}, {Muna}, {Murphy}, {Nelson},
  {Nguyen}, {Ninan}, {N{\"o}the}, {Ogaz}, {Oh}, {Parejko}, {Parley}, {Pascual},
  {Patil}, {Patil}, {Plunkett}, {Prochaska}, {Rastogi}, {Reddy Janga},
  {Sabater}, {Sakurikar}, {Seifert}, {Sherbert}, {Sherwood-Taylor}, {Shih},
  {Sick}, {Silbiger}, {Singanamalla}, {Singer}, {Sladen}, {Sooley},
  {Sornarajah}, {Streicher}, {Teuben}, {Thomas}, {Tremblay}, {Turner},
  {Terr{\'o}n}, {van Kerkwijk}, {de la Vega}, {Watkins}, {Weaver}, {Whitmore},
  {Woillez}, {Zabalza}, \& {Astropy Contributors}}]{2018AJ....156..123A}
{Astropy Collaboration}, {Price-Whelan}, A.~M., {Sip{\H{o}}cz}, B.~M., {et~al.}
  2018, \aj, 156, 123, \dodoi{10.3847/1538-3881/aabc4f}

\bibitem[{{Barchi} {et~al.}(2020){Barchi}, {de Carvalho}, {Rosa}, {Sautter},
  {Soares-Santos}, {Marques}, {Clua}, {Gon{\c{c}}alves}, {de S{\'a}-Freitas},
  \& {Moura}}]{2020A&C....3000334B}
{Barchi}, P.~H., {de Carvalho}, R.~R., {Rosa}, R.~R., {et~al.} 2020, Astronomy
  and Computing, 30, 100334, \dodoi{10.1016/j.ascom.2019.100334}

\bibitem[{{Barden} {et~al.}(2012){Barden}, {H{\"a}u{\ss}ler}, {Peng},
  {McIntosh}, \& {Guo}}]{2012MNRAS.422..449B}
{Barden}, M., {H{\"a}u{\ss}ler}, B., {Peng}, C.~Y., {McIntosh}, D.~H., \&
  {Guo}, Y. 2012, \mnras, 422, 449, \dodoi{10.1111/j.1365-2966.2012.20619.x}

\bibitem[{Binney \& Merrifield(1998)}]{binney_and_merrifield}
Binney, J., \& Merrifield, M. 1998, {Galactic astronomy} (Princeton University
  Press), 796.
\newblock \url{https://press.princeton.edu/titles/6358.html}

\bibitem[{{Brown} {et~al.}(2019){Brown}, {Duncan}, {Landt}, {Kirk}, {Ricci},
  {Kamraj}, {Salvato}, \& {Ananna}}]{2019MNRAS.489.3351B}
{Brown}, M.~J.~I., {Duncan}, K.~J., {Landt}, H., {et~al.} 2019, \mnras, 489,
  3351, \dodoi{10.1093/mnras/stz2324}

\bibitem[{{Chen}(2017)}]{2017arXiv170403924C}
{Chen}, Y.-C. 2017, arXiv e-prints, arXiv:1704.03924.
\newblock \doarXiv{1704.03924}

\bibitem[{{Cheng} {et~al.}(2021{\natexlab{a}}){Cheng}, {Huertas-Company},
  {Conselice}, {Arag{\'o}n-Salamanca}, {Robertson}, \&
  {Ramachandra}}]{2021MNRAS.503.4446C}
{Cheng}, T.-Y., {Huertas-Company}, M., {Conselice}, C.~J., {et~al.}
  2021{\natexlab{a}}, \mnras, 503, 4446, \dodoi{10.1093/mnras/stab734}

\bibitem[{{Cheng} {et~al.}(2021{\natexlab{b}}){Cheng}, {Conselice},
  {Arag{\'o}n-Salamanca}, {Aguena}, {Allam}, {Andrade-Oliveira}, {Annis},
  {Bluck}, {Brooks}, {Burke}, {Carrasco Kind}, {Carretero}, {Choi}, {Costanzi},
  {da Costa}, {Pereira}, {De Vicente}, {Diehl}, {Drlica-Wagner}, {Eckert},
  {Everett}, {Evrard}, {Ferrero}, {Fosalba}, {Frieman}, {Garc{\'\i}a-Bellido},
  {Gerdes}, {Giannantonio}, {Gruen}, {Gruendl}, {Gschwend}, {Gutierrez},
  {Hinton}, {Hollowood}, {Honscheid}, {James}, {Krause}, {Kuehn}, {Kuropatkin},
  {Lahav}, {Maia}, {March}, {Menanteau}, {Miquel}, {Morgan},
  {Paz-Chinch{\'o}n}, {Pieres}, {Plazas Malag{\'o}n}, {Roodman}, {Sanchez},
  {Scarpine}, {Serrano}, {Sevilla-Noarbe}, {Smith}, {Soares-Santos}, {Suchyta},
  {Swanson}, {Tarle}, {Thomas}, \& {To}}]{2021MNRAS.507.4425C}
{Cheng}, T.-Y., {Conselice}, C.~J., {Arag{\'o}n-Salamanca}, A., {et~al.}
  2021{\natexlab{b}}, \mnras, 507, 4425, \dodoi{10.1093/mnras/stab2142}

\bibitem[{{Cho} {et~al.}(2015){Cho}, {Lee}, {Shin}, {Choy}, \&
  {Do}}]{2015arXiv151106348C}
{Cho}, J., {Lee}, K., {Shin}, E., {Choy}, G., \& {Do}, S. 2015, arXiv e-prints,
  arXiv:1511.06348.
\newblock \doarXiv{1511.06348}

\bibitem[{{Cisternas} {et~al.}(2011){Cisternas}, {Jahnke}, {Bongiorno},
  {Inskip}, {Impey}, {Koekemoer}, {Merloni}, {Salvato}, \&
  {Trump}}]{2011ApJ...741L..11C}
{Cisternas}, M., {Jahnke}, K., {Bongiorno}, A., {et~al.} 2011, \apjl, 741, L11,
  \dodoi{10.1088/2041-8205/741/1/L11}

\bibitem[{{Croton} {et~al.}(2006){Croton}, {Springel}, {White}, {De Lucia},
  {Frenk}, {Gao}, {Jenkins}, {Kauffmann}, {Navarro}, \&
  {Yoshida}}]{2006MNRAS.365...11C}
{Croton}, D.~J., {Springel}, V., {White}, S. D.~M., {et~al.} 2006, \mnras, 365,
  11, \dodoi{10.1111/j.1365-2966.2005.09675.x}

\bibitem[{{Di Matteo} {et~al.}(2005){Di Matteo}, {Springel}, \&
  {Hernquist}}]{2005Natur.433..604D}
{Di Matteo}, T., {Springel}, V., \& {Hernquist}, L. 2005, \nat, 433, 604,
  \dodoi{10.1038/nature03335}

\bibitem[{{Dieleman} {et~al.}(2015){Dieleman}, {Willett}, \&
  {Dambre}}]{2015MNRAS.450.1441D}
{Dieleman}, S., {Willett}, K.~W., \& {Dambre}, J. 2015, \mnras, 450, 1441,
  \dodoi{10.1093/mnras/stv632}

\bibitem[{{Dimauro} {et~al.}(2018){Dimauro}, {Huertas-Company}, {Daddi},
  {P{\'e}rez-Gonz{\'a}lez}, {Bernardi}, {Barro}, {Buitrago}, {Caro},
  {Cattaneo}, {Dominguez-S{\'a}nchez}, {Faber}, {H{\"a}u{\ss}ler}, {Kocevski},
  {Koekemoer}, {Koo}, {Lee}, {Mei}, {Margalef-Bentabol}, {Primack},
  {Rodriguez-Puebla}, {Salvato}, {Shankar}, \&
  {Tuccillo}}]{2018MNRAS.478.5410D}
{Dimauro}, P., {Huertas-Company}, M., {Daddi}, E., {et~al.} 2018, \mnras, 478,
  5410, \dodoi{10.1093/mnras/sty1379}

\bibitem[{{Dom{\'\i}nguez S{\'a}nchez} {et~al.}(2018){Dom{\'\i}nguez
  S{\'a}nchez}, {Huertas-Company}, {Bernardi}, {Tuccillo}, \&
  {Fischer}}]{2018MNRAS.476.3661D}
{Dom{\'\i}nguez S{\'a}nchez}, H., {Huertas-Company}, M., {Bernardi}, M.,
  {Tuccillo}, D., \& {Fischer}, J.~L. 2018, \mnras, 476, 3661,
  \dodoi{10.1093/mnras/sty338}

\bibitem[{{Dom{\'\i}nguez S{\'a}nchez} {et~al.}(2019){Dom{\'\i}nguez
  S{\'a}nchez}, {Huertas-Company}, {Bernardi}, {Kaviraj}, {Fischer}, {Abbott},
  {Abdalla}, {Annis}, {Avila}, {Brooks}, {Buckley-Geer}, {Carnero Rosell},
  {Carrasco Kind}, {Carretero}, {Cunha}, {D'Andrea}, {da Costa}, {Davis}, {De
  Vicente}, {Doel}, {Evrard}, {Fosalba}, {Frieman}, {Garc{\'\i}a-Bellido},
  {Gaztanaga}, {Gerdes}, {Gruen}, {Gruendl}, {Gschwend}, {Gutierrez},
  {Hartley}, {Hollowood}, {Honscheid}, {Hoyle}, {James}, {Kuehn}, {Kuropatkin},
  {Lahav}, {Maia}, {March}, {Melchior}, {Menanteau}, {Miquel}, {Nord},
  {Plazas}, {Sanchez}, {Scarpine}, {Schindler}, {Schubnell}, {Smith}, {Smith},
  {Soares-Santos}, {Sobreira}, {Suchyta}, {Swanson}, {Tarle}, {Thomas},
  {Walker}, \& {Zuntz}}]{2019MNRAS.484...93D}
{Dom{\'\i}nguez S{\'a}nchez}, H., {Huertas-Company}, M., {Bernardi}, M.,
  {et~al.} 2019, \mnras, 484, 93, \dodoi{10.1093/mnras/sty3497}

\bibitem[{{Ellison} {et~al.}(2019){Ellison}, {Viswanathan}, {Patton},
  {Bottrell}, {McConnachie}, {Gwyn}, \& {Cuillandre}}]{2019MNRAS.487.2491E}
{Ellison}, S.~L., {Viswanathan}, A., {Patton}, D.~R., {et~al.} 2019, \mnras,
  487, 2491, \dodoi{10.1093/mnras/stz1431}

\bibitem[{{Fabian}(2012)}]{2012ARA&A..50..455F}
{Fabian}, A.~C. 2012, \araa, 50, 455,
  \dodoi{10.1146/annurev-astro-081811-125521}

\bibitem[{Fabian {et~al.}(2006)Fabian, Celotti, \&
  Erlund}]{10.1111/j.1745-3933.2006.00234.x}
Fabian, A.~C., Celotti, A., \& Erlund, M.~C. 2006, Monthly Notices of the Royal
  Astronomical Society: Letters, 373, L16,
  \dodoi{10.1111/j.1745-3933.2006.00234.x}

\bibitem[{Fabian {et~al.}(2008)Fabian, Vasudevan, \&
  Gandhi}]{10.1111/j.1745-3933.2008.00430.x}
Fabian, A.~C., Vasudevan, R.~V., \& Gandhi, P. 2008, Monthly Notices of the
  Royal Astronomical Society: Letters, 385, L43,
  \dodoi{10.1111/j.1745-3933.2008.00430.x}

\bibitem[{{Ferrarese} \& {Merritt}(2000)}]{2000ApJ...539L...9F}
{Ferrarese}, L., \& {Merritt}, D. 2000, \apjl, 539, L9, \dodoi{10.1086/312838}

\bibitem[{Gabor {et~al.}(2009)Gabor, Impey, Jahnke, Simmons, Trump, Koekemoer,
  Brusa, Cappelluti, Schinnerer, Smolčić, \& et~al.}]{Gabor_2009}
Gabor, J.~M., Impey, C.~D., Jahnke, K., {et~al.} 2009, \apj, 691, 705–722,
  \dodoi{10.1088/0004-637x/691/1/705}

\bibitem[{{Gebhardt} {et~al.}(2000){Gebhardt}, {Bender}, {Bower}, {Dressler},
  {Faber}, {Filippenko}, {Green}, {Grillmair}, {Ho}, {Kormendy}, {Lauer},
  {Magorrian}, {Pinkney}, {Richstone}, \& {Tremaine}}]{2000ApJ...539L..13G}
{Gebhardt}, K., {Bender}, R., {Bower}, G., {et~al.} 2000, \apjl, 539, L13,
  \dodoi{10.1086/312840}

\bibitem[{{Ghosh} {et~al.}(2020){Ghosh}, {Urry}, {Wang}, {Schawinski}, {Turp},
  \& {Powell}}]{2020ApJ...895..112G}
{Ghosh}, A., {Urry}, C.~M., {Wang}, Z., {et~al.} 2020, \apj, 895, 112,
  \dodoi{10.3847/1538-4357/ab8a47}

\bibitem[{{Ghosh} {et~al.}(2022){Ghosh}, {Urry}, {Rau}, {Perreault-Levasseur},
  {Cranmer}, {Schawinski}, {Stark}, {Tian}, {Ofman}, {Ananna}, {Auge},
  {Cappelluti}, {Sanders}, \& {Treister}}]{2022ApJ...935..138G}
{Ghosh}, A., {Urry}, C.~M., {Rau}, A., {et~al.} 2022, \apj, 935, 138,
  \dodoi{10.3847/1538-4357/ac7f9e}

\bibitem[{{Glikman} {et~al.}(2015){Glikman}, {Simmons}, {Mailly}, {Schawinski},
  {Urry}, \& {Lacy}}]{2015ApJ...806..218G}
{Glikman}, E., {Simmons}, B., {Mailly}, M., {et~al.} 2015, \apj, 806, 218,
  \dodoi{10.1088/0004-637X/806/2/218}

\bibitem[{Goodfellow {et~al.}(2014)Goodfellow, Pouget-Abadie, Mirza, Xu,
  Warde-Farley, Ozair, Courville, \& Bengio}]{goodfellow2014generative}
Goodfellow, I.~J., Pouget-Abadie, J., Mirza, M., {et~al.} 2014, Generative
  Adversarial Networks, \url{https://arxiv.org/abs/1406.2661}.
\newblock \doarXiv{1406.2661}

\bibitem[{{Harris} {et~al.}(2020){Harris}, {Millman}, {van der Walt},
  {Gommers}, {Virtanen}, {Cournapeau}, {Wieser}, {Taylor}, {Berg}, {Smith},
  {Kern}, {Picus}, {Hoyer}, {van Kerkwijk}, {Brett}, {Haldane}, {del R{\'\i}o},
  {Wiebe}, {Peterson}, {G{\'e}rard-Marchant}, {Sheppard}, {Reddy}, {Weckesser},
  {Abbasi}, {Gohlke}, \& {Oliphant}}]{2020Natur.585..357H}
{Harris}, C.~R., {Millman}, K.~J., {van der Walt}, S.~J., {et~al.} 2020, \nat,
  585, 357, \dodoi{10.1038/s41586-020-2649-2}

\bibitem[{{Harrison}(2017)}]{2017NatAs...1E.165H}
{Harrison}, C.~M. 2017, Nature Astronomy, 1, 0165,
  \dodoi{10.1038/s41550-017-0165}

\bibitem[{{Heckman} \& {Best}(2014)}]{2014ARA&A..52..589H}
{Heckman}, T.~M., \& {Best}, P.~N. 2014, \araa, 52, 589,
  \dodoi{10.1146/annurev-astro-081913-035722}

\bibitem[{{Hewlett} {et~al.}(2017){Hewlett}, {Villforth}, {Wild},
  {Mendez-Abreu}, {Pawlik}, \& {Rowlands}}]{2017MNRAS.470..755H}
{Hewlett}, T., {Villforth}, C., {Wild}, V., {et~al.} 2017, \mnras, 470, 755,
  \dodoi{10.1093/mnras/stx997}

\bibitem[{{Hopkins} {et~al.}(2005){Hopkins}, {Hernquist}, {Cox}, {Di Matteo},
  {Martini}, {Robertson}, \& {Springel}}]{2005ApJ...630..705H}
{Hopkins}, P.~F., {Hernquist}, L., {Cox}, T.~J., {et~al.} 2005, \apj, 630, 705,
  \dodoi{10.1086/432438}

\bibitem[{{Hopkins} {et~al.}(2006){Hopkins}, {Hernquist}, {Cox}, {Di Matteo},
  {Robertson}, \& {Springel}}]{2006ApJS..163....1H}
---. 2006, \apjs, 163, 1, \dodoi{10.1086/499298}

\bibitem[{{Hopkins} {et~al.}(2007){Hopkins}, {Richards}, \&
  {Hernquist}}]{2007ApJ...654..731H}
{Hopkins}, P.~F., {Richards}, G.~T., \& {Hernquist}, L. 2007, \apj, 654, 731,
  \dodoi{10.1086/509629}

\bibitem[{{Huertas-Company} {et~al.}(2015){Huertas-Company}, {Gravet},
  {Cabrera-Vives}, {P{\'e}rez-Gonz{\'a}lez}, {Kartaltepe}, {Barro}, {Bernardi},
  {Mei}, {Shankar}, {Dimauro}, {Bell}, {Kocevski}, {Koo}, {Faber}, \&
  {Mcintosh}}]{2015ApJS..221....8H}
{Huertas-Company}, M., {Gravet}, R., {Cabrera-Vives}, G., {et~al.} 2015, \apjs,
  221, 8, \dodoi{10.1088/0067-0049/221/1/8}

\bibitem[{Hunter(2007)}]{Hunter:2007}
Hunter, J.~D. 2007, Computing in Science \& Engineering, 9, 90,
  \dodoi{10.1109/MCSE.2007.55}

\bibitem[{{Kingma} \& {Ba}(2014)}]{2014arXiv1412.6980K}
{Kingma}, D.~P., \& {Ba}, J. 2014, arXiv e-prints, arXiv:1412.6980.
\newblock \doarXiv{1412.6980}

\bibitem[{{Kocevski} {et~al.}(2012){Kocevski}, {Faber}, {Mozena}, {Koekemoer},
  {Nandra}, {Rangel}, {Laird}, {Brusa}, {Wuyts}, {Trump}, {Koo}, {Somerville},
  {Bell}, {Lotz}, {Alexander}, {Bournaud}, {Conselice}, {Dahlen}, {Dekel},
  {Donley}, {Dunlop}, {Finoguenov}, {Georgakakis}, {Giavalisco}, {Guo},
  {Grogin}, {Hathi}, {Juneau}, {Kartaltepe}, {Lucas}, {McGrath}, {McIntosh},
  {Mobasher}, {Robaina}, {Rosario}, {Straughn}, {van der Wel}, \&
  {Villforth}}]{v2012ApJ...744..148K}
{Kocevski}, D.~D., {Faber}, S.~M., {Mozena}, M., {et~al.} 2012, \apj, 744, 148,
  \dodoi{10.1088/0004-637X/744/2/148}

\bibitem[{{Koekemoer} {et~al.}(2011){Koekemoer}, {Faber}, {Ferguson}, {Grogin},
  {Kocevski}, {Koo}, {Lai}, {Lotz}, {Lucas}, {McGrath}, {Ogaz}, {Rajan},
  {Riess}, {Rodney}, {Strolger}, {Casertano}, {Castellano}, {Dahlen},
  {Dickinson}, {Dolch}, {Fontana}, {Giavalisco}, {Grazian}, {Guo}, {Hathi},
  {Huang}, {van der Wel}, {Yan}, {Acquaviva}, {Alexander}, {Almaini}, {Ashby},
  {Barden}, {Bell}, {Bournaud}, {Brown}, {Caputi}, {Cassata}, {Challis},
  {Chary}, {Cheung}, {Cirasuolo}, {Conselice}, {Roshan Cooray}, {Croton},
  {Daddi}, {Dav{\'e}}, {de Mello}, {de Ravel}, {Dekel}, {Donley}, {Dunlop},
  {Dutton}, {Elbaz}, {Fazio}, {Filippenko}, {Finkelstein}, {Frazer}, {Gardner},
  {Garnavich}, {Gawiser}, {Gruetzbauch}, {Hartley}, {H{\"a}ussler},
  {Herrington}, {Hopkins}, {Huang}, {Jha}, {Johnson}, {Kartaltepe},
  {Khostovan}, {Kirshner}, {Lani}, {Lee}, {Li}, {Madau}, {McCarthy},
  {McIntosh}, {McLure}, {McPartland}, {Mobasher}, {Moreira}, {Mortlock},
  {Moustakas}, {Mozena}, {Nandra}, {Newman}, {Nielsen}, {Niemi}, {Noeske},
  {Papovich}, {Pentericci}, {Pope}, {Primack}, {Ravindranath}, {Reddy},
  {Renzini}, {Rix}, {Robaina}, {Rosario}, {Rosati}, {Salimbeni}, {Scarlata},
  {Siana}, {Simard}, {Smidt}, {Snyder}, {Somerville}, {Spinrad}, {Straughn},
  {Telford}, {Teplitz}, {Trump}, {Vargas}, {Villforth}, {Wagner}, {Wandro},
  {Wechsler}, {Weiner}, {Wiklind}, {Wild}, {Wilson}, {Wuyts}, \&
  {Yun}}]{2011ApJS..197...36K}
{Koekemoer}, A.~M., {Faber}, S.~M., {Ferguson}, H.~C., {et~al.} 2011, \apjs,
  197, 36, \dodoi{10.1088/0067-0049/197/2/36}

\bibitem[{{Kormendy} \& {Ho}(2013)}]{2013ARA&A..51..511K}
{Kormendy}, J., \& {Ho}, L.~C. 2013, \araa, 51, 511,
  \dodoi{10.1146/annurev-astro-082708-101811}

\bibitem[{Krizhevsky {et~al.}(2012)Krizhevsky, Sutskever, \& Hinton}]{Imagenet}
Krizhevsky, A., Sutskever, I., \& Hinton, G. 2012, NeurIPS, 25,
  \dodoi{10.1145/3065386}

\bibitem[{Lecun {et~al.}(1998)Lecun, Bottou, Bengio, \& Haffner}]{726791}
Lecun, Y., Bottou, L., Bengio, Y., \& Haffner, P. 1998, Proceedings of the
  IEEE, 86, 2278, \dodoi{10.1109/5.726791}

\bibitem[{{Liu} {et~al.}(2019){Liu}, {Liu}, {Dong}, {Zhou}, {Wang}, {Lu}, \&
  {Yuan}}]{2019ApJS..243...21L}
{Liu}, H.-Y., {Liu}, W.-J., {Dong}, X.-B., {et~al.} 2019, \apjs, 243, 21,
  \dodoi{10.3847/1538-4365/ab298b}

\bibitem[{{Lotz} {et~al.}(2008){Lotz}, {Jonsson}, {Cox}, \&
  {Primack}}]{2008MNRAS.391.1137L}
{Lotz}, J.~M., {Jonsson}, P., {Cox}, T.~J., \& {Primack}, J.~R. 2008, \mnras,
  391, 1137, \dodoi{10.1111/j.1365-2966.2008.14004.x}

\bibitem[{Madau \& Dickinson(2014)}]{doi:10.1146/annurev-astro-081811-125615}
Madau, P., \& Dickinson, M. 2014, Annual Review of Astronomy and Astrophysics,
  52, 415, \dodoi{10.1146/annurev-astro-081811-125615}

\bibitem[{{Madau} {et~al.}(1998){Madau}, {Pozzetti}, \&
  {Dickinson}}]{1998ApJ...498..106M}
{Madau}, P., {Pozzetti}, L., \& {Dickinson}, M. 1998, \apj, 498, 106,
  \dodoi{10.1086/305523}

\bibitem[{{Marian} {et~al.}(2019){Marian}, {Jahnke}, {Mechtley}, {Cohen},
  {Husemann}, {Jones}, {Koekemoer}, {Schulze}, {van der Wel}, {Villforth}, \&
  {Windhorst}}]{2019ApJ...882..141M}
{Marian}, V., {Jahnke}, K., {Mechtley}, M., {et~al.} 2019, \apj, 882, 141,
  \dodoi{10.3847/1538-4357/ab385b}

\bibitem[{Massey(1951)}]{10.2307/2280095}
Massey, F.~J. 1951, Journal of the American Statistical Association, 46, 68.
\newblock \url{http://www.jstor.org/stable/2280095}

\bibitem[{McKinney(2010)}]{pandas_2010}
McKinney, W. 2010, in Proceedings of the 9th Python in Science Conference,
  56--61, \dodoi{10.25080/Majora-92bf1922-00a}

\bibitem[{{Nishizawa} {et~al.}(2020){Nishizawa}, {Hsieh}, {Tanaka}, \&
  {Takata}}]{2020arXiv200301511N}
{Nishizawa}, A.~J., {Hsieh}, B.-C., {Tanaka}, M., \& {Takata}, T. 2020, arXiv
  e-prints, arXiv:2003.01511.
\newblock \doarXiv{2003.01511}

\bibitem[{{Peng} {et~al.}(2002){Peng}, {Ho}, {Impey}, \&
  {Rix}}]{2002AJ....124..266P}
{Peng}, C.~Y., {Ho}, L.~C., {Impey}, C.~D., \& {Rix}, H.-W. 2002, \aj, 124,
  266, \dodoi{10.1086/340952}

\bibitem[{{Powell} {et~al.}(2017){Powell}, {Urry}, {Cardamone}, {Simmons},
  {Schawinski}, {Young}, \& {Kawakatsu}}]{2017ApJ...835...22P}
{Powell}, M.~C., {Urry}, C.~M., {Cardamone}, C.~N., {et~al.} 2017, \apj, 835,
  22, \dodoi{10.3847/1538-4357/835/1/22}

\bibitem[{{Reed} {et~al.}(2016){Reed}, {Akata}, {Yan}, {Logeswaran}, {Schiele},
  \& {Lee}}]{2016arXiv160505396R}
{Reed}, S., {Akata}, Z., {Yan}, X., {et~al.} 2016, arXiv e-prints,
  arXiv:1605.05396.
\newblock \doarXiv{1605.05396}

\bibitem[{{Ricci} {et~al.}(2017){Ricci}, {Bauer}, {Treister}, {Schawinski},
  {Privon}, {Blecha}, {Arevalo}, {Armus}, {Harrison}, {Ho}, {Iwasawa},
  {Sanders}, \& {Stern}}]{2017MNRAS.468.1273R}
{Ricci}, C., {Bauer}, F.~E., {Treister}, E., {et~al.} 2017, \mnras, 468, 1273,
  \dodoi{10.1093/mnras/stx173}

\bibitem[{{Rowe} {et~al.}(2015){Rowe}, {Jarvis}, {Mandelbaum}, {Bernstein},
  {Bosch}, {Simet}, {Meyers}, {Kacprzak}, {Nakajima}, {Zuntz}, {Miyatake},
  {Dietrich}, {Armstrong}, {Melchior}, \& {Gill}}]{2015A&C....10..121R}
{Rowe}, B.~T.~P., {Jarvis}, M., {Mandelbaum}, R., {et~al.} 2015, Astronomy and
  Computing, 10, 121, \dodoi{10.1016/j.ascom.2015.02.002}

\bibitem[{{Sanders} {et~al.}(1988){Sanders}, {Soifer}, {Elias}, {Madore},
  {Matthews}, {Neugebauer}, \& {Scoville}}]{1988ApJ...325...74S}
{Sanders}, D.~B., {Soifer}, B.~T., {Elias}, J.~H., {et~al.} 1988, \apj, 325,
  74, \dodoi{10.1086/165983}

\bibitem[{{Schawinski} {et~al.}(2012){Schawinski}, {Simmons}, {Urry},
  {Treister}, \& {Glikman}}]{2012MNRAS.425L..61S}
{Schawinski}, K., {Simmons}, B.~D., {Urry}, C.~M., {Treister}, E., \&
  {Glikman}, E. 2012, \mnras, 425, L61,
  \dodoi{10.1111/j.1745-3933.2012.01302.x}

\bibitem[{{Schawinski} {et~al.}(2011){Schawinski}, {Treister}, {Urry},
  {Cardamone}, {Simmons}, \& {Yi}}]{2011ApJ...727L..31S}
{Schawinski}, K., {Treister}, E., {Urry}, C.~M., {et~al.} 2011, \apjl, 727,
  L31, \dodoi{10.1088/2041-8205/727/2/L31}

\bibitem[{{Schawinski} {et~al.}(2014){Schawinski}, {Urry}, {Simmons},
  {Fortson}, {Kaviraj}, {Keel}, {Lintott}, {Masters}, {Nichol}, {Sarzi},
  {Skibba}, {Treister}, {Willett}, {Wong}, \& {Yi}}]{2014MNRAS.440..889S}
{Schawinski}, K., {Urry}, C.~M., {Simmons}, B.~D., {et~al.} 2014, \mnras, 440,
  889, \dodoi{10.1093/mnras/stu327}

\bibitem[{{Shankar} {et~al.}(2009){Shankar}, {Weinberg}, \&
  {Miralda-Escud{\'e}}}]{2009ApJ...690...20S}
{Shankar}, F., {Weinberg}, D.~H., \& {Miralda-Escud{\'e}}, J. 2009, \apj, 690,
  20, \dodoi{10.1088/0004-637X/690/1/20}

\bibitem[{{Simard}(1998)}]{1998ASPC..145..108S}
{Simard}, L. 1998, in Astronomical Society of the Pacific Conference Series,
  Vol. 145, Astronomical Data Analysis Software and Systems VII, ed.
  R.~{Albrecht}, R.~N. {Hook}, \& H.~A. {Bushouse}, 108

\bibitem[{{Simard} {et~al.}(2011){Simard}, {Mendel}, {Patton}, {Ellison}, \&
  {McConnachie}}]{2011ApJS..196...11S}
{Simard}, L., {Mendel}, J.~T., {Patton}, D.~R., {Ellison}, S.~L., \&
  {McConnachie}, A.~W. 2011, \apjs, 196, 11, \dodoi{10.1088/0067-0049/196/1/11}

\bibitem[{{Simmons} \& {Urry}(2008)}]{2008ApJ...683..644S}
{Simmons}, B.~D., \& {Urry}, C.~M. 2008, \apj, 683, 644, \dodoi{10.1086/589827}

\bibitem[{{Stark} {et~al.}(2018){Stark}, {Launet}, {Schawinski}, {Zhang},
  {Koss}, {Turp}, {Sartori}, {Zhang}, {Chen}, \&
  {Weigel}}]{2018MNRAS.477.2513S}
{Stark}, D., {Launet}, B., {Schawinski}, K., {et~al.} 2018, \mnras, 477, 2513,
  \dodoi{10.1093/mnras/sty764}

\bibitem[{{Stemo} {et~al.}(2020){Stemo}, {Comerford}, {Barrows}, {Stern},
  {Assef}, \& {Griffith}}]{2020ApJ...888...78S}
{Stemo}, A., {Comerford}, J.~M., {Barrows}, R.~S., {et~al.} 2020, \apj, 888,
  78, \dodoi{10.3847/1538-4357/ab5f66}

\bibitem[{{Stewart} {et~al.}(2008){Stewart}, {Bullock}, {Wechsler}, {Maller},
  \& {Zentner}}]{2008ApJ...683..597S}
{Stewart}, K.~R., {Bullock}, J.~S., {Wechsler}, R.~H., {Maller}, A.~H., \&
  {Zentner}, A.~R. 2008, \apj, 683, 597, \dodoi{10.1086/588579}

\bibitem[{{Taylor}(2005)}]{2005ASPC..347...29T}
{Taylor}, M.~B. 2005, in Astronomical Society of the Pacific Conference Series,
  Vol. 347, Astronomical Data Analysis Software and Systems XIV, ed.
  P.~{Shopbell}, M.~{Britton}, \& R.~{Ebert}, 29

\bibitem[{{Treister} {et~al.}(2012){Treister}, {Schawinski}, {Urry}, \&
  {Simmons}}]{2012ApJ...758L..39T}
{Treister}, E., {Schawinski}, K., {Urry}, C.~M., \& {Simmons}, B.~D. 2012,
  \apjl, 758, L39, \dodoi{10.1088/2041-8205/758/2/L39}

\bibitem[{{Tuccillo} {et~al.}(2018){Tuccillo}, {Huertas-Company},
  {Decenci{\`e}re}, {Velasco-Forero}, {Dom{\'\i}nguez S{\'a}nchez}, \&
  {Dimauro}}]{2018MNRAS.475..894T}
{Tuccillo}, D., {Huertas-Company}, M., {Decenci{\`e}re}, E., {et~al.} 2018,
  \mnras, 475, 894, \dodoi{10.1093/mnras/stx3186}

\bibitem[{{Ueda} {et~al.}(2014){Ueda}, {Akiyama}, {Hasinger}, {Miyaji}, \&
  {Watson}}]{2014ApJ...786..104U}
{Ueda}, Y., {Akiyama}, M., {Hasinger}, G., {Miyaji}, T., \& {Watson}, M.~G.
  2014, \apj, 786, 104, \dodoi{10.1088/0004-637X/786/2/104}

\bibitem[{{Urrutia} {et~al.}(2008){Urrutia}, {Lacy}, \&
  {Becker}}]{2008ApJ...674...80U}
{Urrutia}, T., {Lacy}, M., \& {Becker}, R.~H. 2008, \apj, 674, 80,
  \dodoi{10.1086/523959}

\bibitem[{Variawa {et~al.}(2022)Variawa, Van~Zyl, \& Woolway}]{9709840}
Variawa, M.~Z., Van~Zyl, T.~L., \& Woolway, M. 2022, IEEE Access, 10, 19539,
  \dodoi{10.1109/ACCESS.2022.3150881}

\bibitem[{{Vega-Ferrero} {et~al.}(2021){Vega-Ferrero}, {Dom{\'\i}nguez
  S{\'a}nchez}, {Bernardi}, {Huertas-Company}, {Morgan}, {Margalef}, {Aguena},
  {Allam}, {Annis}, {Avila}, {Bacon}, {Bertin}, {Brooks}, {Carnero Rosell},
  {Carrasco Kind}, {Carretero}, {Choi}, {Conselice}, {Costanzi}, {da Costa},
  {Pereira}, {De Vicente}, {Desai}, {Ferrero}, {Fosalba}, {Frieman},
  {Garc{\'\i}a-Bellido}, {Gruen}, {Gruendl}, {Gschwend}, {Gutierrez},
  {Hartley}, {Hinton}, {Hollowood}, {Honscheid}, {Hoyle}, {Jarvis}, {Kim},
  {Kuehn}, {Kuropatkin}, {Lima}, {Maia}, {Menanteau}, {Miquel}, {Ogando},
  {Palmese}, {Paz-Chinch{\'o}n}, {Plazas}, {Romer}, {Sanchez}, {Scarpine},
  {Schubnell}, {Serrano}, {Sevilla-Noarbe}, {Smith}, {Suchyta}, {Swanson},
  {Tarle}, {Tarsitano}, {To}, {Tucker}, {Varga}, \&
  {Wilkinson}}]{2021MNRAS.506.1927V}
{Vega-Ferrero}, J., {Dom{\'\i}nguez S{\'a}nchez}, H., {Bernardi}, M., {et~al.}
  2021, \mnras, 506, 1927, \dodoi{10.1093/mnras/stab594}

\bibitem[{Villforth {et~al.}(2014)Villforth, Hamann, Rosario, Santini, McGrath,
  Wel, Chang, Guo, Dahlen, Bell, Conselice, Croton, Dekel, Faber, Grogin,
  Hamilton, Hopkins, Juneau, Kartaltepe, Kocevski, Koekemoer, Koo, Lotz,
  McIntosh, Mozena, Somerville, \& Wild}]{10.1093/mnras/stu173}
Villforth, C., Hamann, F., Rosario, D.~J., {et~al.} 2014, \mnras, 439, 3342,
  \dodoi{10.1093/mnras/stu173}

\bibitem[{Virtanen {et~al.}(2020)Virtanen, Gommers, Oliphant, Haberland, Reddy,
  Cournapeau, Burovski, Peterson, Weckesser, Bright, {van der Walt}, Brett,
  Wilson, Millman, Mayorov, Nelson, Jones, Kern, Larson, Carey, Polat, Feng,
  Moore, {VanderPlas}, Laxalde, Perktold, Cimrman, Henriksen, Quintero, Harris,
  Archibald, Ribeiro, Pedregosa, {van Mulbregt}, \& {SciPy 1.0
  Contributors}}]{2020SciPy-NMeth}
Virtanen, P., Gommers, R., Oliphant, T.~E., {et~al.} 2020, Nature Methods, 17,
  261, \dodoi{10.1038/s41592-019-0686-2}

\bibitem[{{Volonteri} {et~al.}(2003){Volonteri}, {Haardt}, \&
  {Madau}}]{2003ApJ...582..559V}
{Volonteri}, M., {Haardt}, F., \& {Madau}, P. 2003, \apj, 582, 559,
  \dodoi{10.1086/344675}

\bibitem[{{Walmsley} {et~al.}(2020){Walmsley}, {Smith}, {Lintott}, {Gal},
  {Bamford}, {Dickinson}, {Fortson}, {Kruk}, {Masters}, {Scarlata}, {Simmons},
  {Smethurst}, \& {Wright}}]{2020MNRAS.491.1554W}
{Walmsley}, M., {Smith}, L., {Lintott}, C., {et~al.} 2020, \mnras, 491, 1554,
  \dodoi{10.1093/mnras/stz2816}

\bibitem[{{Wilman} {et~al.}(2013){Wilman}, {Fontanot}, {De Lucia}, {Erwin}, \&
  {Monaco}}]{2013MNRAS.433.2986W}
{Wilman}, D.~J., {Fontanot}, F., {De Lucia}, G., {Erwin}, P., \& {Monaco}, P.
  2013, \mnras, 433, 2986, \dodoi{10.1093/mnras/stt941}

\bibitem[{{York} {et~al.}(2000){York}, {Adelman}, {Anderson}, {Anderson},
  {Annis}, {Bahcall}, {Bakken}, {Barkhouser}, {Bastian}, {Berman}, {Boroski},
  {Bracker}, {Briegel}, {Briggs}, {Brinkmann}, {Brunner}, {Burles}, {Carey},
  {Carr}, {Castander}, {Chen}, {Colestock}, {Connolly}, {Crocker}, {Csabai},
  {Czarapata}, {Davis}, {Doi}, {Dombeck}, {Eisenstein}, {Ellman}, {Elms},
  {Evans}, {Fan}, {Federwitz}, {Fiscelli}, {Friedman}, {Frieman}, {Fukugita},
  {Gillespie}, {Gunn}, {Gurbani}, {de Haas}, {Haldeman}, {Harris}, {Hayes},
  {Heckman}, {Hennessy}, {Hindsley}, {Holm}, {Holmgren}, {Huang}, {Hull},
  {Husby}, {Ichikawa}, {Ichikawa}, {Ivezi{\'c}}, {Kent}, {Kim}, {Kinney},
  {Klaene}, {Kleinman}, {Kleinman}, {Knapp}, {Korienek}, {Kron}, {Kunszt},
  {Lamb}, {Lee}, {Leger}, {Limmongkol}, {Lindenmeyer}, {Long}, {Loomis},
  {Loveday}, {Lucinio}, {Lupton}, {MacKinnon}, {Mannery}, {Mantsch}, {Margon},
  {McGehee}, {McKay}, {Meiksin}, {Merelli}, {Monet}, {Munn}, {Narayanan},
  {Nash}, {Neilsen}, {Neswold}, {Newberg}, {Nichol}, {Nicinski}, {Nonino},
  {Okada}, {Okamura}, {Ostriker}, {Owen}, {Pauls}, {Peoples}, {Peterson},
  {Petravick}, {Pier}, {Pope}, {Pordes}, {Prosapio}, {Rechenmacher}, {Quinn},
  {Richards}, {Richmond}, {Rivetta}, {Rockosi}, {Ruthmansdorfer}, {Sandford},
  {Schlegel}, {Schneider}, {Sekiguchi}, {Sergey}, {Shimasaku}, {Siegmund},
  {Smee}, {Smith}, {Snedden}, {Stone}, {Stoughton}, {Strauss}, {Stubbs},
  {SubbaRao}, {Szalay}, {Szapudi}, {Szokoly}, {Thakar}, {Tremonti}, {Tucker},
  {Uomoto}, {Vanden Berk}, {Vogeley}, {Waddell}, {Wang}, {Watanabe},
  {Weinberg}, {Yanny}, {Yasuda}, \& {SDSS Collaboration}}]{2000AJ....120.1579Y}
{York}, D.~G., {Adelman}, J., {Anderson}, John~E., J., {et~al.} 2000, \aj, 120,
  1579, \dodoi{10.1086/301513}

\end{thebibliography}
\bibliographystyle{aa_url}

\end{CJK*}
\end{document}